\documentclass[aps,prd,onecolumn,preprintnumbers,floatfix,showpacs,
nofootinbib,superscriptaddress,notitlepage]{revtex4-1}

\usepackage{amsmath} 
\usepackage{amsfonts} 
\usepackage{slashed}
\usepackage{physics} 
\usepackage{graphicx} 
\usepackage{xcolor} 
\usepackage{subcaption}
\usepackage[justification=raggedright,singlelinecheck=false]{caption}
\usepackage{comment}

\usepackage{hyperref} 

\newcommand{\Wc}{\mathcal{W}}
\newcommand{\Mc}{\mathcal{M}}
\newcommand{\Yc}{\mathcal{Y}}
\newcommand{\Kc}{\mathcal{K}}
\newcommand{\Jc}{\mathcal{J}}
\newcommand{\Ic}{\mathcal{I}}
\newcommand{\Rc}{\mathcal{R}}
\newcommand{\Lc}{\mathcal{L}}
\newcommand{\Gc}{\mathcal{G}}
\newcommand{\Cc}{\mathcal{C}}
\newcommand{\Bc}{\mathcal{B}}

\newcommand{\Ec}{\mathcal{E}}
\newcommand{\Tc}{\mathcal{T}}
\newcommand{\Uc}{\mathcal{U}}
\newcommand{\Ac}{\mathcal{A}}
\newcommand{\df}{{\rm df}}

\usepackage[normalem]{ulem}

\newcommand{\Wb}{\mathbf{W}}

\definecolor{jlab_red}{RGB}{192,39,45}
\definecolor{jlab_orange}{RGB}{249,102,0}
\definecolor{jlab_blue}{RGB}{47,122,121}
\definecolor{jlab_green}{RGB}{65,125,10}

\newcommand{\addReviewer}[2]{
  \expandafter\newcommand\csname #1\endcsname[1]{{\sf \color{#2} {#1}:\,##1}}
  \expandafter\newcommand\csname #1cor\endcsname[2]{{\color{#2} {#1}:\,\st{##1}{\sf ##2}}}
  \expandafter\newcommand\csname #1color\endcsname{#2}
}

\addReviewer{rb}{jlab_blue}
\addReviewer{nl}{orange}
\usepackage{orcidlink}

\newcommand{\ucb}{Department of Physics, 
University of California, 
Berkeley, CA 94720, USA}   
\newcommand{\lbnl}{Nuclear Science Division, 
Lawrence Berkeley National Laboratory, Berkeley, 
CA 94720, USA}
\newcommand{\bochum}{Institut f\"ur Theoretische Physik II, Fakult\"at f\"ur Physik und Astronomie, Ruhr-Universität Bochum, 44780 Bochum, Germany}
\newcommand{\ific}{Instituto de F\'{i}sica Corpuscular (IFIC), CSIC‐Universitat de Val\'{e}ncia, 46071 Val\'{e}ncia, Spain}

\begin{document}

\title{
Analytic decomposition of
two-body electroweak processes with left-hand cuts
}

\author{Andr\'e Bai\~ao-Raposo~\orcidlink{0009-0006-5692-2671}}
\email[e-mail: ]{andre.baiaoraposo@ruhr-uni-bochum.de}
\affiliation{\bochum}

\author{Ra\'ul A. Brice\~no~\orcidlink{0000-0003-1109-1473}}
\email[e-mail: ]{rbriceno@berkeley.edu}
\affiliation{\ucb}
\affiliation{\lbnl}

\author{Nicolas~Lang~\orcidlink{0000-0003-2344-4838}}
\email[e-mail: ]{nicolas.lang@ific.uv.es}
\affiliation{\ific}

\author{Felipe G. Ortega-Gama~\orcidlink{0000-0001-8453-9481}}
\email[e-mail: ]{fgortegagama@berkeley.edu}
\affiliation{\ucb}
\affiliation{\lbnl}

\author{Bianca Pol~\orcidlink{0000-0002-9015-8965}}
\email[e-mail: ]{bpol@berkeley.edu}
\affiliation{\ucb}
\affiliation{\lbnl}

\date{\today}

 \begin{abstract}

We derive an on-shell representation for electroweak $2+\Jc\to2$ transition amplitudes involving two hadrons in both the initial and final state for systems where there are left-hand singularities generated by light-particle exchange, e.g. one-pion exchange. The derivation treats the insertion of the electroweak current perturbatively, keeping only the leading-order contribution in the external field, while the hadronic interactions are treated to all orders, including one-particle exchange effects. We find that, in such processes, the amplitude can have logarithmic singularities, as well as one-particle pole singularities, due to these exchanges. We isolate those contributions, as well as previously identified triangle singularities. The result is expressed in terms of the purely hadronic amplitude, exchange-current kernels, triangle functions, and a class of short-distance transition functions that are real and smooth below unaccounted-for thresholds. While we consider only transitions with spinless particles, this work is an important step toward constraining form factors of systems involving nucleons or vector mesons in the heavy quark sector, where the one-pion-exchange singularity is quite close to threshold.

\end{abstract}

\maketitle

\section{Introduction}
One can view theoretical hadron spectroscopy as being driven by four increasingly refined questions:
\begin{enumerate}
\item Are the new particle candidates observed in experiments genuine states of quantum chromodynamics (QCD)?
\item If so, what are their internal structures?
\item Given this structural information, what is the nature of these states?\footnote{Here ``nature'' refers to the dominant long-distance degrees of freedom and binding mechanism, not to an immutable classification of the state.}
\item What general principles of nonperturbative QCD can be inferred from the spectrum?
\end{enumerate}
The advent of lattice QCD is making it possible to address the first of these questions from first principles for an increasingly broad set of systems. Representative examples include the light scalar sector, with the elusive $\sigma$ resonance~\cite{Rodas:2023nec} and the near-threshold $f_0(980)$~\cite{Briceno:2017qmb}, and $a_0(980)$~\cite{Dudek:2016cru}, as well as channels with exotic quantum numbers such as the lightest $\pi_1$~\cite{Woss:2020ayi}.

Nowhere has this program generated more excitement than in the heavy-quark sector~\cite{Lebed:2016hpi, Brambilla:2019esw, Francis:2024fwf}. Over the last two decades, experiments have revealed a rapidly growing spectrum of charmoniumlike and open-charm structures, including candidates with minimal tetraquark and pentaquark quark content. A striking feature of many of these states is their proximity to hadron-hadron thresholds. Since they often appear as enhancements in complicated production and decay processes, it is essential to determine whether they correspond to poles of QCD scattering amplitudes, or simply to kinematic effects associated with thresholds or triangle singularities.

A prominent example of a state lying near threshold, that has been confirmed by multiple lattice QCD calculations~\cite{Padmanath:2022cvl,Whyte:2024ihh,Lyu:2023xro, PitangaLachini:2026lyd, Nagatsuka:2025szy}, is the $T_{cc}$, observed by the LHCb Collaboration~\cite{LHCb:2021auc, LHCb:2021vvq}. This state has the quantum numbers of $cc\overline{u}\overline{d}$, where $u$, $d$, $c$ are the up, down, and charm quarks, and it decays to a three-body final state via $T_{cc}\to D D^\star\to DD\pi$. It lies just below the nearest $DD^\star$ threshold, making it a compelling tetraquark candidate. Yet, this proximity to threshold leaves open the question of whether $T_{cc}$ is best described as a compact tetraquark or as a loosely bound $DD^\star$ molecular state, analogous to the deuteron. 

Given that we have some level of confidence that this state is indeed a resonance of QCD, we can turn to the second question listed above: what is its internal structure? One could always use a preferred model or effective field theory (EFT) to fit lattice QCD and/or experimental data to infer the nature of a given state, as has been done in, for example, Ref.~\cite{Abolnikov:2024key} for the $T_{cc}$. The downside of this is that one's conclusions are only as strong as the model or EFT chosen. Here, instead, we consider the possibility of resolving the internal structure of such states without having to resort to effective models. 

If one could resolve its internal structure, one could begin to rule out certain scenarios. For example, constraints on the charge radius of this state could be discriminating: an unnaturally small radius would favor the compact tetraquark picture, while a large radius would support the molecular interpretation. A study of the analogous bottom-quark system, $T_{bb}$, has recently been performed~\cite{Vujmilovic:2025czt}, finding a charge radius significantly smaller than the combined charge radii of the $B$ and $B^\ast$ mesons, providing evidence in favor of a compact diquark picture for that state. An analogous study of the $T_{cc}$ remains an important open problem.

Generally, constraining the structural information of resonances that decay on timescales of $\mathcal{O}(10^{-23}~\text{s})$ is virtually impossible to do experimentally. However, some structural information may be accessed directly via lattice QCD. A framework for doing so has been developed in Refs.~\cite{Briceno:2015tza, Baroni:2018iau, Briceno:2020vgp}, which introduced a new class of reactions involving the coupling of two-particle scattering states via the insertion of a current that is local in time. These reactions, which we denote as $2+\Jc\to 2$, can provide structural constraints on bound states (deep and shallow) and resonances alike. The current can be an electroweak probe or something more complicated, such as those being used to access the partonic structure of stable hadrons~\cite{Hackett:2023rif,HadStruc:2021qdf,HadStruc:2024rix,Lin:2021brq,Alexandrou:2019ali,Bhattacharya:2023ays,Bhattacharya:2024wtg,Chu:2025kew}. This formalism builds on earlier correspondences derived for simpler reactions of the type $\Jc\to 2$ or $1+\Jc\to 2$~\cite{Lellouch:2000pv,Christ:2005gi,Briceno:2014uqa,Briceno:2015csa,Agadjanov:2016fbd}, which have already been applied in several lattice QCD studies~\cite{Briceno:2016kkp,Alexandrou:2018jbt,Feng:2014gba,Andersen:2018mau,Erben:2019nmx,Radhakrishnan:2022ubg,Ortega-Gama:2024rqx}. Non-trivial checks of the $2+\Jc\to2$ formalism have been performed in Refs.~\cite{Briceno:2019nns,Briceno:2020xxs}, and first steps toward its implementation in lattice calculations have recently been taken~\cite{Moscoso:2026wmz, Wang:2026kab}.

However, the formalism of Refs.~\cite{Briceno:2015tza, Baroni:2018iau, Briceno:2020vgp} is technically not applicable to study the structure of states like $T_{cc}$, even if we restrict our attention to unphysically heavy quark masses where the $D^\star$ is stable and $T_{cc}$ couples only to two-particle states. In fact, it is not even applicable to a state as simple as the deuteron. This is due to two limitations: Firstly, the formalism is at present only applicable to spinless hadrons. This is technically challenging but conceptually straightforward to address. Secondly, and more subtly, these states appear as dynamically generated poles in the vicinity of, if not directly on top of, logarithmic branch cuts commonly known as left-hand cuts. These cuts arise from the exchange of light particles (e.g.\ the pion in QCD) and have introduced complications in the determination of resonance and bound state poles from the finite-volume spectrum in lattice QCD studies, as seen in the case of the $T_{cc}$ \cite{Du:2023hlu}, $H$-dibaryon \cite{Green:2021qol}, and other systems. The treatment of left-hand cuts in purely hadronic $2\to2$ amplitudes has recently been addressed formally in Refs.~\cite{Raposo:2023oru, Raposo:2025dkb, Meng:2023bmz, Bubna:2024izx,  Hansen:2024ffk, Dawid:2024oey, Yu:2025gzg}, with some of these approaches having already been used to re-analyse lattice data \cite{Meng:2024kkp,
Dawid:2024dgy, Abolnikov:2026sik,
Alharazin:2026lno,
Rodas:2026zmm, PitangaLachini:2026lyd}. ~\footnote{A recent application~\cite{PitangaLachini:2026lyd} of the formalism presented in Ref.~\cite{Raposo:2025dkb} showed that in describing the $T_{cc}$ the explicit inclusion of the left-hand cut led to a significant statistical improvement in the description of lattice QCD generated data.} The corresponding problem for $2+\Jc\to2$ amplitudes has not yet been solved, and is the focus of the present work.

We derive an on-shell representation of the $2+\Jc\to 2$ amplitude in the presence of light-particle exchanges, extending the range of applicability of Ref.~\cite{Briceno:2020vgp}. This serves as the first step towards a formalism that can be applied in lattice QCD calculations, as we do not yet attempt to derive a correspondence between these amplitudes and the finite-volume observables determinable via lattice QCD. We leave this for future work. This work can thus be understood as a natural merger of Ref.~\cite{Briceno:2020vgp}, which isolated the analytic structure of $2+\Jc\to 2$ amplitudes, and Ref.~\cite{Raposo:2025dkb}, which isolated the singularities of $2\to 2$ amplitudes in the presence of left-hand cuts.

In this work, we assume that the asymptotic states are composed of spinless particles and that the exchanged particle is also spinless. The current is local but is allowed to take on any Lorentz structure. We make no further assumptions about the states or the current, and in this sense, the formalism is universal. As such, beyond studies of exotic hadrons such as the already-mentioned $T_{cc}$, it may be relevant for future investigations of a wide class of processes such as deuteron breakup $d+\nu \to np$, coherent neutrino-deuteron scattering $\nu d \to \nu d$, or photodisintegration of the deuteron $\gamma^\star d \to np$, as well as long-range contributions to neutrinoless double beta decay \cite{Nicholson:2016byl}. The current insertion is treated at leading order, i.e., perturbatively, while the purely hadronic interactions are treated to all orders in perturbation theory, including all possible exchanges. As in the case of the formalism of Ref.~\cite{Briceno:2020vgp}, where the objects appearing in the on-shell representation of the $2 + \Jc \to 2$ amplitude can be related to EFT-derived one- and two-body currents, one could expect to obtain similar correspondences between the objects in the representation presented in this work and EFT results, e.g.~for the deuteron electromagnetic form factors \cite{Kaplan:1998sz, Park:1995pn, Pastore:2011ip, Kolling:2011mt}.

The rest of this work is organized as follows. In Sec.~\ref{sec:main}, we summarize the final on-shell representation and the ingredients that must be determined. In Sec.~\ref{sec:Mc_review}, we review the purely hadronic construction with explicit one-particle exchange, and in Sec.~\ref{sec:W_review} we recall the OPE-free $2+\Jc\to2$ formalism. In Sec.~\ref{sec:2J2_wOPE}, we provide a detailed derivation of the extension of the $2+\Jc\to2$ formalism in the presence of exchanges.
We describe in Sec.~\ref{sec:triangles_2d} how the azimuthal integrals appearing in triangle-type 3D integrals can be evaluated analytically.
Our result includes the presence of two new classes of diagrams, the most singular of which is associated with the exchange particle coupling to the current.
We discuss in detail the singularities associated with these diagrams. This is supplemented by a discussion of the criteria for singularities according to Landau conditions of the partial-wave projection of these diagrams in Appendix~\ref{app:PWA}. In that same appendix, we present the partial wave projection of the new exchange kernels that appear for the $\ell=0$ case. For the most generic case, where the initial/final states are in different reference frames, we leave our expressions in terms of 2D integrals. In the special case that they are in the same reference frame, we find exact analytic results for these. 

\section{Main result}
\label{sec:main}

\begin{figure}
    \centering
    \includegraphics[width=0.8\linewidth]{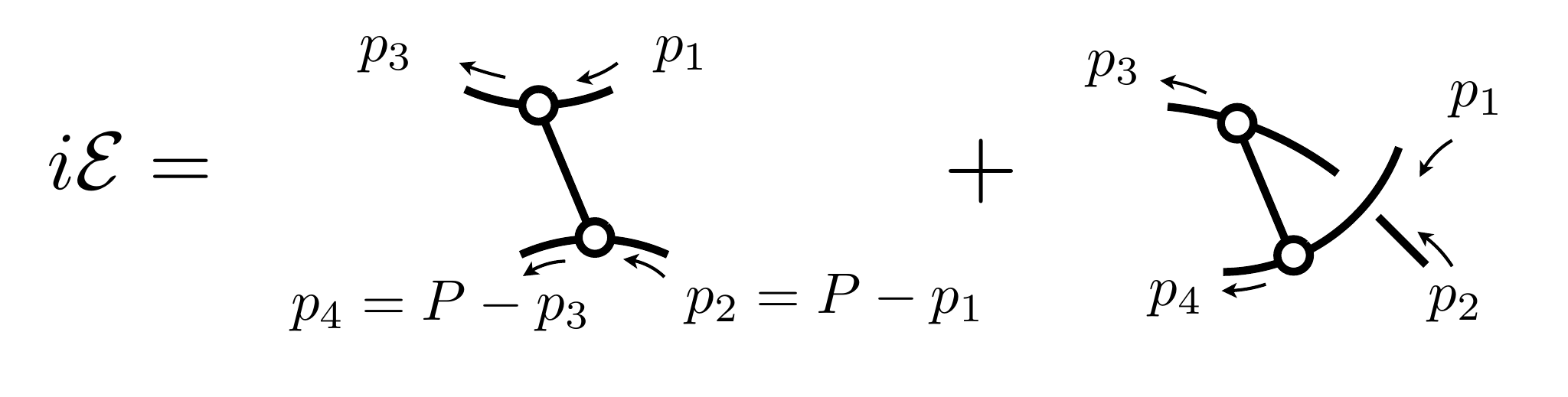}
    \caption{Shown is a diagrammatic definition of the $t$-channel and $u$-channel one-particle exchange diagrams. For a detailed definition of $\Ec$, see Eq.~\eqref{eq:Ec_def}.}
    \label{fig:single-exchange}
\end{figure}

In this work, we consider the insertion of a generic local current,  $\mathcal{J}^A$, where the index $A$ denotes its Lorentz structure. For a vector current, $A=\mu$; for a rank-two tensor current, $A=\mu\nu$; and so forth. The amplitude of interest is the $2+\Jc^A\to 2$ amplitude, which we denote by $\Wc^A$ and define as the fully connected contribution to the following matrix element, 
\begin{equation}
    \Wc^A(P_f, P_i, q) = \mel{P_f,p_3;{\rm out}}{\mathcal{J}^A(0)}{P_i,p_1 ; {\rm in}}_{\text{conn.}}\,,
\end{equation}
where $|P_i, p_1 ;{\rm in}\rangle$ is an incoming asymptotic two-particle state defined by one particle carrying four-momentum $p_1$ and the other $p_2=P_i-p_1$. The outgoing state is similarly defined. Throughout, the initial and final two-particle invariants are denoted by $s_i=P_i^2$ and $s_f=P_f^2$. The subscript ``conn." reminds us that this amplitude only includes the fully connected contribution. Note also that the current insertion has been placed at the spacetime origin~\footnote{This is done to ensure that the matrix element has no additional four-momentum conserving delta function.}.

We consider a kinematic region bounded from above by the first inelastic threshold that is not included explicitly, such as a three- or four-particle threshold, and from below by the first left-hand singularities that are not treated explicitly; for example, two-particle exchange cuts. The singularities retained in the formalism are only those associated with two-particle $s$-channel propagation, as well as one-particle $t$- and $u$-channel exchanges. We collectively refer to the latter category as one-particle exchange (OPE) contributions. To simplify the derivation, the external particles are taken to be spinless with arbitrary masses. The exchanged particle is also taken to be spinless, with mass $m_e$.

As in the case of the previously derived formalism for these amplitudes~\cite{Briceno:2020vgp}, which we review in Sec.~\ref{sec:W_review}, the full amplitude can be written as the sum over two terms,
\begin{equation}
    i\Wc^A = i\Wc_{\rm pole}^A + i\Wc_\df^A\,.
\end{equation}
The first of these, $\Wc_{\rm pole}^A$, contains single-particle poles associated with the external particles coupling to the current. 
The pole term is fixed by the on-shell one-body current and the purely hadronic amplitude. The second term, $\Wc_\df^A$, is labeled by the subscript $\df$, which stands for divergence-free. This amplitude can be decomposed into a triangle, or one-body-current, contribution and a two-body-current contribution,
\begin{equation}
        \label{eq:main_v0}
        i\Wc_\df^A = i\Wc_{\df,{\rm 1B}}^A + i\Wc_{\df,{\rm 2B}}^A\,.
\end{equation}
The first term, $\Wc_{\df,{\rm 1B}}^A$, contains only diagrams in which the current couples to one of the particles propagating in an internal two-particle loop. The second term, $\Wc_{\df,{\rm 2B}}^A$, includes everything else, such as all diagrams in which the current couples to a two-particle irreducible kernel.

\begin{figure}
    \centering
    \includegraphics[width=0.8\linewidth]{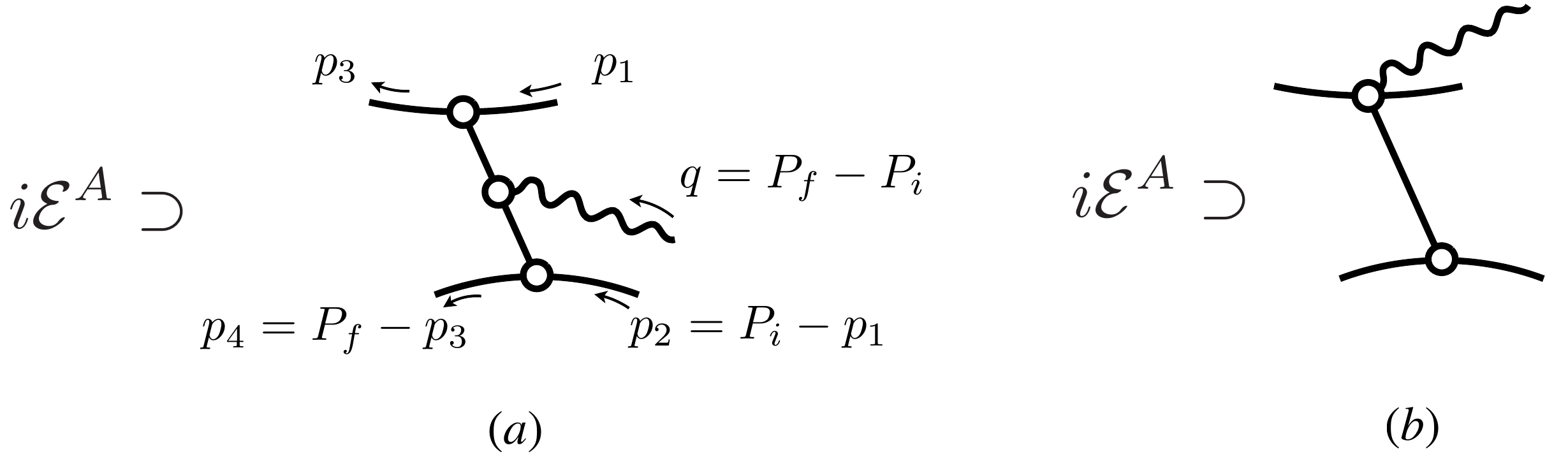}
    \caption{  ($a$) Shown is the tree-level $t$-channel diagram contribution associated with the current coupling to the exchange particle. ($b$) Shown is one of the two tree-level $t$-channel contributions associated with short-distance insertions to the vertices. The other contribution is due to the current coupling to the bottom-most vertex.}
    \label{fig:iEA}
\end{figure}

To give a detailed description of these individual quantities, it is important to begin by emphasizing the key difference between the formalism derived here and that presented in Ref.~\cite{Briceno:2020vgp}, namely the OPE contribution. The first OPE contribution is purely hadronic, which, following Refs.~\cite{Raposo:2023oru, Raposo:2025dkb}, we denote by $\Ec$, and depict in Fig.~\ref{fig:single-exchange}.  Reference~\cite{Raposo:2025dkb} showed that the purely hadronic amplitude $\Mc$ can be written as 
\begin{align}
    \Mc=\Mc_\Ec+
    \left(1 + i\Lc \right) 
\,
\,
\big[ \Mc_{0}^{-1} + \Cc \big]^{-1} \left(1 + i\Rc \right),
\end{align}
where $\Mc_0$ is independent of $\Ec$, while $\Mc_\Ec$, $\Cc$, $\Lc $, and $\Rc$ are all fully determinable given $\Ec$. In particular, $\Mc_\Ec$ satisfies an integral equation for which $\Ec$ is the inhomogeneous term. Meanwhile, $\Cc$, $\Lc $, and $\Rc$ can be written as integrals of $\Mc_\Ec$. The rest of the functions of $\Mc_\Ec$, described below, will require the evaluation of, at most, 3-dimensional integrals of $\Mc_\Ec$, without requiring the solution of any additional integral equations. We give detailed definitions of these, as well as a review of the work presented in Ref.~\cite{Raposo:2025dkb}, in Sec.~\ref{sec:Mc_review}.

With this, we can write down the one-body contribution to $\Wc_\df^A$ as,
\begin{align}
 i\Wc_{\df,{\rm 1B}}^A
 =\sum_j f_j\left[
 \widetilde\Lc\,i\widetilde\Gc_j^A\,\widetilde\Rc
 +\widetilde\Lc\,i\widetilde{\mathbf G}_{\Rc,j}^A
 +i\widetilde{\mathbf G}_{\Lc,j}^A\,\widetilde\Rc
 +i\mathbf G_{\Ec,j}^A
 \right] \,,
\label{eq:main_Wdf1B}
\end{align}
where the summed index $j$ runs over all allowed one-body form factors $f_j$ and their associated kinematic tensors.
The $\widetilde{\Lc} $ and $\widetilde{\Rc} $ functions can be determined from $\Mc_0$ and $\Ec$, 
\begin{align}
\widetilde{\Lc} &\equiv \left(1+i\Lc\right)\big[\Mc_0^{-1}+\Cc\big]^{-1}\,
\label{eq:main_Ltilde}
\\
\widetilde{\Rc} &\equiv \big[\Mc_0^{-1}+\Cc\big]^{-1}\left(1+i\Rc\right)\,.
\label{eq:main_Rtilde}
\end{align}
 The remaining functions appearing in Eq.~\eqref{eq:main_Wdf1B} [$\widetilde\Gc_j^A\,$, $\widetilde{\mathbf G}_{\Rc,j}^A$, $\widetilde{\mathbf G}_{\Lc,j}^A\,$, $\mathbf G_{\Ec,j}^A$] are 3D triangle integrals, defined explicitly in Sec.~\ref{sec:Wdf1B},  which in general depend on $\Ec$.

The two-body contribution, meanwhile, requires the introduction of a new kernel, $\Ec^A$, depicted in Fig.~\ref{fig:iEA} and defined in Sec.~\ref{sec:2J2_wOPE}. As can be seen from Fig.~\ref{fig:iEA}, this new kernel has a single- and a double-pole contribution. The latter arises from the possibility of the exchanged particle coupling to the external current. It is important, and worth emphasizing, that $\Ec^A$ can be determined from relevant subprocesses. These include $1+\Jc^A \to 1$ and $1+\Jc^A\to 2$ processes, which are well understood~\cite{Lellouch:2000pv, Briceno:2014uqa, Briceno:2015csa}. 

Given this additional input, the two-body-current contribution can be written as 
\begin{align}
   i\Wc_{\df,{\rm 2B}}^A =
    i\Wb^A_\Ec
    +i\widetilde{\Lc}\, i\Wb^A_{\Rc}
    +i\Wb^A_{\Lc}\,i\widetilde{\Rc}
    + \widetilde{\Lc}\left(i\Ac^A-i\Wb^A_{\Cc}\right)\widetilde{\Rc}\,,
\label{eq:main_Wdf2B}
\end{align}
where the quantities $\Wb^A_\Ec$, $\Wb^A_{\Rc}$, $\Wb^A_{\Lc}$, and $\Wb^A_{\Cc}$ are functions of $\Ec^A$ and $\Mc_\Ec$. The only new short-distance object is $\Ac^A$, which is smooth and real in the elastic region.

\subsubsection{Discussion on the implementation of this formalism}

Given the different moving parts of the formalism summarized above and described in detail in the remainder of this work, it is useful to separate the construction into a sequence of inputs and numerical tasks. The first step is the purely hadronic one: determine the partial-wave-projected amplitude $\Mc_\Ec$ from Eq.~\eqref{eq:MEt}, together with the exchange-free amplitude $\Mc_0$. This is the only step that requires solving an integral equation. Once this has been done, the quantities $\Lc$, $\Rc$, and $\Cc$, and therefore the endcaps $\widetilde\Lc$ and $\widetilde\Rc$, follow from the one-loop integrals in Eqs.~\eqref{eq:RL}-\eqref{eq:CL} and Eqs.~\eqref{eq:Ltilde}-\eqref{eq:Rtilde}.

The second step is to construct the exchange-current kernel $\Ec^A$. This requires the on-shell exchange couplings, the one-body form factors of the exchanged and external particles, and the on-shell $1+\Jc^A\to2$ transition amplitudes that enter through $h^A_{\rm on}$. With this input fixed, the functions $\Wb^A_\Ec$, $\Wb^A_\Rc$, $\Wb^A_\Lc$, and $\Wb^A_\Cc$ are obtained from the 3D integrals defined in Sec.~\ref{sec:Wdf2B}. The third step is the evaluation of the triangle-type functions entering Eq.~\eqref{eq:main_Wdf1B}. Again, these are just 3D integrals that must be evaluated. 

Although in practice the integral equation needed to be evaluated in Eq.~\eqref{eq:MEt} is a 3D equation, after partial wave projection, which can be done analytically, this reduces to a 1D integral. This was explained in detail in the context of three-hadron systems in Refs.~\cite{Jackura:2020bsk, Dawid:2023jrj, Dawid:2024oey}, and Ref.~\cite{Raposo:2025dkb} showed that these tools developed for three-body systems are immediately transferable to the two-body system with an OPE. 

Similarly, the remaining 3D integrals can be reduced to lower-dimensional integrals. For example, App.~B.6 of Ref.~\cite{Baroni:2018iau} shows how the azimuthal dependence of triangle-type integrals can be isolated by choosing one of the two-body center-of-momentum frames (CMFs) and aligning the boost between the initial and final CMFs with the $z$ axis. We review the corresponding reduction for the present notation in Sec.~\ref{sec:triangles_2d}. A separate trick that has proven useful in solving related integral equations is contour deformation~\cite{Jackura:2020bsk, Dawid:2023jrj, Dawid:2024oey}. A future point of investigation will need to be how to adopt or modify these techniques for the more complicated integrals that will need to be evaluated here, which will involve triangle singularities.

The only remaining unknown in Eq.~\eqref{eq:main_Wdf2B} is the smooth real function $\Ac^A$. This function can be parameterized directly in the invariants $s_f$, $s_i$, and $q^2$, after the relevant tensor decomposition is chosen. In an eventual lattice-QCD application, the coefficients of this parameterization would be determined by matching the finite-volume matrix elements to the infinite-volume representation derived here. This would require further extension of the formalism presented in Refs.~\cite{Briceno:2015tza, Baroni:2018iau}.

As an example, let us consider the implementation of this for the eventual determination of the $T_{cc}$ electromagnetic form factor(s), in the oversimplified scenario where we ignore the spin of the $D^\star$. In such a case, this is the list of input parameters needed for a complete analysis:
\begin{enumerate}
    \item the purely hadronic $D D^\star$ amplitude, including the $D^\star \to D\pi$ coupling,
    \item single-particle form factors for the $\pi$, $D $, and $D^\star$, 
    \item $D^\star +\gamma^\star \to D \pi $ \&  $D +\gamma^\star \to D^\star \pi $ transition amplitude(s)~\footnote{These can be accessed in the physical region via lattice QCD~\cite{Lellouch:2000pv, Briceno:2015csa}, and can be related to one another via analytic continuation.},
    \item $D +\gamma^\star \to D \pi $ transition amplitude,
    \item $D^\star \to D^\star \pi$ coupling,
    \item analytically continue $\Wc_\df$ to the $T_{cc}$ pole and obtain the form factor(s) from the residue. 
\end{enumerate}
The $D^\star \to D^\star \pi$ coupling is not needed in the $2\to2$ scattering, as the $D\to D\pi$ transition is forbidden, but is necessary in this case as the vertex $D+\gamma^\star\to D\pi$ is allowed. Note the $D \to D \pi$ vertex is prohibited for a parity conserving theory. This also implies that one does not need to determine the $D^\star +\gamma^\star \to D^\star \pi $ amplitude for this analysis.

\section{Review of existing formalisms for $\Mc$ and $\Wc$}
\label{sec:review}
Before presenting the derivation of our main result, it is useful to review the existing formalisms that we build on, namely, those presented in Refs.~\cite{Briceno:2020vgp, Raposo:2025dkb}. We first sketch the derivation of the purely-hadronic $2\to2$ amplitude, which introduces much of the notation used throughout, and consider the left-hand cut caused by OPEs. We then move to the case of amplitudes with current insertions but without explicit consideration of possible left-hand singularities. For simplicity, we assume there is at most a single channel open composed of scalar particles of masses $m_1$ and $m_2$, which can in general be either identical or distinct. From there, generalizing to an arbitrary number of two-body channels is straightforward.

\subsection{Purely hadronic $2\to2$ amplitude}
\label{sec:Mc_review}


In the absence of any left-hand cuts and inelastic thresholds, the purely hadronic partial-wave projected $2\to2$ scattering amplitude $\mathcal{M}$ can in general be written as
\begin{equation}
    \label{eq:kmatrix2to2}
    i\mathcal{M}(s) = i\mathcal{K}(s)\,\frac{1}{1-i\rho(s)\,\mathcal{K}(s)}\;,
\end{equation}
where $\mathcal{K}$ is a K matrix. In this simple case, $\mathcal{K}$ is a purely real function of the Mandelstam invariant $s$, defined in terms of the total four-momentum $P$ as $s=P^2$. The other quantity that appears is $\rho$, the two-body phase space,
\begin{align}
\rho(s) = \frac{\xi q^\star } { 8 \pi \sqrt{s}},
\end{align}
where $\xi$ is a symmetry factor, equal to $1/2$ if the particles are identical and $1$ otherwise, and $q^\star$ is the on-shell relative momentum, which can be defined using the K\"allen triangle function $\lambda(a,b,c)=a^2+b^2+c^2-2(ab+ac+bc)$ as,
\begin{align}
    q^\star = \frac{\lambda^{1/2}(s, m_1^2 , m_2^2) }{2\sqrt{s}}.
    \label{eq:qstar}
\end{align}

The representation of the scattering amplitude given in Eq.~\eqref{eq:kmatrix2to2} is an analytic function of energy in the physical region except at threshold, where the phase space features a square root singularity.
This singularity gives rise to a branch cut along the real energy axis starting, for degenerate particles, at $s=4m^2$ with the discontinuity across the cut given by
\begin{equation}
    \text{Disc}\,\Mc(s) = 2i \Mc^*(s) \rho(s) \Mc(s)\,.
\end{equation}

When the energy region is extended below the two-particle threshold, diagrams with single light particle exchanges, like those shown in Fig.~\ref{fig:single-exchange}, contribute non-analytic behavior that is not captured by the representation of Eq.~\eqref{eq:kmatrix2to2} if the K matrix is constrained to be real. In general, the K matrix is a scheme-dependent function that we want to ensure is purely real. Thus, following Ref.~\cite{Raposo:2025dkb}, we make the OPEs explicit and introduce a new K matrix, $\Kc_0$, which is ensured to be real, even when one hits the singularity associated with such exchanges. 

Let us denote the exchanged particle type by $e$ and its mass by $m_e$. If we consider processes where there can be both $t$- and $u$-channel OPEs, the tree-level contributions can be written as
\begin{align}
i\Tc = ig_{13} \frac{i}{(p_3-p_1)^2 - m_e^2 + i\epsilon} ig_{24} \,, \\
i\Uc = ig_{14} \frac{i}{(p_1-p_4)^2 - m_e^2 + i\epsilon} ig_{23} \,,
\end{align}
where $g_{xy}$ is the $x\to y+e$ on-shell coupling.
Once the energy is low enough below threshold, the inverse of the propagator of particle $e$ can vanish, giving rise to pole singularities. If we partial-wave project the corresponding amplitudes, these pole singularities become logarithmic cuts. Cuts of this type are typically referred to as left-hand cuts; in the case of degenerate external particles, the associated left-hand cuts for both $t$- and $u$-channel tree-level contributions begin at $s=4m^2-m^2_e$.


Let us turn our attention to contributions beyond tree level, following some of the key steps presented in Ref.~\cite{Raposo:2025dkb}. We begin by writing the integral equations that the off-shell amplitude must satisfy.
For that, we write the scattering amplitude, $\Mc(p_3,p_1)$, as a function of the in/out-going off-shell particle momentum $p_{1/3}$, and we leave the dependence on the total momentum $P$ implicit.
Let $\Delta^{(j)}$ be the fully dressed propagator for the $j$-th particle with mass $m_j$. With this, we can write the Bethe-Salpeter equation that the off-shell scattering amplitude must satisfy as
\begin{align}
    \label{eq:2to2int}
    i \mathcal{M}(p_3,p_1) &= i \Bc(p_3,p_1) + \xi \int \frac{d^4k}{(2\pi)^4}i\Bc(p_3,k) i\Delta^{(1)}(k) i\Delta^{(2)}(P-k) i\Mc(k,p_1)\;
    \\
    &\equiv i\Bc^{(1)}(p_3,p_1) + i\Tc(p_3,p_1) + i\Uc(p_3,p_1) \nonumber\\
    &\hspace{1cm}+ \xi \int \!\frac{d^4k}{(2\pi)^4} \, [i\Bc^{(1)}(p_3,k) + i\Tc(p_3,k) + i\Uc(p_3,k)]  i\Delta^{(1)}(k) i\Delta^{(2)}(P-k) i\Mc(k,p_1)
\end{align}
where $\Bc \equiv \Bc^{(1)} + \Tc + \Uc$ is the Bethe-Salpeter kernel containing the sum of all two-particle irreducible diagrams in the $s$ channel. Note that the subtracted kernel $\Bc^{(1)}$ is now ensured to be real, not just above threshold, but also for energies along the left-hand cut.

Following the steps presented in Ref.~\cite{Briceno:2020vgp}, we iterate this integral equation and isolate the singular contribution order by order in the loop expansion.
The singular contributions above the two-particle threshold and below the first inelastic threshold come from the $s$-channel two-particle loops.
Below the two-particle threshold, they come from single-particle exchanges in the $t$ and $u$ channels. 

We first address a generic example of a two-particle loop, which we label as $\Ic_0(p_3,p_1)$, following the same notation as above.
We write this loop with two generic kernels as \emph{endcap} functions $\mathfrak{L}(p_3,k)$ and $\mathfrak{R}(k,p_1)$, where $k$ is the loop momentum.
The only assumption at this stage is that these functions have no singularities in the above-threshold kinematic region. This generic loop diagram can be written as,
\begin{equation}
    \Ic_0 (p_3,p_1) = \xi \int \frac{d^4k}{(2\pi)^4} i\mathfrak{L}(p_3,k) i\Delta^{(1)}(k) i\Delta^{(2)}(P-k) i \mathfrak{R}(k,p_1) \;,
\end{equation}
where, without loss of generality, we evaluate this in the CMF, $P^\mu= (\sqrt{s},0)$.

We then recognize that the singular contribution in this expression requires that the $k^0$ integral picks up the on-shell pole, $k^0=\omega_{1}(\mathbf{k}^{\star}) = \sqrt{m_1^2+\mathbf{k}^{\star 2}}$, where the ${\star}$ is there to remind us that we have chosen to restrict our attention to the CMF. This is a minor distinction in this case, but when considering the insertion of currents, where there are different CMFs, this notation becomes rather useful. After isolating the contribution from the aforementioned pole, one can write $\Ic_0$ as,
\begin{equation}
\Ic_0 (p_3,p_1) =
    \Ic_1 (p_3,p_1)
    + \xi \int \frac{d^3\mathbf{k}^\star}{(2\pi)^3}\frac{1}{2\omega_{1}(\mathbf{k}^{\star})} \left[ i\mathfrak{L}(p_3,k) i\Delta^{(2)}(P-k) i \mathfrak{R}(k,p_1) \right]_{k^0=\omega_{1}(\mathbf{k}^{\star})}
    \;,
    \label{eq:I0_v2}
\end{equation}
Here, we introduced $\Ic_1$, which, by definition, is the difference between the original integral and the new integral being considered, and, most importantly, is ensured to be a non-singular function.

One can expand all quantities around the on-shell point where $k^\star\equiv\abs{\mathbf{k}^{\star}}$ is equal to the on-shell relative momentum $q^\star$ of Eq.~\eqref{eq:qstar}, including the $\Delta^{(2)}$ propagator, which has a pole at this on-shell point.
To do this, we first project the angular dependence on the spatial loop momentum $\mathbf k^\star$ to spherical harmonics, as follows:
\begin{align}
\mathfrak{L}(p_3,{k})\big|_{k^0=\omega_{1}(\mathbf{k}^{\star})}
     &= \sum_{\ell m }
     \mathfrak{L}_{\ell m}(p_3,k^\star) \, \Yc_{\ell m}^T(\mathbf{k}^\star,q^\star) \,,
     \label{eq:L_pw}
     \\
\mathfrak{R}({k},p_1)\big|_{k^0=\omega_{1}(\mathbf{k}^{\star})}
     &= \sum_{\ell m }
     \mathfrak{R}_{\ell m} (k^\star,p_1) \, \Yc_{\ell m}(\mathbf{k}^\star,q^\star) \,,
     \label{eq:R_pw}
\end{align}
where $\Yc$ and $\Yc^T$ are modified spherical harmonics defined to cancel singularities of the spherical harmonics at threshold,
\begin{align}
\Yc_{\ell m}(\mathbf{k}^\star,q^\star) 
&\equiv \sqrt{4\pi} \left(\frac{\vert \mathbf{k}^\star \vert}{q^\star}\right)^{\ell} Y_{\ell m}(\hat{\mathbf{k}}^\star) \,,
\label{eq:Yharm_def}
\\
\Yc_{\ell m}^T(\mathbf{k}^\star,q^\star) 
&\equiv \sqrt{4\pi} \left(\frac{\vert \mathbf{k}^\star \vert}{q^\star}\right)^{\ell} Y^*_{\ell m}(\hat{\mathbf{k}}^\star) \,.
\label{eq:YharmT_def}
\end{align}
We have also used the fact that the partial-wave projected endcaps only depend on the total angular momentum $\ell$ and not its azimuthal component $m$ in their labels. 

Although we have not made any assumptions about the endcaps so far, we proceed to only expand around the on-shell value $k^\star=q^\star$ those associated with $\Bc^{(1)}$, and leave the OPE kernels $\Tc/\Uc$ with an off-shell dependence on $k^\star$.
For those cases in which we on-shell expand the endcaps, we use the following identity, 
\begin{align}
    \mathfrak{R}_{\ell m}({k}^\star,p_1) 
    &\equiv \mathfrak{R}_{\ell m}({q}^\star,p_1)
    +
    \delta \mathfrak{R}_{\ell m}({k}^\star,p_1) \,,
    \label{eq:Rell_on}
\end{align}
where we have introduced $\delta$ as an operation that removes the on-shell point from a function.
An identical identity can be written for the $\mathfrak{L}_{\ell m}$ endcap.
In the case of the OPEs, we simply use Eqs.~\eqref{eq:L_pw} and \eqref{eq:R_pw} to partial-wave project, and leave the integral over their generally off-shell kinematic dependence.

The rationale for not projecting the OPEs on shell is to prevent the introduction of spurious left-hand singularities in our final expression for the scattering.
When an OPE kernel is dressed by a loop integral as in Eq.~\eqref{eq:I0_v2}, the integration smooths over the left-hand singularity, with the final result missing the left-hand cut single-particle exchange of the on-shell OPE kernel.
This does not prevent these loop integrals from having other left-hand cut singularities in general; e.g.\ loop integrals containing multiple light-particle exchanges will feature cuts associated with the corresponding thresholds.
However, the cut closest to threshold, the one considered in this work, is the singularity associated with the single particle exchange.

This leads to our final expression for the generic two-particle loop, in terms of a real non-singular kernel $\Ic_2$, and a 3D loop integral
\begin{align}\label{eq:I0_I2pdot}
    \Ic_0 (p_3,p_1)
     &=
     \Ic_2 (p_3,p_1) +  [i\mathfrak{L} \cdot i\Delta_{2} \cdot i \mathfrak{R}](p_3,p_1) \,,
\end{align}
where we have now introduced a compact dot ``$\cdot$'' product notation and a symbol for the pole contribution of the two-particle propagator, $\Delta_2$:
\begin{align}
    [i\mathfrak{L} \cdot i\Delta_{2} \cdot i \mathfrak{R}] 
    &\equiv 
    \sum_{\ell m} \xi\int\frac{d^3\mathbf{k^\star}}{(2\pi)^3} 
    i\mathfrak{L}_{\ell m}(k^\star) \,
    \Yc_{\ell m}^T(\mathbf{k}^\star,q^\star)
    \,
    i\Delta_{2}(\mathbf{k^\star})
    \,
    i\mathfrak{R}_{\ell m}(k^\star)\,
    \Yc_{\ell m}(\mathbf{k}^\star,q^\star) \,,
    \\
    &\equiv    \sum_{\ell m} \xi\int\frac{d^3\mathbf{k^\star}}{(2\pi)^3} 
    i\mathfrak{L}_{\ell m}(k^\star) \,
    \Yc_{\ell m}^T(\mathbf{k}^\star,q^\star)
    \,
    \frac{\omega_1(q^\star)}{\omega_1(\mathbf{k^\star})}\frac{iH({k^\star})}{2E[q^{\star2} - \mathbf{k}^{\star 2} + i\epsilon]}
    \,
    i\mathfrak{R}_{\ell m}(k^\star)\,
    \Yc_{\ell m}(\mathbf{k}^\star,q^\star).
             \label{eq:dot_prod}
\end{align}
Here we left implicit the kernel dependence on $p_{3/1}$, and we introduced $H$ as a cutoff function that must vanish as ${k}^\star\to \infty$ to ensure that the integral is finite and must be equal to $1$ as ${k}^\star\to q^\star$. An example of a function that satisfies these criteria is $H({k^\star}) = e^{-\alpha ({k}^{\star 2}- q^{\star2} )}$, but even a step function, as long as the cutoff is energy is much higher than all masses and thresholds, would keep the appropriate analytic structure of the loop within our energy region of interest.
Kernels within this dot product, except for OPE kernels, are projected on shell, behaving as constant functions of $k^\star$, and evaluated at the on-shell point $k^\star = q^\star$.
The OPE kernels are left off-shell, as a function of the integrated momentum magnitude $k^\star$. This compact notation makes it straightforward to generalize our result to arbitrary channels and spins by replacing the subscript $\ell$ with some index that accommodates all relevant degrees of freedom. 

Next, we turn our attention to the OPE, prior to it being projected to a given partial wave. Here, we adopt a definition of the OPE that is ensured to not have spurious poles, as first flagged in Ref.~\cite{Raposo:2023oru}, but is also ensured to be symmetric under the interchange of the initial/final state,~\footnote{It is worth noting that Ref.~\cite{Raposo:2023oru} originally used a prescription for $\Ec$ that was not explicitly symmetric under the interchange of the initial/final states. This was changed in the Erratum of the published article. }
\begin{align}
\Ec (\mathbf k'; \mathbf k)
&\equiv
\left[ \Tc (k',P-k';k,P-k) \right]_{k'^0 = \omega_{1} (\mathbf k'),\, k^0 = \omega_{1} (\mathbf k)} 
\nonumber\\
&\hspace{1cm} + \frac{1}{2}\left[ \Uc (k',P-k';k,P-k) \right]_{k'^0 = \omega_{1} (\mathbf k'),\, k^0 = E - \omega_{2} (\mathbf P - \mathbf k)},
\nonumber\\
&\hspace{1cm} + \frac{1}{2}\left[ \Uc (k',P-k';k,P-k) \right]_{k'^0 = E-\omega_{2} (\mathbf P - \mathbf k'),\, k^0 = \omega_{1} (\mathbf k)}.
\label{eq:Ec_def}
\end{align}
It is easy to see that $\Ec$ is equal to $\Tc +\Uc$ in the limit that all particles are on-shell, but differs away from the on-shell limit. This allows us to replace $\Tc+\Uc$ with this definition of $\Ec$ at all orders in the definition of $\Mc$, at the cost of absorbing the difference into non-singular kernels. Note that the partial-wave projected $\Ec$ can be written in terms of Legendre Q functions, as explained in Ref.~\cite{Raposo:2025dkb}. The partial-wave projected version of $\Ec$ is what would be featured in the compact dot-product notation in Eq.~\eqref{eq:dot_prod}, with the important distinction that it will not be placed on-shell inside of integrals.

\begin{figure}
    \centering
    \includegraphics[width=0.95\linewidth]{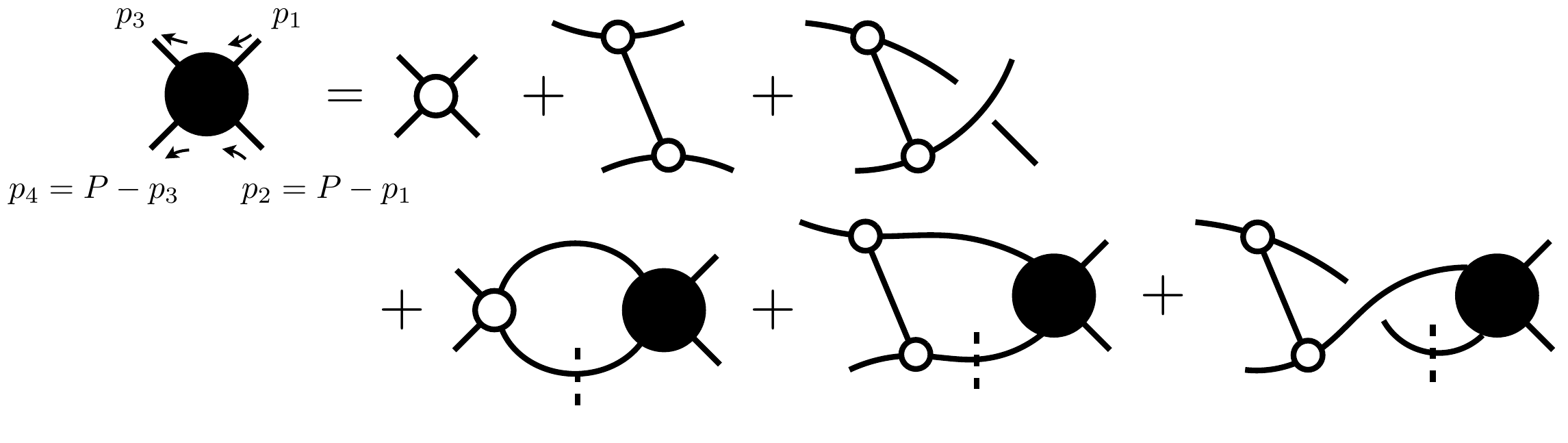}
    \caption{Diagrammatic representation of the Bethe-Salpeter equation with the on-shell projection of the kernels in the loops. The scattering amplitude $\Mc$ is shown as a black circle, the kernel $\widetilde{\Bc}$ is an open/white circle, while the $\Ec$ kernels were introduced in Fig.~\ref{fig:single-exchange}. The dashed line cutting one of the loop propagators indicates the projection of the $k^0$ integral as well as of the $\widetilde{\Bc}$ kernel to their on-shell value.}
    \label{fig:bs_MBtilde}
\end{figure}

We can now return to rewrite Eq.~\eqref{eq:2to2int} after applying all of the steps outlined above. Namely, we reduce the four-dimensional integrals down to 3D integrals, replace $\Tc+\Uc$ by $\Ec$, isolate the pole piece of the two-body propagator, perform partial-wave projection, and place the smooth kernels on shell, while keeping $\Ec$ off shell. All non-singular contributions are then absorbed into a new kernel $ \widetilde{\Bc}$, which is fully on-shell. Following these steps, and using the dot-product notation of Eq.~\eqref{eq:dot_prod}, we can then rewrite Eq.~\eqref{eq:2to2int} compactly as, 
\begin{align}
i\Mc &
= i\widetilde{\Bc} + i\Ec +  [(i\widetilde{\Bc} + i\Ec) \cdot i\Delta_{2} \cdot i\Mc] \,.
\end{align}
This integral equation provides a more adequate starting point than Eq.~\eqref{eq:2to2int} to derive a representation of the amplitude with an explicit contribution from the threshold singularity as well as the closest left-hand cut singularity. This integral equation is shown diagrammatically in Fig.~\ref{fig:bs_MBtilde}.
We define the auxiliary amplitudes through the (3-dimensional) integral equations
\begin{align}
i\Mc_{0} & =
i\widetilde{\Bc} + i\widetilde{\Bc}\,\cdot i \Delta_{2} \cdot \,  \, i\Mc_{0} \,, \label{eq:M0t} \\ 
i\Mc_{\Ec} &= i\Ec + i\Ec\,
\cdot i \Delta_{2} \cdot \, 
i\Mc_{\Ec} \,.
\label{eq:MEt}
\end{align}
It is easy to show that $\Mc_0$ satisfies an equation of the form of Eq.~\eqref{eq:kmatrix2to2}. More explicitly, we can partial-wave project $\Mc_0$ and rewrite it in terms of a K matrix, $\Kc_0$, as
\begin{align}
    \Mc_0 = \left[\Kc_0^{-1}-i\rho \right]^{-1}.
\end{align}
In other words, $\Mc_0$ only has kinematic singularities above threshold. Above threshold, one can show that $\Mc_\Ec$ also can be written in this form, although that serves no practical use in this context. More importantly, $\Mc_\Ec$ has the logarithmic left-hand cut.

With $\Mc_0$ and $\Mc_\Ec$, one can further rewrite the full amplitude. Skipping several of the key steps, we present the main result for the partial-wave projected on-shell amplitude,
\begin{align}
i\Mc &=
i\Mc_{\Ec}
+ \big[ 1 +  i\Mc_{\Ec} \cdot i\Delta_{2} \cdot {} \big]
\sum_{n=0}^\infty i\Mc_{0} \left({} \cdot i\Delta_{2} \cdot i\Mc_{\Ec} \cdot i\Delta_{2} \cdot i\Mc_{0}
\right)^n \big[{} \cdot i\Delta_{2} \cdot i\Mc_{\Ec} + 1 \big] \,
\label{eq:Mfull_integral}\\
&= i\Mc_{\Ec}
+ \left(1 + i\Lc \right) 
\,
i\,
\big[ \Mc_{0}^{-1} + \Cc \big]^{-1} \left(1 + i\Rc \right) \,,
\label{eq:Mfull_final}
\end{align}
where we have introduced the following auxiliary kernels, given as integrals of the form of Eq.~\eqref{eq:dot_prod},
\begin{align}
i\Rc
&\equiv
\Yc^T \cdot i\Delta_{2} \cdot i\Mc_{\Ec} \,,
\label{eq:RL}
\\
i\Lc
& \equiv 
 i\Mc_{\Ec} \cdot i\Delta_{2} \cdot \Yc  \,,
\label{eq:LL}
\\
i\Cc
&\equiv 
\Yc^T \cdot i\Delta_{2} \cdot i\Mc_{\Ec} \cdot i\Delta_{2} \cdot \Yc  \,.
\label{eq:CL}
\end{align}
In general, the auxiliary kernels are matrices in the degrees of freedom not fixed by the energy. In this case, the only such degree of freedom is orbital angular momentum, which is conserved for spinless particles, rendering these quantities as diagonal matrices in $\ell$.

Throughout our derivation, the dot-product notation is used for intermediate steps, e.g.~Eq.~\eqref{eq:Mfull_integral}, and to define auxiliary kernels like Eqs.~\eqref{eq:RL},~\eqref{eq:LL},~\eqref{eq:CL}.
The kernels appearing in these integral equations should be understood as functions of 3-dimensional momentum of one of the particles in the incoming and outgoing states of the kernels.
However, final results without dot products, such as Eq.~\eqref{eq:Mfull_final}, should be read as matrix equations in angular-momentum space, where all kernels have been partial-wave projected.
Furthermore, when all the kernels are diagonal in orbital angular momentum, as in Eq.~\eqref{eq:Mfull_final} for spinless particles, the final result can be interpreted as an algebraic equation for each partial wave.

\subsection{Electroweak $2+\mathcal{J}\to2$ amplitude above the two-particle threshold}
\label{sec:W_review}

\begin{figure}
    \centering
    \includegraphics[width=0.7\linewidth]{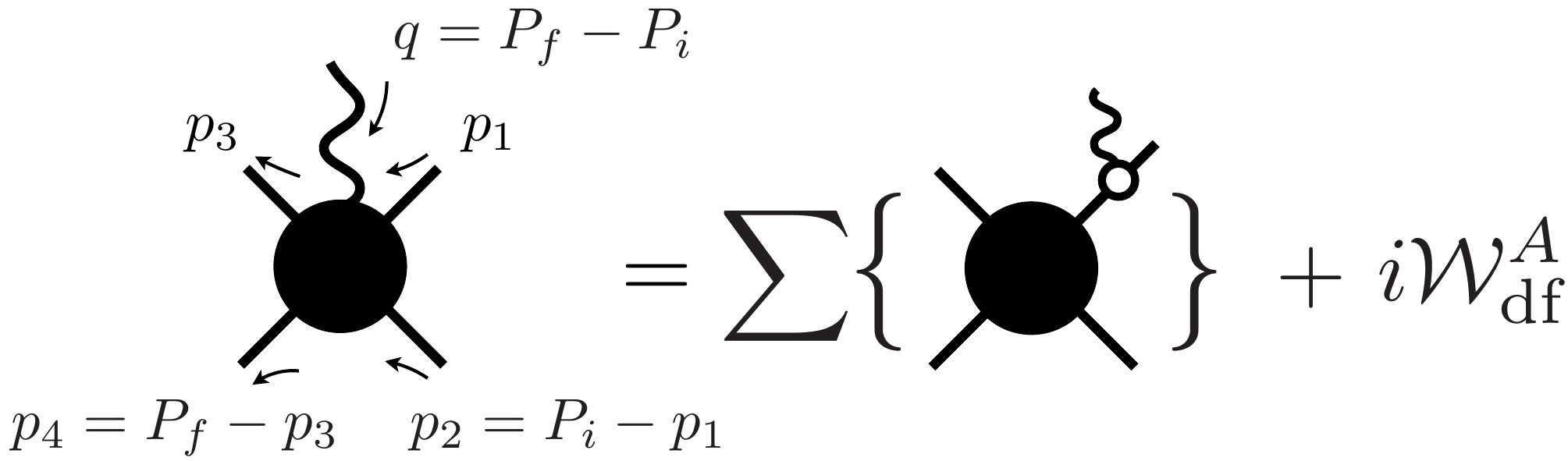}
    \caption{The all-orders $2 \to 2$ hadronic transition amplitude, with the divergent pole piece subtracted out from the remaining divergence-free amplitude. The summation symbol over the external leg divergences represents the current coupling to each different external leg of the purely hadronic amplitude.}
    \label{fig:iWA}
\end{figure}

We now turn our attention to the $2+\mathcal{J}\to2$ amplitude.
Given that in Sec.~\ref{sec:2J2_wOPE} we present a detailed derivation of the analytic expression for the $2+\mathcal{J}\to2$ amplitude while taking into account the OPEs, here we give a very brief summary of the formalism for scenarios where the OPE contribution is either exactly $0$ or when one is only interested in kinematics above threshold. 

\begin{figure}
    \centering
    \includegraphics[width=\linewidth]{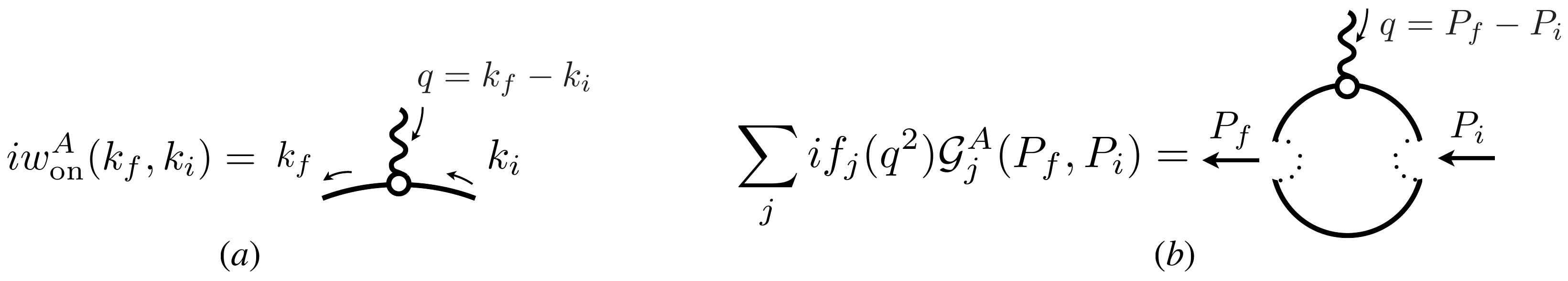}
    \caption{
    Diagram $(a)$ shows the on-shell projected $1+\Jc \rightarrow 1$ transition. The main body provides an explicit definition of $w^A_{a,\rm on}$.
    Diagram (b) shows the triangle diagram. The solid internal lines represent the propagator pole pieces $D$. Dashed semicircles represent the spherical harmonics left by the partial-wave projected on-shell endcaps.}
    \label{fig:triangle}
\end{figure}

For a current with a generic Lorentz index structure $A$, we label the corresponding amplitude by $\Wc^A$. 
Before we review the less trivial contributions of $\Wc^A$, we are reminded that this amplitude has long-range contributions featuring simple poles~\cite{Briceno:2015tza, Briceno:2020vgp}. 
These contributions, depicted diagrammatically in Fig.~\ref{fig:iWA}, are due to the direct coupling of the external individual particles to the current. 
Isolating these contributions is generally a scheme-dependent prescription. Following the choice made in Refs.~\cite{Briceno:2015tza, Briceno:2020vgp}, we can write the amplitude as, 
\begin{align}
    i \Wc^A = i\Wc_{\rm df}^A + \sum_a\bigg(iw_{a,\rm on}^A \, iD_a  \, \sum_{\ell m} \Yc_{\ell m} \,  i\Mc_\ell\,  \Yc^T_{\ell m} \bigg), 
    \label{eq:Wdf_def}
\end{align}
where we have left the kinematic dependence implicit, the sum runs over all possible external legs, $D_a(k) = 1 /(k^2 - m^2_a) $ is the pole of the propagator of the $a$-th particle, $w_{a,\rm on}^A $, depicted in Fig.~\ref{fig:triangle}~($a$) is the matrix element of the single-particle state that has been partly placed on-shell and is defined more carefully below, and the modified spherical harmonics $\Yc$ and $\Yc^T$ ensure no spurious singularities appear. 
As discussed in Sec.~\ref{sec:main}, we label the difference between the full amplitude and these single pole terms with a subscript $\rm df$ to denote ``\emph{divergence free}".~\footnote{A subtle point addressed in Ref.~\cite{Baroni:2018iau}, is that the initial/final spherical harmonics need to be evaluated using different kinematics.}

To define $w_{a,\rm on}^A $, we need to first define some kinematics that are generally off-shell. 
Let $P_{i/f}$ be the four-momentum of the initial/final two-particle state. 
The momentum entering the one-particle matrix element is $k_i$, and the outgoing momentum is $k_f$, where $k_{i/f} = P_{i/f}-k$, and $k$ is the momentum of the particle without the current insertion.
For the scenario depicted in the diagram within the summation of Fig.~\ref{fig:iWA}, where the current insertion is attached to the particle with momentum $p_1$, the momentum $k$ would be equal to $p_2$. In this case the momenta $k$ and $k_i$ would be on shell, while $k_f$ would be off shell for generic external kinematics.
With this, we can generally write the off-shell matrix element in terms of kinematic Lorentz tensors $K^A_{j,a}$ and off-shell form factors, $f_{j,a}(q^2,k_f^2,k_i^2)$, where $q^2 = (P_i-P_f)^2$, as follows, 
\begin{equation}
    iw^A_a(k_f,k_i) = \sum_j K^A_{j,a}(k_f,k_i) i f_{j,a}(q^2,k_f^2,k_i^2). 
\end{equation}
Following a similar set of steps as in Eq.~\eqref{eq:Rell_on}, one can write these as,
\begin{align}
    w^A_a(k_f,k_i) 
    &\equiv \sum_j K^A_{j,a} (k_f,k_i)  f_{j,a} (q^2)
    +
    \sum_j K^A_{j,a}(k_f,k_i)  \,  \delta f_{j,a}(q^2,k_f^2,k_i^2)
    \\
    &\equiv w^A_{a,\rm on} (k_f,k_i)
    +
    \sum_j K^A_{j,a}(k_f,k_i) \,  \delta f_{j,a}(q^2,k_f^2,k_i^2)
    ,
\end{align}
where we project the form factor on shell, and in the last equality we define $w^A_{a,\rm on}$. In all that follows, we will leave implicit the $a$ index denoting which particle the current couples to, and assume any sum over $j$ also runs over distinguishable particles.

Having defined the pole contribution of $\Wc^A$, we now turn to the remainder, $\Wc^A_\df$, which encodes the dynamical contributions unique to this amplitude.
For the case of identical particles, far from singularities associated with exchange diagrams, $\Wc^A_\df$ can be partial-wave projected and represented as 
\begin{equation}
    i\Wc_{\df}^A(P_f,P_i) = \mathcal{M}(s_f)  \left[ i\mathcal{A}^A(P_f,P_i) + \sum_{j} if_{j}(q^2)\mathcal{G}^A_{j}(P_f,P_i) \right]  \mathcal{M}(s_i),
    \label{eq:iWdf0}
\end{equation}
where $\mathcal{A}^A$ is smooth and real in the elastic scattering region, and in general can be a dense matrix in angular momentum, only restricted by selection rules associated with the symmetries of the current.
The $\Gc_j^A$ functions multiplying the form factors, $f_j$, are kinematic triangle functions, illustrated in Fig.~\ref{fig:triangle}~(b).
These triangle functions are also dense matrices in angular momentum with analogous restrictions to those of $\mathcal{A}^A$.
If we assume identical particles, these triangle functions can be compactly written as, 
\begin{align}\label{eq:GA_4D}
    \mathcal{G}^A_{j;\ell' m' ;\ell m}(P_f,P_i) =
    i\int\frac{d^4k}{(2\pi)^4} 
    \frac{
    \Yc^T_{\ell' m' }(\mathbf{k}^\star_f,q^\star_f)   K_j^A(k_f,k_i) 
     \, 
      \Yc_{\ell m }(\mathbf{k}^\star_i,q^\star_i)
      }
      {(k^2 - m^2 + i\epsilon) 
      (k_f^2 - m^2 + i\epsilon) 
      (k_i^2 - m^2 + i\epsilon)}.
\end{align}
For non-zero current momentum insertion, the initial and final states, in general, have different center of mass frames, so that the modified spherical harmonics are evaluated with respect to the relative momentum $k_{i(f)}^\star$ evaluated in the initial(final) frame, as well as with respect to the initial(final) on shell relative momentum $q_{i(f)}^\star$.
These functions have two different kinds of singularities in the kinematic region of interest.
One type of singularity is associated with the initial or final state thresholds, and is proportional to the two-body phase space.
The second type of singularity, which is not present in the elastic scattering case, is the class of logarithmic singularities associated with regions in the loop integral where the three particles within the diagram can simultaneously go on-shell. For example, if we consider the scenario where the current is a scalar and fix $\ell = \ell' = 0$,  one finds~\cite{Briceno:2020vgp},
\begin{equation}\label{eq:trianglelog}
    \mathcal{G}_{00,00} \propto  \log\left( \frac{1+z^\star_f + i\epsilon}{1-(z_f^\star +i\epsilon)}\right) + \log\left( \frac{1+z^\star_i + i\epsilon}{1-(z_i^\star +i\epsilon)}\right) + \ldots
\end{equation}
where $z^{\star}_{i(f)}$ is a function of the external masses and kinematics.

\section{Electroweak $2+\mathcal{J}\to2$ amplitude including exchange contributions}\label{sec:2J2_wOPE}

Having reviewed the previous formalism presented in Ref.~\cite{Briceno:2020vgp} for $2+\mathcal{J}\to2$ amplitudes where the OPE plays no role, we now proceed to extend this work to make the OPE contributions explicit. The light exchange particle can, in general, couple to the external current. This means that there are now possible double-pole contributions at tree level. An example, assuming a $t$-channel type of exchange, is shown in Fig.~\ref{fig:iEA}($a$). More specifically, following the notation of Ref.~\cite{Briceno:2020vgp}, if we denote $i\textbf{W}^A_{0|0} $ as the 2PI kernel with a coupling to the current, this will include a term of the form,
\begin{align}
   i\textbf{W}^A_{0|0} \supset  ig_{13}\frac{i}{(p_3-p_1)^2-m_e^2}iw^A_{e,{\rm on}}(p_3-p_1,p_2-p_4)
   \frac{i}{(p_2-p_4)^2-m_e^2}ig_{24}\,.
\label{eq:Wamp_exch_1}
\end{align}
This is the most singular contribution, obtained from placing on-shell the generally off-shell vertex function and the light-particle matrix element. 

Note, for simplicity, we have only shown the $t$-channel contribution. Given this, it is straightforward to obtain the $u$-channel analogue. With this in mind, in what follows, we will proceed to only explicitly show the $t$-channel contributions, but as will become evident, we also sum over the $u$-channel contributions. 

\begin{figure}
    \centering
    \includegraphics[width=0.8\linewidth]{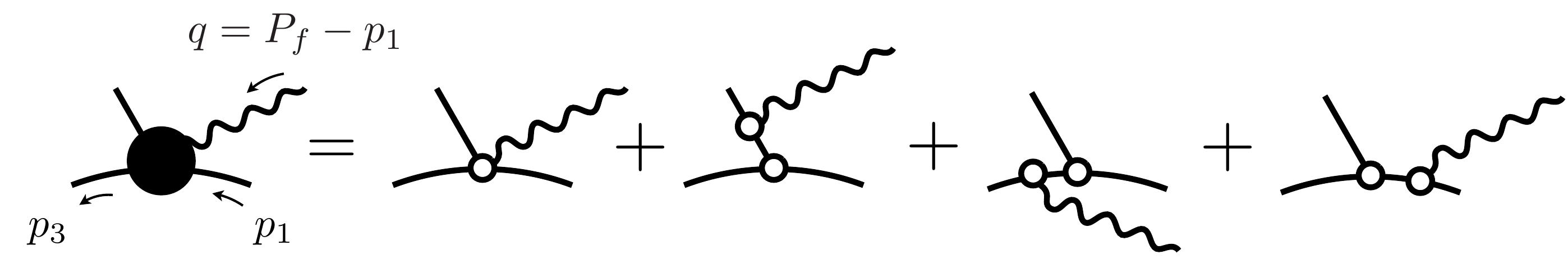}
    \caption{Diagrammatic representation of the $1+\Jc\to 2$ transition. The transition can be decomposed into a short-range contribution, $h^A$, and three long-range contributions each featuring the $g_{13}$ vertex and their corresponding single particle current insertion and propagator. For this work, we are interested in the final two-particle state energy below threshold, so that the re-scattering contributions are purely real and absorbed into the short-range kernel $h^A$.}
    \label{fig:ih}
\end{figure}

This procedure of placing the vertex on-shell leads to differences that cancel one or both of the propagators and inevitably generates a new class of short-range kernels. These, which we label as $h^A$, are associated with the unphysical $1+\Jc \to 2$ process, where one of the particles in the two-particle final state is the exchange particle. The $h^A$ kernels also include all remaining short-range contributions leading to the single-pole contributions depicted in Fig.~\ref{fig:iEA}($b$), akin to the OPE driving terms of $\Mc_\Ec$, discussed in Sec.~\ref{sec:Mc_review},
\begin{align}
   i\textbf{W}^A_{0|0} \supset  ig_{13}\frac{i}{(p_3-p_1)^2-m_e^2} ih^A_{\rm on}(p_4,p_2,q)
   + ih^A_{\rm on}(p_3,p_1,q)\frac{i}{(p_4-p_2)^2-m_e^2} ig_{24},
\label{eq:Wamp_exch_2}
\end{align}
where we have gone ahead and replaced all vertex functions with their on-shell values. 
This is evident in $g$, since it is just a coupling; meanwhile, the on-shell projection of $h^A_{\rm on}$ is, in general, a function that has the same Lorentz structure as the current. Consequently, we use the same on-shell prescription introduced in Ref.~\cite{Briceno:2020vgp} for the one-particle matrix elements appearing inside the triangle loops. Namely, $h^A_{\rm on}$ can generally be written as
\begin{equation}
    h^A_{\rm on}(p_3,p_1,q) = \sum_i K^A_i(p_3,p_1,q) f_i(q^2,(p_1+q)^2),
\end{equation}
where $f_i$ are on-shell transition form factors, and $K^A_i$ are generally off-shell kinematic tensors.
To describe the on-shell $1+\Jc \to 2$ transition associated with $h^A$ in its kinematically allowed region, i.e.\ if $(p_1+q)^2\geq (m_e+m_3)^2$, this matrix element would contain long distance contributions in addition to the $h^A$ function,
\begin{multline}
    \mel{P_f,p_3;{\rm out}}{\Jc^A(0)}{p_1} = h^A_{\rm on}(p_3,p_1,q)
    + ig_{13}iD_e(p_3-p_1)w^A_{e,{\rm on}}(p_3-p_1,P_f-p_3)
    \\+ ig_{13}iD_1(P_f-p_1)w^A_{1,{\rm on}}(p_3-p_1,P_f-p_3)
    + w^A_{3,{\rm on}}(p_3,p_3-P_f+p_1)iD_3(p_3-P_f+p_1)ig_{13}\,.
    \label{eq:1J2_hdec}
\end{multline}
This decomposition is illustrated in Fig.~\ref{fig:ih}.
The analytic continuation of $1+\Jc \to 2$ below threshold, $(p_1+q)^2< (m_e+m_3)^2$, together with Eq.~\eqref{eq:1J2_hdec}, fixes our prescription to determine $h^A_{\text{on}}$, which in this kinematic region remains purely real.

Putting everything together, we can write down a new OPE kernel associated with the insertion of the current, which we label as $\Ec^A_t$. This is just the sum of the three terms highlighted above, namely the double-pole diagram and the two single-pole diagrams. Figure~\ref{fig:iEA} shows two of these contributions; the third is the mirror image of Fig.~\ref{fig:iEA}($b$). Explicitly,
\begin{align}
    i\Ec^A_t &\equiv 
    ig_{13}\,iD_e(p_3-p_1)\,iw^A_{e,{\rm on}}(p_3-p_1,p_2-p_4)\,
   iD_e(p_2-p_4)\,ig_{24}
   \nonumber\\
   &\hspace{.5cm}
   + ig_{13}\,iD_e(p_3-p_1) \,ih^A_{\rm on}(p_4,p_2,q)
   + ih^A_{\rm on}(p_3,p_1,q)\,iD_e(p_4-p_2) \,ig_{24}
   \,.
\label{eq:Ec_At}
\end{align}
If the channel of interest supports a $u$-channel contribution of this type, then there is an additional class of contributions, which we label as $i\Ec^A_u$.

An advantage of introducing these quantities explicitly is that we immediately recognize that all remaining 2PI kernels with a current insertion within $\mathbf W^A_{0|0}$ will be free of the singularities that ultimately lead to left-hand cuts. This is in close analogy to the separation of the full Bethe-Salpeter kernel $\Bc$ in the $2\to2$ case, into singular OPE contributions, $\Tc$ and $\Uc$, and a subtracted non-singular kernel, $\Bc^{(1)}$. In that derivation, outlined in Sec.~\ref{sec:Mc_review}, we end up with an effective replacement $\Bc \to \widetilde{\Bc}+\Ec$ when reducing loop integrals from 4D to 3D integrals, which is then used to re-organize the skeleton expansion of the amplitude $\Mc$. In short, the first step in this procedure is to pick up the pole term of the $k^0$ loop integrals. The second step is to partial-wave project all kernels and, except for the OPE kernels $\Ec$, place them fully on shell, so that the skeleton expansion can be written in terms of our dot-product notation defined above. Note that the OPE kernels $\Ec$ follow from $\Tc$ and $\Uc$ being set to specific partially off-shell kinematics by the residue of the pole picked up in the $k^0$ integral, as defined in Eq.~\eqref{eq:Ec_def}.

For the $2 + \Jc \to 2$ case, we proceed in a parallel manner and arrive at a similar result for the 2PI transition kernel $\mathbf{W}^A_{0|0}$, namely
\begin{align}\label{eq:WA00_split}
\mathbf{W}^A_{0|0} \to  \widetilde\Bc^A + \Ec^A \,,
\end{align}
where $\Ec^A$ is defined from $\Ec^A_t$ and $\Ec^A_u$ with the appropriate kinematic configurations imposed by the $k^0$ integrations, in analogy to $\Ec$. For a single-channel case where $m_1 = m_3$ and $m_2 = m_4$, we can write
\begin{align} \nonumber
\Ec^A (\mathbf k', \mathbf k) &\equiv \Ec^A_t (k', P_f-k'; k, P_i-k) \Big|_{k'^0 = \omega_{1} (\mathbf k'),\, k^0 = \omega_{1} (\mathbf k)} \\ \nonumber
& \qquad {}
+ \frac 12 \Ec^A_u  (k', P_f-k'; k, P_i-k) \Big|_{k'^0 = \omega_{1} (\mathbf k'),\, k^0 = E_i - \omega_{2} (\mathbf P_i - \mathbf k)} \\
& \qquad {}
+ \frac12 \Ec^A_u  (k', P_f-k'; k, P-k) \Big|_{k'^0 = E_f -\omega_{1} (\mathbf P_f - \mathbf k'),\, k^0 = \omega_1 ( \mathbf k)}\,.
\label{eq:EcA_def}
\end{align}
The first term on the right-hand side comprises the $t$-channel contributions, while the remaining two terms are the $u$-channel contributions, symmetrized over initial/final two-hadron states, cf.~\eqref{eq:Ec_def}.

The arguments given in Sec.~\ref{sec:Mc_review}, in the paragraphs around Eq.~\eqref{eq:Rell_on}, for keeping the $\Ec$ kernel partially off shell apply also now to the $\Ec^A$ kernels. Moreover, similar to how $\widetilde\Bc$ absorbs smooth contributions associated with the $2\to 2$ process, e.g.~the kernel $\Ic_2$ of Eq.~\eqref{eq:I0_I2pdot}, the kernel $\widetilde\Bc^A$ absorbs smooth contributions associated with the $2+\Jc\to 2$ process.

In the remainder of this section, we focus our attention on deriving the on-shell description of the full amplitude, including the contribution from the OPE kernels. We leave details associated with the partial-wave projection of $\Ec^A$ to Appendix~\ref{app:PWA}. For simplicity, we assume there is a single two-particle channel that can go on-shell in the initial and final state. This is an assumption that is easy to lift to accommodate any number of two-particle channels.

\subsection{Skeleton expansion of the amplitude}

The introduction of exchange diagrams adds new classes of singularities that we will need to consider explicitly. The pole singularities appearing in $\Wc^A$ and depicted in the first term of the right-hand side of Fig.~\ref{fig:iWA}, which arise from a current attaching to one of the external legs, are still present. These are exactly removed from $\Wc_\df^A$, defined by Eq.~\eqref{eq:Wdf_def}. Consequently, we focus our attention on the quantity $\Wc_\df^A$.

In prior work, the reduction to 3D integrals was done for all diagrams except those of the triangle type. The advantage of leaving a four-dimensional representation is that it is easier to evaluate analytically. As we will see in Sec.~\ref{sec:Wdf1B}, the proliferation of triangle-type diagrams seems to suggest that there is no major advantage in leaving these integrals as 4D integrals. In particular, triangle-type diagrams can now couple to OPE exchanges, which need to be summed to all orders. As a result, it seems more convenient at this stage to reduce these down to 3D integrals that can be evaluated numerically. The difference between this prescription and the prior one can be absorbed into the definition of the smooth function $\Ac^A$.

Following the steps described above, we reduce all loop integrals in the skeleton expansion for the full amplitude, $\Wc_\df^A$, down to 3D integrals. Having done this, we can distinguish two classes of contributions:
\begin{align}
\Wc_\df^A = \Wc_{\df, \rm 1B}^A + \Wc_{\df, \rm 2B}^A \,.
\label{eq:WAdf_decomp}
\end{align}
The object $\Wc_{\df, 1B}^A$ collects all diagrams in which the current couples to one of the propagating particles in two-particle bubbles, i.e.~what we call triangle-type diagrams. The second term, $\Wc^A_{\df, \rm 2B}$, corresponds to diagrams in which the current couples to a two-to-two kernel, giving rise to $\textbf{W}^A_{0|0}$ kernels. We split it further into
\begin{align}
\Wc_{\df, \rm 2B}^A = \Wc_{\df, \Ec^A}^A + \Wc_{\df, \slashed{\Ec}^A}^A \,,
\label{eq:WA2B_decomp}
\end{align}
with the first term including contributions where the current couples to an OPE, thus containing the $\Ec^A$ kernel. The second term includes the remaining diagrams, where the current couples to subtracted OPE-less kernels $\widetilde{\Bc}^A$.
We proceed now to analyze each of these contributions to the amplitude in turn, starting with $\Wc_{\df, \rm 2B}^A$.

It is worth remarking that we made a choice to identify $\Wc_{\df, \Ec^A}^A$ as part of $\Wc_{\df, \rm 2B}^A$, even though $\Ec^A$ is defined, in part, in terms of single-particle matrix elements. In particular, the double-pole contribution of $\Ec^A$ depends on the form factor(s) of the exchanged particle. The reasoning behind this choice is that this contribution is not two-particle reducible in the $s$-channel of the asymptotic particles. That said, there is no significance in this choice, since we include all possible contributions to $\Wc_{\df}^A$ in the end. 

\subsubsection{Analytic decomposition of $\Wc_{\df, \rm 2B}^A$}
\label{sec:Wdf2B}
Let us start by taking a closer look at $\Wc_{\df, \Ec^A}^A$ because it is the most distinct term from those considered in prior derivations. Its skeleton expansion takes the form
\begin{equation}
    i\Wc_{\df, \mathcal{E}^A}^A
    =
    \big[ 1 +  i\Mc \cdot i\Delta_{2} \cdot {} \big]\,
     i\Ec^A 
      \,\big[{} \cdot i\Delta_{2} \cdot i\Mc + 1 \big] \,,
    \label{eq:WEIntEq}
\end{equation}

Following the same steps presented in Ref.~\cite{Raposo:2025dkb} for the purely hadronic amplitude with an OPE, reviewed in Sec.~\ref{sec:review}, one can rewrite Eq.~\eqref{eq:WEIntEq} as a geometric series in the insertions of $\widetilde{\Bc}$ and $\Ec$,
\begin{equation}
    i\Wc_{\df, \mathcal{E}^A}^A
    = \sum_{n,m=0}^\infty [(i\widetilde{\Bc} + i\Ec)\cdot i\Delta_2 \, \cdot   ]^n
     \, i\Ec^A\, [\cdot\, i\Delta_2\, \cdot(i\widetilde{\Bc} + i\Ec) ]^m\,.
\end{equation}
This has the advantage of letting us isolate the singularities of these contributions in terms of $\Mc_0$ and $\Mc_\Ec$. 

To see the pattern, it is useful to consider a couple of terms in isolation. First, let us sum all contributions that have no insertions of $\widetilde{\Bc}$. We can write these in terms of $\Mc_\Ec$ as follows, 
\begin{align}
    i\mathbf{W}^A_{\Ec}
     &\equiv\sum_{n,m=0}^\infty [i\Ec\cdot i\Delta_2 \, \cdot   ]^n
     i\Ec^A\, [\cdot\, i\Delta_2\, \cdot i\Ec ]^m\,,
     \\ &= 
    \big[ 1 +  i\Mc_{\Ec} \cdot i\Delta_{2} \cdot {} \big]\,
     i\Ec^A 
      \,\big[{} \cdot i\Delta_{2} \cdot i\Mc_{\Ec} + 1 \big]\, . \label{eq:WEt}
\end{align}

Now, we just have to consider all possible insertions of $\widetilde{\Bc}$. For this, it is again useful to consider some examples. First, consider contributions where $\widetilde{\Bc}$ is inserted at all orders to the right of $i \Ec^A$ with a $\Mc_\Ec$ in the rightmost endcap,
\begin{align}
    \sum_{n=0}^\infty i\Ec^A\,\cdot i\Delta_2\, \cdot i\widetilde{\Bc} \, [\cdot \, i\Delta_2\, \cdot i\widetilde{\Bc}  ]^n \cdot i\Delta_2\, \cdot i\Mc_\Ec 
    &=i\Ec^A\,\cdot i\Delta_2\, \cdot i\Mc_0 \cdot i\Delta_2\, \cdot i\Mc_\Ec.
\end{align}
This class of contributions will be written in terms of $\Lc$, defined in Eq.~\eqref{eq:LL}. There will be a symmetric contribution in terms of $\Rc$, defined in Eq.~\eqref{eq:RL}. 

Another class of contributions arise when you have the $\widetilde{\Bc}$ sandwiching an all-orders insertion of $\Ec$, 
\begin{align}
    &i\widetilde{\Bc}\cdot i\Delta_2\,\cdot \,\sum_{n=0}^\infty i\mathcal{E} [{}\cdot i\Delta_2\, \cdot i\Ec  ]^n \cdot i\Delta_2\, \cdot i\widetilde{\Bc} 
    =i\widetilde{\Bc}\cdot i\Delta_2 \,\cdot i\Mc_{\Ec}\cdot i\Delta_2\,\cdot i\widetilde{\Bc}\,.
\end{align}
This class of contributions can be written in terms of $\Cc$, defined in Eq.~\eqref{eq:CL}. In fact, we see that whenever $\Cc$ appears, it appears adjacent to a $i\Mc_0$. This means that the quantity that emerges is 
\begin{align}
    i \Mc_0 \frac{1}{1 + \Cc \, \Mc_0}
    =
    i \left[\Mc_0^{-1} + \Cc \right]^{-1}.
\end{align}
With these properties and examples at hand, we can write down the final expression for $\Wc_{\df,\Ec^A}^A$ as
\begin{align}
    i\Wc_{\df,\Ec^A}^A &=
    i\mathbf{W}^A_{\Ec}
    +\left(1 + i\Lc \right) \, i \big[ \Mc_{0}^{-1} + \Cc \big]^{-1} i\Wb^A_{\Rc}
    +i\Wb^A_{\Lc} \, i \big[ \Mc_{0}^{-1} + \Cc \big]^{-1} \left(1+ i\Rc \right) 
    \\
    &\hspace{2cm}+\left(1 + i\Lc \right) \, i \big[ \Mc_{0}^{-1} + \Cc \big]^{-1} i\Wb^A_{\Cc}
    \, i \big[ \Mc_{0}^{-1} + \Cc \big]^{-1} \left(1 + i\Rc \right) 
\end{align}
where we have defined the following functions
\begin{align}
i\Wb^A_{\Rc}
&\equiv
\Yc^T \cdot i\Delta_{2} \cdot i\mathbf{W}^A_{\Ec}  \,,
\label{eq:RL_AE}
\\
i\Wb^A_{\Lc}
& \equiv 
i\mathbf{W}^A_{\Ec} \cdot i\Delta_{2} \cdot \Yc  \,,
\label{eq:LL_AE}
\\
i\Wb^A_{\Cc}
&\equiv 
\Yc^T \cdot i\Delta_{2} \cdot i\mathbf{W}^A_{\Ec} \cdot i\Delta_{2} \cdot \Yc \,.
\label{eq:CL_AE}
\end{align}

One can simplify the notation here by introducing
\begin{align}
\widetilde{\Lc} &\equiv \left(1 + i\Lc \right) \,  \big[ \Mc_{0}^{-1} + \Cc 
\big]^{-1} ,
\label{eq:Ltilde}
\\
\widetilde{\Rc} &\equiv  \big[ \Mc_{0}^{-1} + \Cc \big]^{-1} 
\left(1 + i\Rc \right) \, .
\label{eq:Rtilde}
\end{align}
With these quantities, one can write $\Wc_{\df, \Ec^A}^A $ as
\begin{align}
   i\Wc_{\df, \Ec^A}^A =
    i\mathbf{W}^A_\Ec
    +i\widetilde{\Lc}\, i\Wb^A_{\Rc}
    +i\Wb^A_{\Lc} \, i\widetilde{\Rc} 
    + i\widetilde{\Lc} \,  i\Wb^A_{\Cc}
    \, i\widetilde{\Rc}.
    \label{eq:WdfE}
\end{align}
This result should be understood as a matrix equation in angular momentum space, where all kernels have been partial-wave projected.

The remaining contribution to $\Wc_{\df, \rm 2B}^A$ is that arising from a $\widetilde\Bc^A$ insertion, $\Wc_{\df, \slashed{\Ec}^A}^A$. The skeleton expansion for this term can be written as
\begin{equation}
i\Wc_{\df, \slashed{\Ec}^A}^A
    =i\widetilde\Bc^A
    +i\mathcal{M}\cdot i\Delta_2 \cdot i\widetilde\Bc^A
    +i\widetilde\Bc^A \cdot i\Delta_2 \cdot i\mathcal{M}
    +i\mathcal{M}\cdot i\Delta_2 \cdot i\widetilde\Bc^A \cdot i\Delta_2 \cdot i\mathcal{M}\,,
    \label{eq:WslashEIntEq}
\end{equation}
which we can again rewrite as a geometric series in insertions of $\widetilde \Bc$ and $\Ec$:
\begin{equation}
i\Wc_{\df, \slashed{\Ec}^A}^A
    = \sum_{n,m=0}^\infty [(i\widetilde{\Bc} + i\Ec)\cdot i\Delta_2 \, \cdot {}]^n
     \, i\widetilde\Bc^A\, [{}\cdot\, i\Delta_2\, \cdot(i\widetilde{\Bc} + i\Ec) ]^m\,.
     \label{eq:WdfEslash}
\end{equation}

We begin by considering the simpler case with no OPEs, with all $\Ec$ insertions in the series above dropping out. Since the $\widetilde\Bc$ and $\widetilde\Bc^A$ kernels are on shell, we bring their partial-wave projections outside any loop integrations, leaving only the modified spherical harmonics as endcaps. As a result, the loop integrals collapse to
\begin{equation}
\Yc_{\ell'm'}^T \cdot i\Delta_2 \cdot \Yc_{\ell m} = i( \Ic_{\rm pv} - i\rho ) \, \delta_{\ell'\ell}\delta_{m'm} \,,
\end{equation}
where the right-hand side is the standard result that follows from decomposing the two-particle pole $\Delta_2$ into a principal value piece and a Dirac delta. This leads to a smooth real contribution $\Ic_{\rm pv}$ and a phase space factor. Applying this result and dressing the kernel $\widetilde\Bc$ with $\Ic_{\rm pv}$ insertions to all orders through the relation $\Kc_0^{-1} = \widetilde\Bc^{-1} - \Ic_{\rm pv}$, we can bring the series above to the form
\begin{align}
i\Wc_{\df, \slashed{\Ec}^A}^A \Big\vert_{\Ec = 0} 
&= \sum_{n,m=0}^\infty [i\Kc_0 \, \rho ]^n \, (1 - \Kc_0 \Ic_{\rm pv}) \, i\widetilde\Bc^A \, (1 - \Ic_{\rm pv} \Kc_0 ) \, [\rho \, i\Kc_0 ]^m,
\\
&= \Mc_0 \, i\Ac^A \Mc_0 ,
\label{eq:WdfEslash_noE}
\end{align}
where we have introduced a new nonsingular quantity $\Ac^A \equiv (\Kc_0 ^{-1} - \Ic_{\rm pv}) \widetilde\Bc^A (\Kc_0^{-1}  - \Ic_{\rm pv})$ in the last equality. It is worth noting that this is identical in structure to Eq.~\eqref{eq:iWdf0}, but with $\Mc_0$ replacing the full hadronic amplitude $\Mc$. This correspondence is, of course, expected in the absence of OPEs.

We now move to incorporating the OPEs into the final- and initial-state rescatterings. Recalling the skeleton expansion of Eq.~\eqref{eq:WdfEslash}, we can start by grouping all combinations of consecutive $\widetilde\Bc$ which neighbor or sandwich an insertion of $\widetilde\Bc^A$. This grouping simply reproduces the $\Wc_{\df, \slashed{\Ec}^A}^A\big\vert_{\Ec = 0}$ object we have already examined. Doing just this, Eq.~\eqref{eq:WdfEslash} can be rewritten as,
\begin{equation}
i\Wc_{\df, \slashed{\Ec}^A}^A
    = \left[ 1 + \sum_{n=0}^\infty [(i\widetilde{\Bc} + i\Ec)\cdot i\Delta_2 \, \cdot {}]^n i\Ec \cdot\, i\Delta_2\cdot {} \right]
     \left(i\Wc_{\df, \slashed{\Ec}^A}^A\big\vert_{\Ec = 0}\right)
     \left[ {}\cdot i\Delta_2 \cdot i\Ec \sum_{m=0}^\infty [{}\cdot i\Delta_2 \cdot (i\widetilde{\Bc} + i\Ec)]^m + 1 \right]
     \,.
\label{eq:WdfEslash_v1}
\end{equation}
Looking at each of the endcaps in brackets, we can group together consecutive $\Ec$ insertions to obtain $\Mc_\Ec$ auxiliary amplitudes, and consecutive $\widetilde\Bc$ insertions to obtain $\Mc_0$ amplitudes. Taking the bracket on the left as an example, we can show that the following combination arises
\begin{equation}
\left[ 1 + \sum_{n=0}^\infty [(i\widetilde{\Bc} + i\Ec)\cdot i\Delta_2 \, \cdot {}]^n i\Ec \cdot\, i\Delta_2\cdot {} \right] 
= (1 + i\Mc_\Ec\cdot i\Delta_2  \cdot {} ) \sum_{n=0}^\infty [i\Mc_0 \cdot i\Delta_2 \cdot i\Mc_\Ec\cdot i\Delta_2  \cdot {}]^n \,.
\end{equation}
From the result of Eq.~\eqref{eq:WdfEslash_noE}, we see the combination above is followed on the right by an $\Mc_0$. Projecting all $\Mc_0$ to spherical harmonics and grouping the resulting harmonics together with the $\Mc_\Ec$ amplitudes to form $\Lc$ and $\Cc$ factors, we can then show
\begin{align}
(1 + i\Mc_\Ec\cdot i\Delta_2  \cdot {} ) \sum_{n=0}^\infty [i\Mc_0 \cdot i\Delta_2 \cdot i\Mc_\Ec\cdot i\Delta_2  \cdot {}]^n \Mc_0 = (1 + i\Lc) \, [\Mc_0^{-1} + \Cc]^{-1} = \widetilde \Lc \,,
\end{align}
where $\widetilde \Lc$ is the endcap defined in Eq.~\eqref{eq:Ltilde}. We proceed analogously for the right-hand brackets in Eq.~\eqref{eq:WdfEslash_v1}, which yields an endcap $\widetilde\Rc$ and leads to the final result
\begin{equation}
i\Wc_{\df, \slashed{\Ec}^A}^A
    = \widetilde\Lc \, i\Ac^A \,\widetilde \Rc
     \,.
\label{eq:WdfEslash_vf}
\end{equation}

Putting this and Eq.~\eqref{eq:WdfE} together, we get the total 2B amplitude
\begin{align}
   i\Wc_{\df, \rm 2B}^A =
    i\mathbf{W}^A_\Ec
    +i\widetilde{\Lc}\, i\Wb^A_{\Rc}
    +i\Wb^A_{\Lc} \, i\widetilde{\Rc} 
    + \widetilde{\Lc} (i\Ac^A -i\Wb^A_{\Cc})
    \, \widetilde{\Rc}\,.
    \label{eq:Wdf2B}
\end{align}
We emphasize that all $\Wc_{\df, \Ec^A}^A$ terms and the endcap factors in $\Wc_{\df, \slashed\Ec^A}^A$ are fixed once the OPE couplings and the on-shell subprocess input entering $\Ec^A$ are specified.
The object $\Ac^A$, on the other hand, is unique to this type of transition and not known, and must be determined either from experimental data or through non-perturbative methods such as lattice QCD.

\subsubsection{Derivation of $\Wc_{\df, \rm 1B}^A$}
\label{sec:Wdf1B}
\begin{figure}
    \centering
    \includegraphics[width=0.8\linewidth]{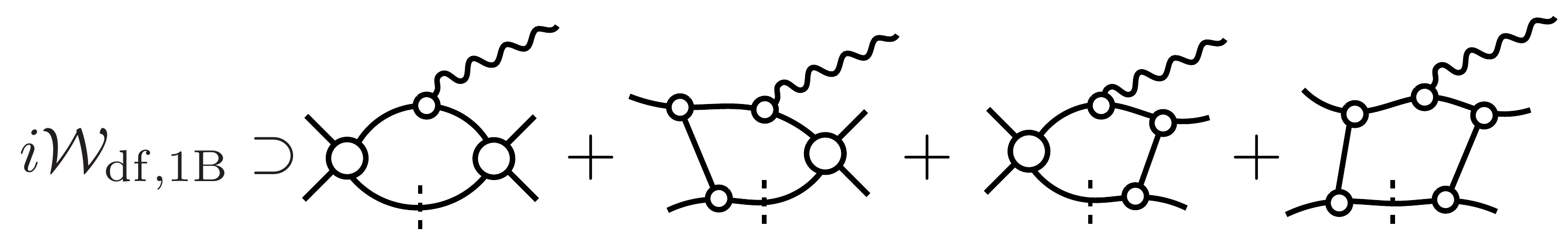}
    \caption{Shown are the four types of diagrams that can have triangle-type singularities.}
    \label{fig:gen_triangles}
\end{figure}

Having considered the contributions due to $\Ec^A$ and $\widetilde \Bc^A$ insertions, we now turn our attention to $\Wc_{\df, \rm 1B}^A$. This comprises all diagrams in which the current attaches to one of the particles, which are not being exchanged, in a two-particle loop. We refer to these as ``triangle-type'' diagrams. Having already performed the reduction from four- to 3D loop integrals in the skeleton expansion, $\Wc_{\df, \rm 1B}^A$ can be expressed as a series in insertions of $\Ec$ and $\widetilde\Bc$ kernels:
\begin{align}
i\Wc_{\df, \rm 1B}^{A} &= \sum_j if_j ~ i\Mc \cdot [-K^A_j \, i\Delta_3] \cdot i\Mc  \,, \\
& = \sum_j if_j \sum_{m,n=0}^\infty
\left[ (i \widetilde\Bc + i\Ec) \cdot i\Delta_2 \cdot {} \right]^n (i \widetilde\Bc + i\Ec)
\cdot [-K^A_j \, i\Delta_3] \cdot
(i \widetilde\Bc + i\Ec)
\left[ {} \cdot  i\Delta_2 \cdot (i \widetilde\Bc + i \Ec) \right]^m \,, 
\label{eq:Wdf1B_skel}
\end{align}
where we use the shorthand dot product between generic left- and right-hand kernels $\mathfrak{L}\cdot [K^A_j \, i\Delta_3] \cdot \mathfrak{R}$ to indicate integration over the loop with a current insertion, which arises from 4D integrals with at least three propagators. As before, $K^A_j$ is the kinematic Lorentz structure associated with the form factor $f_j$, while $i\Delta_3$ is the product of the three propagators in the loop after the integral reduction,
\begin{align}
i\Delta_3(P_f, P_i, \mathbf{k})
\equiv \frac{1}{2 \omega_1(\mathbf{k })}
\frac{H_f(k^\star_f)}{(P_f-k)^2-m_2^2+i \epsilon}
\frac{H_i({k}^\star_i)}{(P_i-k)^2-m_2^2+i \epsilon}\bigg|_{k_0 = \omega_1(\mathbf{k })},
\end{align}
where we specify that each cutoff function, $H_{i/f}$, must be equal to 1 as its argument tends to $q^\star_{i/f}$.
In contrast to the $\Wc^\mu$ result derived in Ref.~\cite{Briceno:2020vgp} and reviewed in Sec.~\ref{sec:W_review}, where a single class $\Gc^A_j$ of triangle diagram emerges, the distinct treatment of $\Ec$ and $\widetilde\Bc$ in the present derivation, owing to their different kinematic configurations, leads to a more complicated structure: we must distinguish four classes of triangle-type diagram, based on which of these two types of kernels flanks the triangle loop. The origin of the four different classes is depicted in Fig.~\ref{fig:gen_triangles}.

Let us start by looking at the simplest case, a loop with $\widetilde\Bc$ kernels on both sides, which corresponds to the first term in Fig.~\ref{fig:gen_triangles}. Because these kernels are on shell, we can project them to spherical harmonics and bring them out of the loop integration altogether, 
\begin{align}
i\widetilde\Bc \cdot [-K^A_j \, i\Delta_3] \cdot i\widetilde\Bc 
    &\equiv 
    \sum_{\ell m,\ell' m'}
    \widetilde\Bc_{\ell}(q^\star_f)
    \left[\int\frac{d^3\mathbf{k^\star}}{(2\pi)^3} 
     \,
    \Yc_{\ell m}^T(\mathbf{k}^\star_f,q^\star_f)
    \,
    K_j^A\,i\Delta_{3}(\mathbf{k^\star})
    \,
    \Yc_{\ell' m'}(\mathbf{k}^\star_i,q^\star_i)
    \right]
     \widetilde\Bc_{\ell'}(q^\star_i),
\end{align}
and we define the first class of triangle function as a matrix in partial-wave space, with elements given by 
\begin{align}
\left( \Gc_{0,j}^A \right)_{\ell'm', \ell m} &\equiv \Yc^T_{\ell'm'} \cdot [K^A_j \, i\Delta_3] \cdot \Yc_{\ell m} \,, \\
&= \int\! \frac{d^3\mathbf{k}}{(2\pi)^3} ~
     \Yc^T_{\ell' m' }(\mathbf{k}^\star_f,q^\star_f)   \, K_j^A(k_f,k_i) \, i\Delta_{3}(P_f, P_i, \mathbf{k})
     \, 
      \Yc_{\ell m }(\mathbf{k}^\star_i,q^\star_i) \,.
      \label{eq:GcalA}
\end{align}
This matrix has the same singularity structure as that of Eq.~\eqref{eq:GA_4D} and differs only up to an analytic function of the external kinematics.
This first kind of loop, once initial and final states are partial-wave projected, can be written as the product $i\widetilde\Bc \, i\Gc_{0,j}^A \,i\widetilde\Bc $ of matrices in angular momentum space.
The triangle diagram that emerges here is essentially identical to that introduced in previous work and reviewed in Sec.~\ref{sec:W_review}. However, instead of the 3D integral above, $\Gc^A_j$ was previously written as a four-dimensional Feynman integral, which is more amenable to analytic evaluations. In general, one can use any representation for this function, as long as it has the same analytic structure. The difference between one choice and another is absorbed into the definitions of the short-distance $\Ac^A$ functions. For this function, reducing the number of integrals from four to three seems a slightly unnecessary exercise that could complicate the evaluation of the remaining integrals, but for the other three classes of triangle-type function shown in Fig.~\ref{fig:gen_triangles}, this reduction leads to a more useful representation.

To see why, we turn now to these other classes. As previously mentioned, these arise from having an $\Ec$ insertion on one side, or both, of the triangle loop. Unlike the $\widetilde\Bc$, the OPE kernels are only partially on shell and cannot be taken out of loop integrations even after spherical harmonic projections, since they carry dependence on the magnitude of loop momenta. Note also that, if one exchange is present on either side of the loop, one must also sum all possible exchanges to all orders, which amounts to considering $\Mc_\Ec$ instead of $\Ec$ on the side of the loop. These integrals can be done by evaluating $\Mc_\Ec$ using the integral equation given in Eq.~\eqref{eq:MEt} first. This class of integral equations can currently only be solved numerically. Consequently, the remaining triangle functions are ones that we only know how to evaluate numerically. Reducing the integrals from four to three dimensions analytically is therefore advantageous.

With this in mind, let us look now at each of the triangle diagrams involving $\Mc_\Ec$ more concretely. Using the building blocks introduced above, we can write these compactly as
\begin{align}
    \mathbf G^A_{\Rc,j} &\equiv \Yc^T \cdot \left[K_j^A \, i\Delta_3 \right]  
    \cdot  \Mc_\Ec \,,
     \\
  \mathbf G^A_{\Lc, j} &\equiv   \,
         \Mc_\Ec  \cdot \left[K_j^A \, i\Delta_3 \right]  \cdot \Yc \,,
     \\
 \mathbf G^{A}_{\Ec,j} &\equiv    \,
         \Mc_\Ec \cdot \left[K_j^A \, i\Delta_3 \right]   
    \cdot  \Mc_\Ec \,.
\end{align}
Here, $\mathbf G^A_{\Rc,j}$ and $\mathbf G^A_{\Lc, j}$ originate from triangles with a $\widetilde\Bc$ on the left or right, respectively, that has been removed from the loop integral. In addition, $\mathbf G^{A}_{\Ec,j}$ originates from a loop with $\Mc_\Ec$ amplitudes on both sides. We adopt a different notation compared to $\Gc_{0,j}^A$ here, and the reason for this should become apparent below.

The four types of triangle function described must be dressed with all possible initial and final state rescatterings. For the first type of triangles, $\Gc_{0,j}^A$, we start by investigating the case with no OPEs, where this in fact the only class of triangle that contributes. Given the skeleton expansion in Eq.~\eqref{eq:Wdf1B_skel} and the results below it, it is easy to obtain
\begin{align}
i\Wc_{\df, \rm 1B}^A \Big|_{\Ec = 0} = \sum_j f_j \,\Mc_0 \, i\Gc_{0,j}^A \, \Mc_0 \,,
\end{align}
which is the expected result in the absence of OPEs, in line with the triangle contribution in Eq.~\eqref{eq:iWdf0}. To accommodate all possible further rescattering with OPEs, we note that we can piggyback off the result of Eq.~\eqref{eq:WdfEslash_v1}. This is because an analogous structure emerges relating the general contribution of the $\Gc_{0,j}^A$ triangle, which we denote $i\Wc_{\df, \rm 1B,0}^{A}$ below, to the contribution with no OPEs. Skipping to the final result, which mirrors Eq.~\eqref{eq:WdfEslash_vf}, we have
\begin{equation}
i\Wc_{\df, \rm 1B, 0}^{A} = \sum_j f_j \, \widetilde\Lc \, i\Gc_{0,j}^A \, \widetilde\Rc \,,
\end{equation}
using the endcap factors defined in Eqs.~\eqref{eq:Ltilde} and \eqref{eq:Rtilde}.

The dressing patterns for the remaining triangle functions are slightly more involved, but we can again take advantage of previous results. Let us look first at the $\mathbf G^A_{\Rc, j}$ diagram first. By definition, this must be followed on the left in a skeleton diagram by a $\widetilde\Bc$ kernel, or, summing to all orders in $\widetilde\Bc$, an $\Mc_0$ amplitude. This reproduces the dressing pattern of $\Gc^A_{0, j}$ on the left, and we can similarly accommodate OPE insertions through the replacement $\Mc_0\, \mathbf G^A_{\Rc, j} \to \widetilde\Lc\, \mathbf G^A_{\Rc, j}$. On the right-hand side, we need to account separately for diagrams where $\mathbf G^A_{\Rc, j}$ appears at the end of the string of diagrams, and those where it attaches to initial state rescatterings. The latter must again begin with a $\widetilde\Bc$, and hence $\Mc_0$, insertion, i.e.~$\mathbf G^A_{\Rc, j} \cdot i\Delta_2 \cdot i\Mc_0$. We can then resort to the previously-defined endcaps to include all rescattering with OPEs: $\mathbf G^A_{\Rc, j} \cdot i\Delta_2 \cdot i\Mc_0 \to \mathbf G^A_{\Rc, j} \cdot i\Delta_2 \cdot i\widetilde \Rc$.
The resulting contribution is
\begin{align}
i\Wc_{\df, \rm 1B, \Rc}^{A} &= \sum_j f_j \, i\widetilde\Lc \left( \mathbf G_{\Rc,j}^A +  i\mathbf G^A_{\Rc\Lc, j}\widetilde\Rc\right) \,,\\
i\Wc_{\df, \rm 1B, \Lc}^{A} &= \sum_j f_j \, (\, \mathbf G_{\Lc,j}^A + \widetilde\Lc \, i\mathbf G_{\Lc\Rc,j}^A)  \, i\widetilde\Rc \,,
\end{align}
where we also generalized for $\mathbf G^A_{\Lc, j}$ case as it has an identical structure, once the order of kernels is reversed and we exchange $\Lc \leftrightarrow \Rc$.
For these equations we have introduced the following kernels
\begin{align}
    i\mathbf G^A_{\Rc\Lc, j} &\equiv i\mathbf G^A_{\Rc, j} \cdot i\Delta_2 \cdot \Yc\,,\\
    i\mathbf G^A_{\Lc\Rc, j} &\equiv \Yc^T \cdot i\Delta_2 \cdot i\mathbf G^A_{\Lc, j} \,,
\end{align}
The dressing of $i\mathbf G^A_{\Ec, j}$ diagrams follows in a parallel manner, leading to
\begin{align}
i\Wc_{\df, \rm 1B, \Ec}^{A} &= \sum_j f_j \, (
 i\mathbf G_{\Ec,j}^A+
  i\widetilde\Lc \,i\mathbf G_{\Ec\Rc,j}^A+
   i\mathbf G_{\Ec\Lc,j}^A\,i\widetilde\Rc+
    i\widetilde\Lc\,i\mathbf G_{\Ec\Cc,j}^A\,i\widetilde\Rc )\,,
\end{align}
where we used the following generalizations of the $\mathbf G_{\Ec,j}^A$ kernel,
\begin{align}
  i\mathbf G_{\Ec\Rc,j}^A &\equiv \Yc^T \cdot i\Delta_2 \cdot i\mathbf G_{\Ec,j}^A \,,\\
    i\mathbf G_{\Ec\Lc,j}^A  &\equiv i\mathbf G_{\Ec,j}^A\cdot i\Delta_2 \cdot \Yc\,,\\
    i\mathbf G_{\Ec\Cc,j}^A & \equiv\Yc^T \cdot i\Delta_2 \cdot i\mathbf G_{\Ec,j}^A\cdot i\Delta_2 \cdot \Yc\,.
\end{align}

Adding the four classes of triangle contributions together, we can write the overall $i\Wc_{\df, \rm 1B}^A$ in the form
\begin{align}
i\Wc_{\df, \rm 1B}^A 
= \sum_j f_j \left[
\widetilde\Lc \, i\widetilde\Gc_j^A \, \widetilde\Rc 
+ \widetilde\Lc \, i\widetilde{\mathbf G}_{\Rc,j}^A
+ i\widetilde{\mathbf G}_{\Lc,j}^A \, \widetilde\Rc
+ i\mathbf G_{\Ec, j}^A
\right]
\label{eq:Wdf1B}
\end{align}
by combining the following terms
\begin{align}
\widetilde\Gc_j^A &\equiv \Gc_{0,j}^A + i\mathbf G^A_{\Rc\Lc, j} + i\mathbf G^A_{\Lc\Rc, j} - \mathbf G_{\Ec\Cc,j}^A \,, \\
\widetilde{\mathbf G}_{\Rc,j}^A &\equiv  \mathbf G_{\Rc,j}^A + i\mathbf G_{\Ec\Rc,j}^A \,, \\
\widetilde{\mathbf G}_{\Lc,j}^A &\equiv \mathbf G_{\Lc,j}^A + i\mathbf G_{\Ec\Lc,j}^A \,.
\end{align}

\subsection{Summary of the main result }

We collect the results of the previous sections to get our final result for the analytic decomposition of the $2+\Jc\to 2$ amplitude, including the effect of left-hand singularities,
\begin{align}
   i\Wc_{\df}^A =
    i\mathbf{W}^A_\Ec + [f\cdot i\mathbf G_{\Ec}^A]
    +\widetilde{\Lc}\, \left([f\cdot i\widetilde{\mathbf G}_{\Rc}^A]-\Wb^A_{\Rc}\right)
    +\left( [f\cdot i\widetilde{\mathbf G}_{\Lc}^A]-\Wb^A_{\Lc} \right)\, \widetilde{\Rc} 
    + \widetilde{\Lc} \left( i\Ac^A- i\Wb^A_{\Cc}
    + [f\cdot i\widetilde\Gc^A ] \right)\,
    \, \widetilde{\Rc}\,,
    \label{eq:Wdftot}
\end{align}
where in this case we are using the dot ``$\cdot$'' product to simplify the notation regarding the sum over form factors $f_j$ and kernel functions with different kinematic tensors.
All the quantities in this expression, except for $\Ac^A$, are given in terms of sub-processes, e.g.\ the $2\to 2$ scattering, the single particle form factors, and on-shell couplings between the external particles and the light exchange particles.
Each of the entries of the angular momentum matrix $\Ac^A$ is a dynamical function of the external kinematics unique to this process, and most importantly, free of singularities both in the elastic region and below threshold in the vicinity of left-hand singularities.
Once the tensor structure of $\Ac^A$ has been taken into account, it can be parameterized in terms of any smooth function in the $s_f$, $s_i$, and $q^2$ variables that could be constrained using experimental or lattice data.

The result of Eq.~\eqref{eq:Wdftot} is especially relevant when the two-particle system couples to a bound state, with energy in the vicinity of the left-hand cut of the $2\to2$ scattering, providing a clear distinction between the dynamical and kinematic singularities of the amplitude.
In that case, Eq.~\eqref{eq:Wdftot} can be used to rigorously extract the bound state form factors, while rigorously controlling the effect of left-hand singularities on the overall amplitude.

\subsection{Reducing 3D triangular integrals down to 2D}
\label{sec:triangles_2d}
Taking steps akin to those presented in Ref.~\cite{Baroni:2018iau}, the azimuthal dependence in the triangle-type integrals can be isolated and evaluated analytically. This would then reduce the dimensionality of the numerical integrals that must be evaluated from 3D down to 2D. We illustrate this idea with the integral appearing in Eq.~\eqref{eq:GcalA} for $\Gc_{0,j}^A $. 

Without loss of generality, we work in the final-state CMF and choose the $z$ axis to be parallel to the boost relating the initial and final CMFs. This implies that the two four-momenta in this frame can be written as,
\begin{align}
    P_f^{\mu}&=(\sqrt{s_f}, 0, 0, 0),
\\
P_i^{\mu}&=(\gamma\sqrt{s_i},0,0,\gamma\beta\sqrt{s_i}) \,,
\end{align}
where $\gamma$ and $\beta$ are related in the standard way, $\gamma = (1-\beta^{2})^{-1/2}$ and $\gamma$ can be written in terms of $P_f$ and $P_i$ as $\gamma = P_f\cdot P_i /  \sqrt{s_fs_i}=(s_f+s_i-q^2)/(2\sqrt{s_fs_i})$.

Different terms in the integral of $\Gc_{0,j}^A $ require us to evaluate the loop momentum in the initial and final frames.
We continue to choose the ``lab" frame to be the CMF of the final state, meaning that the loop momentum is already evaluated in the final-state frame, and we can perform an inverse boost to evaluate it in the initial frame. Given this, the loop momentum is
\begin{align}    
k^{\mu}&\equiv k^{\mu\star}_f 
\\
&= \big(\omega_1,k\sqrt{1-z^2}\cos\phi,
    k\sqrt{1-z^2}\sin\phi,kz\big),
\end{align}
where $z=\cos\theta$. Performing an inverse boost along the negative z-axis, we get $k^{\mu\star}_i$
\begin{align}    
 k^{\mu\star}_i 
= \big(\gamma(\omega_1-\beta kz),k\sqrt{1-z^2}\cos\phi,
    k\sqrt{1-z^2}\sin\phi,\gamma ( kz-\beta\omega_1)\big).
    \label{eq:kmu_i}
\end{align}
The dependence of this variable appears in two places. The first is the modified spherical harmonics, $\Yc$ and $\Yc^T$ given in Eqs.~\eqref{eq:Yharm_def} and \eqref{eq:YharmT_def}, and the second is the cutoff function, $H$. In both of these, only the spatial components appear. From Eq.~\eqref{eq:kmu_i}, we see that the magnitude of the spatial momentum, $\mathbf k_i^\star$, is equal to, 
\begin{align}
    k_i^\star(k,z)&=\sqrt{k^2(1-z^2)+\gamma^2(kz-\beta\omega_1)^2}\,,
     \,.
\end{align}
This only depends on $k$ and $z$. The spherical harmonics also depend on the angle $z_i$ and $\phi_i$. From Eq.~\eqref{eq:kmu_i}, it is easy to find  
\begin{align}
        z_i&=\frac{\gamma(kz-\beta\omega_k)}{k_i^\star(k,z)}\,,
    \\
    \phi_i&=\phi\,,
\end{align}
where in the first equality we have used the fact that $k_i^\star z_i$ is by definition the component of $ k^{\mu\star}_i$ along the boost direction.
The key observation is that given the boost we have chosen, $\phi_i$ is independent of $k$ and $z$. 
The cutoff function, $H$, only depends on the magnitude of these momenta and therefore does not introduce any additional azimuthal dependence.

We can then write the two spectator denominators in $i\Delta_3$ in terms of these variables,
\begin{align}
    D_f(k)&=s_f+m_1^2-m_2^2-2\sqrt{s_f}\,\omega_1+i\epsilon\,,
    \\
    D_i(k,z)&=s_i+m_1^2-m_2^2
    -2\gamma\sqrt{s_i}\left(\omega_1-\beta k z\right)+i\epsilon\,,
\end{align}
where we have made the $z$ dependence explicit. Again, these do not depend on $\phi$. 

This then motivates the azimuthal average of all quantities that depend explicitly on $\phi$
\begin{equation}
    \overline{\mathcal K}^{A}_{j;\ell' m',\ell m}(k,z)
    \equiv
    \frac{1}{2\pi}\int_0^{2\pi} d\phi\,
    \Yc^T_{\ell' m'}(\mathbf k_f^\star,q_f^\star)
    K_j^A(k_f,k_i)
    \Yc_{\ell m}(\mathbf k_i^\star,q_i^\star)\, .
    \label{eq:KAbar}
\end{equation}
The key observation is that given some Lorentz tensor $K^A_j$, one can evaluate the integral over $\phi$ in Eq.~\eqref{eq:KAbar} analytically.

Equation~\eqref{eq:GcalA} can be written as a two-dimensional integral
\begin{align}
\left( \Gc_{0,j}^A \right)_{\ell'm',\ell m}
    =\frac{1}{(2\pi)^2}
    \int_0^\infty dk\,k^2\int_{-1}^{1} dz\,
    \frac{H_f(k)H_i(k_i^\star)}{2\omega_1\,D_f(k)D_i(k,z)}
    \,\overline{\mathcal K}^{A}_{j;\ell' m',\ell m}(k,z)\, .
\label{eq:G0_2D}
\end{align}
 As a result, the only non-trivial integrals to evaluate numerically are those over $k$ and $z$.

To make this more explicit, let us consider the first non-trivial example, where $K_j^A$ is a rank-1 tensor, e.g. $K_j^A = k^\mu$. If $\ell=\ell'=0$, it is easy to see that only the $\mu =0,3$ components of $ \overline{\mathcal K}^{\mu}_{00,00}$ would be non-zero and the average of $\phi$ would give 1. Therefore, the first non-trivial example would require that either $\ell$ or $\ell'$ is $1$. With this in mind, let $\ell'=0$, and $\ell, |m| = 1,1$. Using the fact that $Y_{1, \pm 1}(\theta,\phi) = \mp\frac{1}{2}\sqrt{\frac{3}{2\pi}}e^{\pm i \phi}\sin\theta$, it is easy to show that, 
\begin{align}
    \overline{\mathcal K}^{\mu}_{00,1\pm1}(k,z)
    = \sqrt{\frac{3}{8}} \, k \,\frac{k_i^\star(k,z)}{q_i^\star}  \left(1-z^2 \right) \, \times (0,\mp1,-i,0).
\end{align}
For the $\ell ,m = 1,0$, the spherical harmonic has no $\phi$ dependence, and as a result only the $\mu = 0,3$ components are non-zero and the $\phi$ average again gives 1 for these components. 

\subsection{Singularities of $\Ec^A$}
\label{sec:Sing_ECA_v0}

The single- and double-pole diagrams (right and left panel, respectively, of figure~\ref{fig:iEA}) contributing to $\Ec^A$, Eq.~\eqref{eq:Ec_At}, have singularities when $s_i$, $s_f$, or both are below the two-particle threshold. In Appendix~\ref{app:PWA}, we give a more detailed account of the partial-wave projection and singularity structure of these functions. Here we summarize some of the key takeaways. For convenience, we consider the same kinematics discussed in the previous subsection, where the initial and final CMFs are related by a boost along the $z$-axis. 

\begin{figure}
    \centering
    \begin{subfigure}[b]{\textwidth}
        \includegraphics[width=0.95\linewidth]{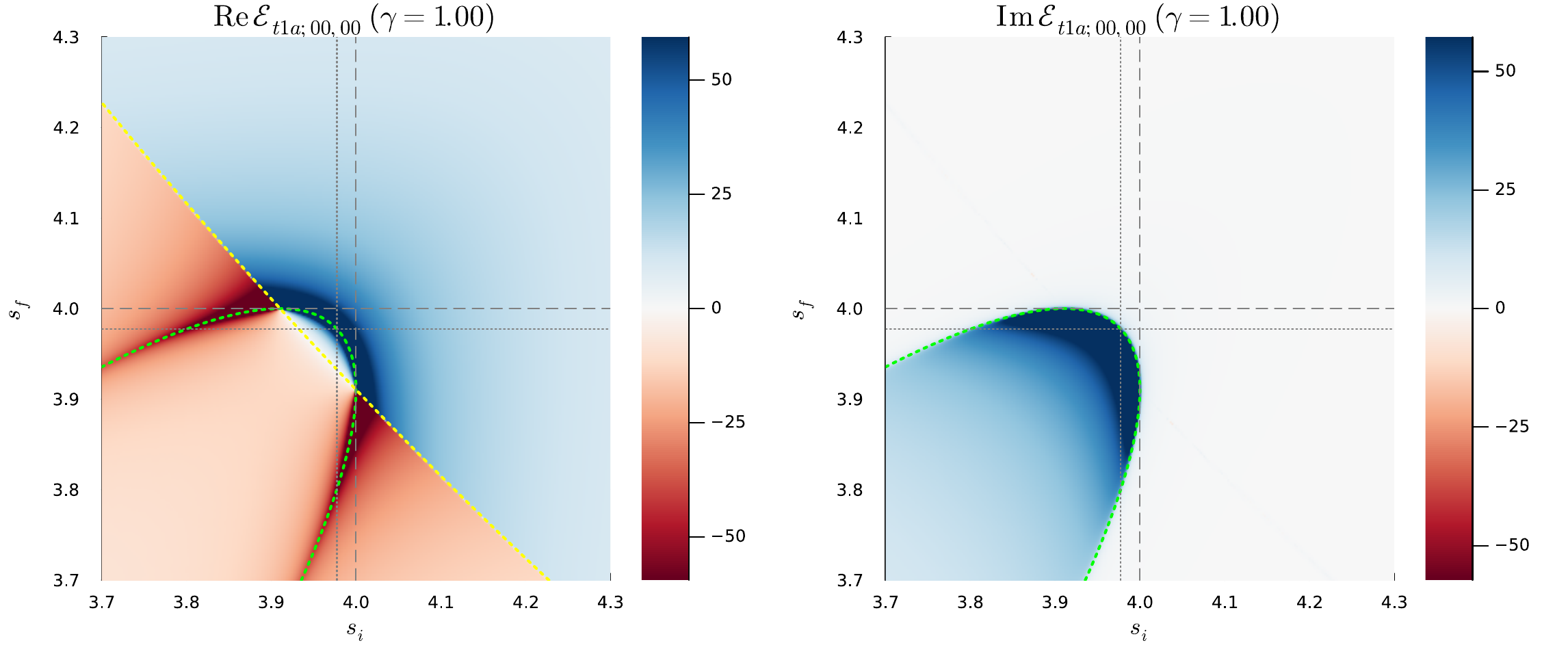}
    \end{subfigure}
    \begin{subfigure}[b]{\textwidth}
        \includegraphics[width=0.95\linewidth]{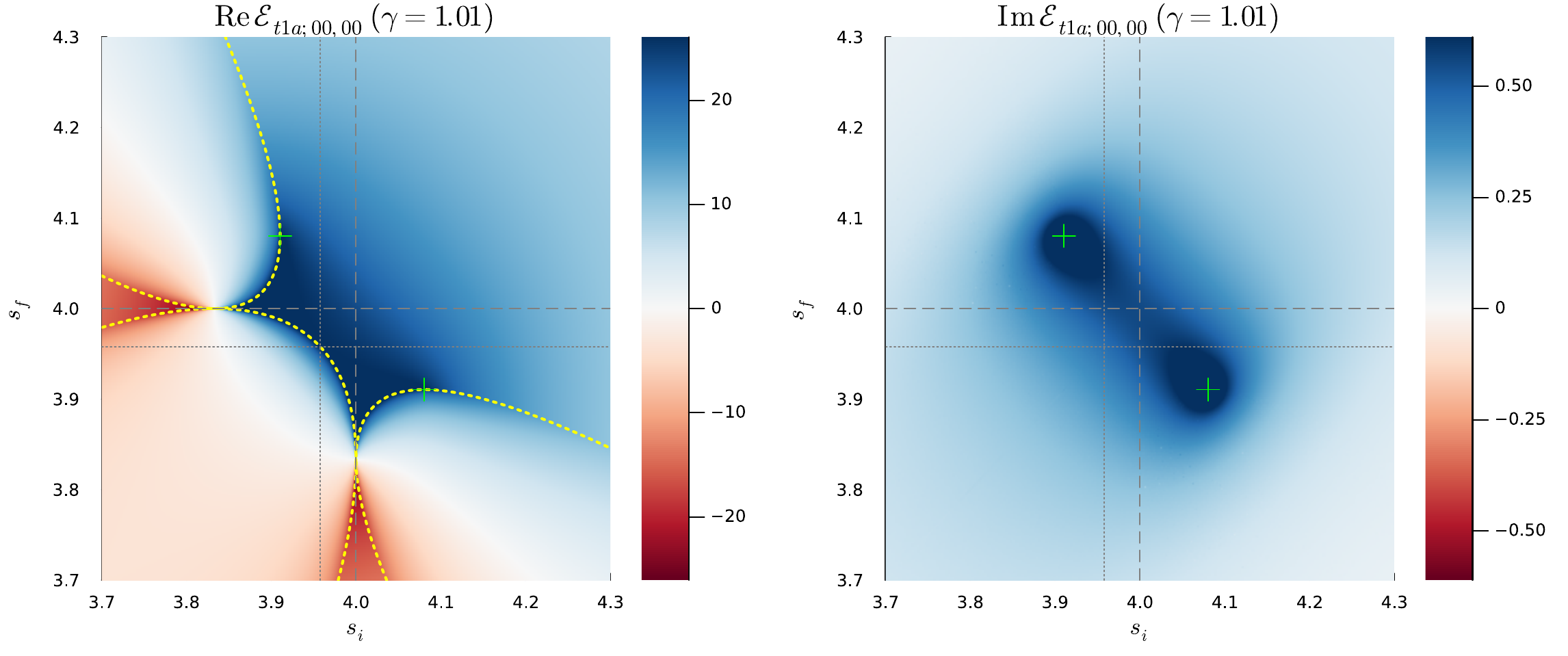}
    \end{subfigure}
    \begin{subfigure}[b]{\textwidth}
        \includegraphics[width=0.95\linewidth]{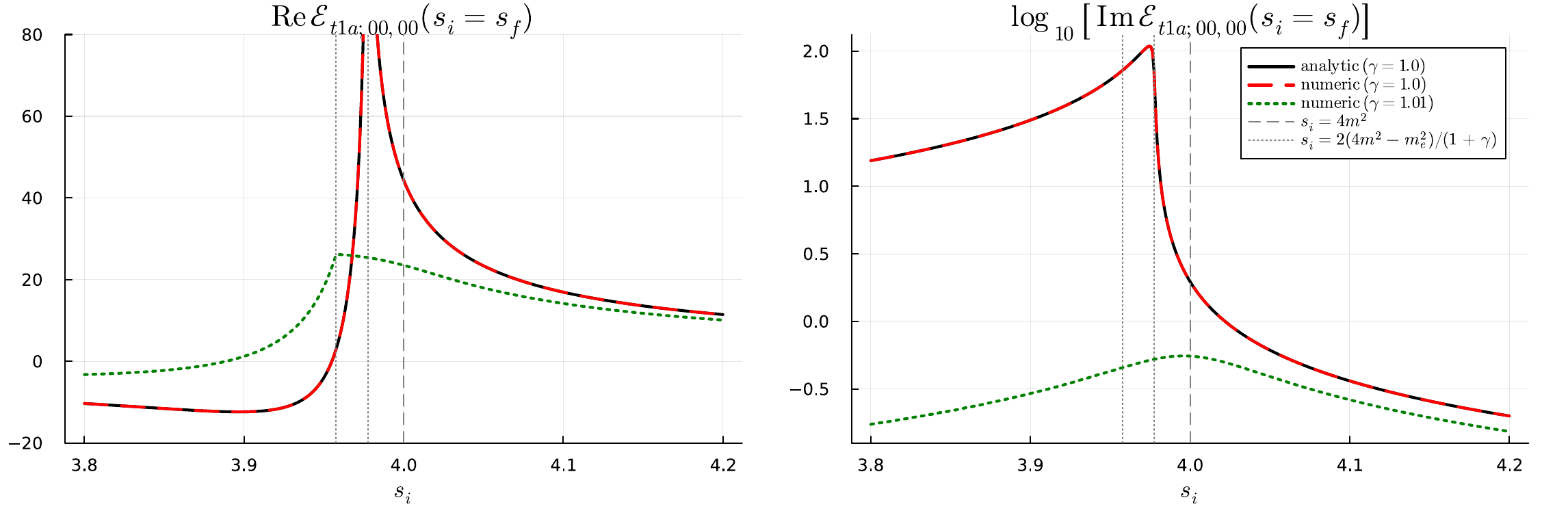}
    \end{subfigure}
    \caption{Heat maps on the $(s_i, s_f)$-plane of the real (left) and imaginary (right) parts of the Lorentz scalar $\mathcal{E}_{t1a;00,00}$ amplitude at $\gamma = 1$ (top panels) and $\gamma = 1.01$ (middle panels), and the amplitude on the line $s_i = s_f$ at those same values of $\gamma$ (bottom panels), obtained from a numerical solution of the integral (see Appendix~\ref{app:PW_2D}). We set $\epsilon = 10^{-3}$, $m=1$ and $m_e = 0.15$, corresponding roughly to the ratio of proton and pion mass. The dashed lines mark the two-particle thresholds $s_i,s_f=4m^2$. For $\gamma=1$, the dot-dashed guide lines at $s_i,s_f=4m^2-m_e^2$ mark the left-hand branch point. For $\gamma>1$, $s_i,\; s_f = 2(4m^2 - m_e^2)/(\gamma + 1)$ is also a branch point in $\mathcal{E}_{t2;00,00}$ but not $\mathcal{E}_{t1a;00,00}$. On the heat maps, singular points and contours are marked in green, other non-analytic contours in yellow. In the bottom panels, the analytic result of the partial-wave projection at $\beta = 0$ is overlaid on top of the numeric one.}
    \label{fig:singularities_single}
\end{figure}

\begin{figure}
    \centering
    \begin{subfigure}[b]{\textwidth}
        \includegraphics[width=0.95\linewidth]{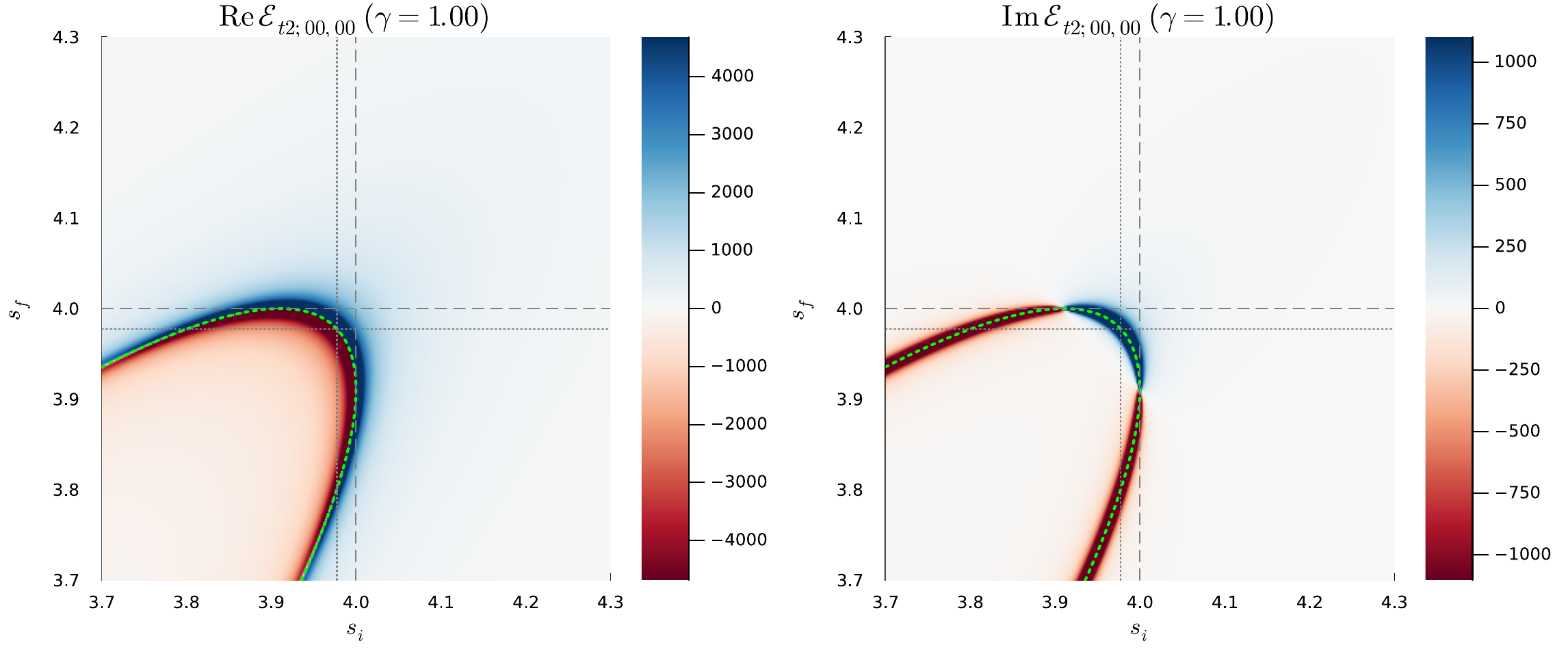}
    \end{subfigure}
    \begin{subfigure}[b]{\textwidth}
        \includegraphics[width=0.95\linewidth]{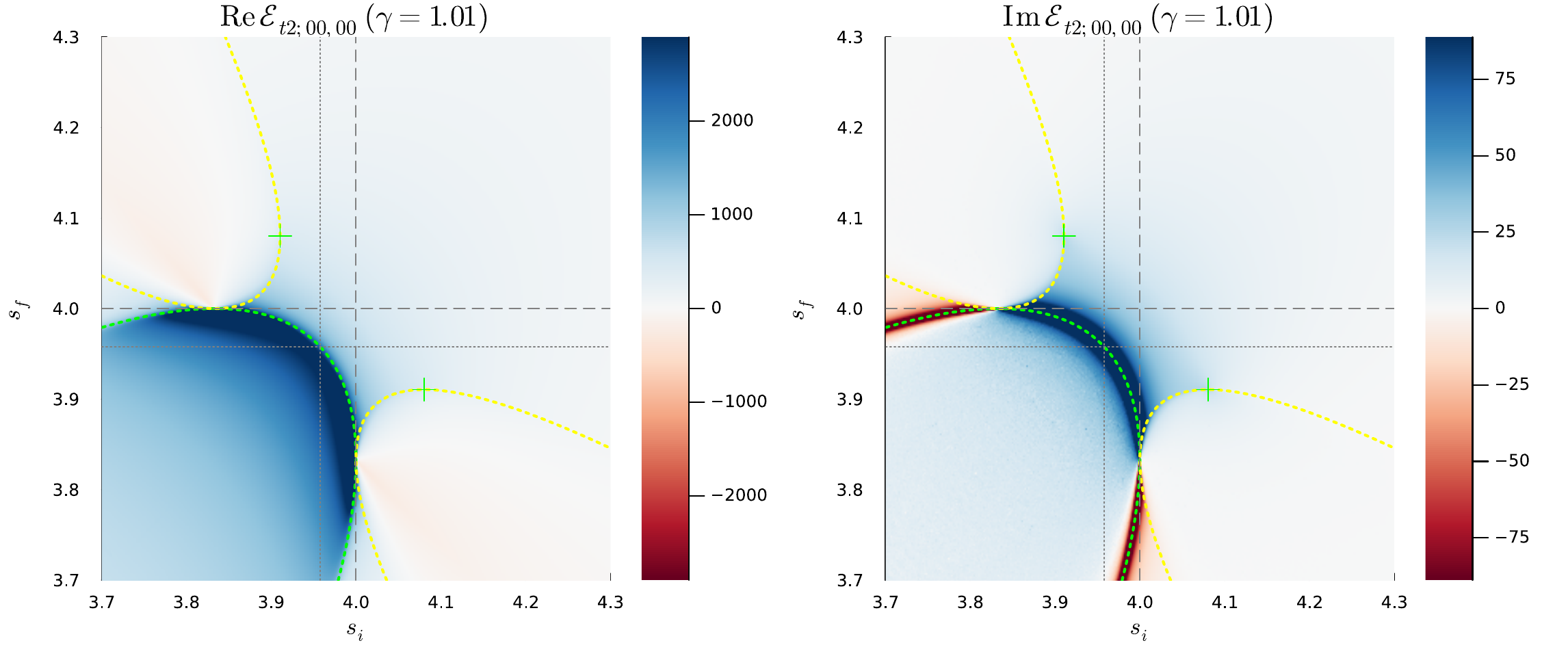}
    \end{subfigure}
    \begin{subfigure}[b]{\textwidth}
        \includegraphics[width=0.95\linewidth]{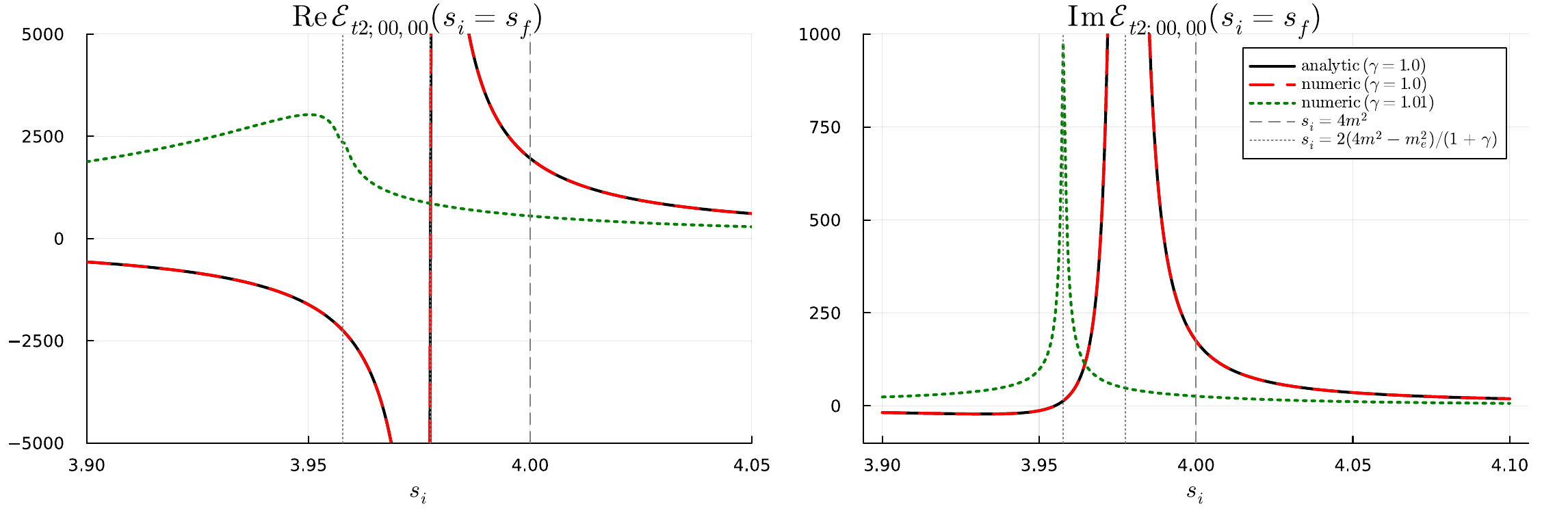}
    \end{subfigure}
    \caption{Like figure~\ref{fig:singularities_single} but for the Lorentz scalar $\mathcal{E}_{t2;00,00}$ amplitude.}
    \label{fig:singularities_double}
\end{figure}

The location and class of singularities are due to the pole singularities of the integrand of the partial-wave projected $\Ec^A$ functions.~\footnote{This is the same as in OPEs appearing in three-body processes, where the singularity of all partial wave OPEs can be written in terms of the Legendre Q function~\cite{Jackura:2023qtp}.} The pole of the integrand is independent of the angular momentum being considered. It is also independent of the Lorentz structure of the injected current. Consequently, the discussion of the singularities presented here applies broadly to the partial-wave projections of $\mathcal{E}^{A}$. As a point of reference, 
we may consider the simplest case of the S-wave-to-S-wave transition amplitude in the $t$ channel,
\begin{equation}
    i\mathcal{E}^{A}_{t;00,00} =
   \frac{1}{4\pi} \int d\Omega_f^\star d\Omega_i^\star 
    \, Y_{00}(\Omega_f^\star)Y^\star_{00}(\Omega_i^\star)
    i\mathcal{E}^{A}_{t}\,,
\end{equation}
where the additional subscript denotes $\ell m; \ell' m'$. This amplitude is computed explicitly for some special cases in Appendix~\ref{app:PW_2D}, and we cross-check our analytic results by also performing this projection numerically for toy values of the hadron masses. Heatmap and line plots of the results are shown in Figures~\ref{fig:singularities_single} and~\ref{fig:singularities_double}, which illustrate the singularities discussed below. In the figures, we label the single-pole contribution, $\Ec_{t1a}$, and the double-pole contribution $\Ec_{t2}$.~\footnote{The subscript $a$ refers to the fact that there are two single-pole contributions, and the second would be labeled $\Ec_{t1b}$. For identical particles, these two contributions are identical only after partial wave projection. } 

In the following, we distinguish cases where the current injects spatial momentum ($\gamma > 1$, $|\beta| > 0$) and a purely time-like current ($\gamma = 1$, $\beta = 0$).
Given the CMF momenta
\begin{align}
    q_{i}^\star &= \frac{\lambda^{1/2}(s_i, m_1^2 , m_2^2) }{2\sqrt{s_{i}}}\;, \hspace{3cm}
    q_{f}^\star = \frac{\lambda^{1/2}(s_f, m_3^2 , m_4^2) }{2\sqrt{s_{f}}}
\end{align}
and energies
\begin{align}
    \omega_{1}^\star &= \frac{s_i+m_1^2-m_2^2 }{2\sqrt{s_{i}}}\;, \hspace{3cm}
    \omega_{3}^\star = \frac{s_f+m_3^2-m_4^2 }{2\sqrt{s_{f}}} \;, \\
    \omega_{2}^\star &= \frac{s_i+m_2^2-m_1^2 }{2\sqrt{s_{i}}}\;, \hspace{3cm}
    \omega_{4}^\star = \frac{s_f+m_4^2-m_3^2 }{2\sqrt{s_{f}}} \;.
\end{align}
we find that $i\mathcal{E}^{A}_{t;lm,l'm'}$ has branch points on the contour
\begin{equation}
    \omega_{a}^\star \omega_{b}^\star \pm q_i^\star q_f^\star = \frac{1}{2} (m_a^2 + m_b^2 - m_e^2).
	\label{eq:main_betazero}
\end{equation}
when  $s_i$ and $s_f$ are below threshold and $\beta = 0$, where $(a,b) = (1,3)$ and $(2,4)$ respectively for the contributions by the two single-pole diagrams. This contour is marked in the top panels of Figure~\ref{fig:singularities_single} for the special case of degenerate particles. The standard t-channel threshold $s = 4m^2 - m_e^2$ is recovered on the line $s = s_i = s_f$ for degenerate masses. 

When $\beta > 0$, this branch point moves into the complex $(s_i,s_f)$ hyperplane, and the amplitude is therefore no longer singular at this point for real energies. However, when one channel lies above its two-particle threshold and the other lies below, distinct endpoint singularities can emerge on the real-energy plane. As shown in Appendix~\ref{app:sing}, the leading nonanalytic behavior near these points scales as $\sqrt{\delta_{t1a}}\log\delta_{t1a}$ (see Eq.~\eqref{eq:endpoint_log}). These singularities are also highlighted in Figure~\ref{fig:singularities_single}.

When initial and final-state particles are degenerate, the projected double-pole diagram has a pole singularity when $\beta = 0$ and a logarithmic singularity for $\beta\neq 0$. The location of these singularities, in general, satisfies
\begin{equation}
    \gamma \omega_i^\star \omega_f^\star \pm q_i^\star q_f^\star = \frac{1}{2} (m_i^2 + m_f^2 - m_e^2) \;,
\end{equation}
where we defined $m_i = m_1 = m_2$, $m_f = m_3 = m_4$.  
This singularity overlaps with the previously discussed branch point when $\beta = 0$. The contour can be seen in Figure~\ref{fig:singularities_double} in the top panels for the case of zero boost, and the middle panels for a small boost.
For degenerate masses, $m_i=m_f=m$, and equal initial and final energies, $s=s_i=s_f$, this condition becomes
\begin{equation}
    s=\frac{2 \left(4 m^2-m_e^2\right)}{\gamma+1}\,,
\end{equation}
which we show with a dotted vertical line in the bottom panels of Figs.~\ref{fig:singularities_single}~and~\ref{fig:singularities_double}.

Lastly there are isolated branch points at non-zero boosts given by
\begin{equation}
    \beta \omega^\star_{a} = q_i^\star \hspace{1cm} \text{and} \hspace{1cm} \beta \omega^\star_{b} = q_f^\star
\end{equation}
together with the condition
\begin{equation}
    \omega_{a}^\star \omega_{b}^\star = \frac{1}{2}\gamma (m_a^2 + m_b^2 - m_e^2) \;.
\end{equation}
In the case of degenerate particles this yields two mirror solutions $\left(s_i^{(1)}, s_f^{(1)}\right)$ and $\left(s_i^{(2)}, s_f^{(2)}\right)$ located at
\begin{equation}
	\begin{aligned}
		s_i^{(1)} = s_f^{(2)} &= \frac{1}{m^2}(2m^2 - m_e^2)^2 \;,\\
		s_f^{(1)} = s_i^{(2)} &=  4 \gamma^2 m^2 \;,
	\end{aligned}
    \label{eq:main_branchpoints}
\end{equation}
i.e., when one channel is above threshold, and the other is below threshold. These are visible in the middle-right panel of Figure~\ref{fig:singularities_single}.
Apart from singularities, the amplitudes have further regions of non-analytic behavior, which we discuss in more detail in the appendix. Some of these are marked in the aforementioned figures.

\section{Conclusion}

We have derived an infinite-volume representation for on-shell $2+\Jc\to2$ transition amplitudes in a kinematic region where left-hand singularities from light-particle exchange must be kept explicit.  The resulting expression separates the amplitude into four classes of singularities of increasing complexity. The first of these is an uninteresting single-particle pole, associated with the external particles being struck by the current. The second class consists of the square root threshold singularities associated with initial/final state re-scattering. The third contains triangular singularities. This class was previously identified in Ref.~\cite{Briceno:2020vgp}, but here we see the possible inclusion of two more kinds of triangle singularity due to the new types of vertices considered. The final type are possible pole singularities associated with the exchange particles coupling to the exchanged current.

The main takeaway from this formalism is that after the isolation of all these singularities, which are, in principle, tractable, all remaining unknown short-distance effects in the two-body-current sector can be collected into a single real function, $\Ac^A$. Although this function is not known, it could plausibly be determined via lattice QCD. This would require a parallel/complementary formalism to match the finite-volume matrix element determined via lattice QCD to the infinite-volume formalism presented here. Although rather technical, one can envision following steps similar to those presented here to study the finite-volume correlator as has been done in, for example, Refs.~\cite{Briceno:2015tza, Baroni:2018iau, Bernard:2012bi, Meissner:2026zos} for $2+\Jc\to 2$ processes where the left-hand cut effects were ignored. 

The present work is restricted to spinless external particles, a spinless exchange particle, and a single elastic two-particle channel. The extension to coupled two-body channels is not conceptually difficult: the quantities appearing here become matrices in channel space. Incorporating spin is more involved because the angular-momentum labels must be enlarged to include helicity or spin-coupled bases~\cite{Raposo:2025dkb}. Once these steps have been taken, along with the parallel necessary finite-volume formalism, one will have a clear pathway to access the structural information of near-threshold states whose dynamics are controlled by light-particle exchange, including deuteronlike systems~\cite{BaryonScattering:2025ziz, Horz:2020zvv, Amarasinghe:2021lqa, Detmold:2024iwz} and exotic hadron candidates such as the $T_{cc}$~\cite{Padmanath:2022cvl,Whyte:2024ihh,Lyu:2023xro, PitangaLachini:2026lyd}.

\section*{Acknowledgments}

ABR is supported by the European Research Council (ERC) consolidator grant StrangeScatt-101088506. FGO and RAB were partly supported by the U.S. Department of Energy, Office of Science, Office of Nuclear Physics under Award No. DE-SC0025665. RAB also acknowledges support from the U.S. Department of Energy, Office of Science, Office of Nuclear Physics under Award No. DE-AC02-05CH11231. NL acknowledges support by the EU H2020 research and innovation programme under the staff exchange grant agreement No-101086085-ASYMMETRY, as well as by the Spanish Ministerio de Ciencia e Innovacion project PID2020-113644GB-I00  and by Generalitat Valenciana through the grant PROMETEO/2019/083.

\bibliography{bib}

@article{Nagatsuka:2025szy,
    author = "Nagatsuka, Masato and Sasaki, Shoichi",
    title = "{Lattice study of scattering phase shifts for DD* and BB* systems using twisted boundary conditions: Search for bound state formation}",
    eprint = "2507.20712",
    archivePrefix = "arXiv",
    primaryClass = "hep-lat",
    doi = "10.1103/ksmm-8d7r",
    journal = "Phys. Rev. D",
    volume = "112",
    number = "11",
    pages = "114510",
    year = "2025"
}

@article{Abolnikov:2024key,
    author = "Abolnikov, Michael and Baru, Vadim and Epelbaum, Evgeny and Filin, Arseniy A. and Hanhart, Christoph and Meng, Lu",
    title = "{Internal structure of the Tcc(3875)+ from its light-quark mass dependence}",
    eprint = "2407.04649",
    archivePrefix = "arXiv",
    primaryClass = "hep-ph",
    doi = "10.1016/j.physletb.2024.139188",
    journal = "Phys. Lett. B",
    volume = "860",
    pages = "139188",
    year = "2025"
}

@article{PitangaLachini:2026lyd,
    author = "Pitanga Lachini, Nelson and Brice{\~n}o, Ra{\'u}l A.",
    title = "{Resolving the $T_{cc}^+$ with $π$ Exchange from Lattice QCD}",
    eprint = "2607.19009",
    archivePrefix = "arXiv",
    primaryClass = "hep-lat",
    month = "7",
    year = "2026"
}

@article{Landau:1959fi,
    author = "Landau, L. D.",
    editor = "ter Haar, D.",
    title = "{On the Analytic Properties of Vertex Parts in Quantum Field Theory}",
    doi = "10.1016/B978-0-08-010586-4.50103-6",
    journal = "Zh. Eksp. Teor. Fiz.",
    volume = "37",
    number = "1",
    pages = "62--70",
    year = "1960"
}

@article{Meissner:2026zos,
    author = "Mei{\ss}ner, Ulf-G. and Rusetsky, Akaki and Sakthivasan, Ajay S. and Schierholz, Gerrit and Wu, Jia-Jun",
    title = "{Form factors of the {\ensuremath{\rho}} meson from effective field theory and the lattice}",
    eprint = "2602.23044",
    archivePrefix = "arXiv",
    primaryClass = "hep-lat",
    reportNumber = "DESY-26-030",
    doi = "10.1007/JHEP06(2026)164",
    journal = "JHEP",
    volume = "06",
    pages = "164",
    year = "2026"
}

@article{Bernard:2012bi,
    author = "Bernard, V. and Hoja, D. and Meissner, U. G. and Rusetsky, A.",
    title = "{Matrix elements of unstable states}",
    eprint = "1205.4642",
    archivePrefix = "arXiv",
    primaryClass = "hep-lat",
    doi = "10.1007/JHEP09(2012)023",
    journal = "JHEP",
    volume = "09",
    pages = "023",
    year = "2012"
}

@article{Rodas:2023nec,
    author = "Rodas, Arkaitz and Dudek, Jozef J. and Edwards, Robert G.",
    collaboration = "Hadron Spectrum",
    title = "{Determination of crossing-symmetric {\ensuremath{\pi}}{\ensuremath{\pi}} scattering amplitudes and the quark mass evolution of the {\ensuremath{\sigma}} constrained by lattice QCD}",
    eprint = "2304.03762",
    archivePrefix = "arXiv",
    primaryClass = "hep-lat",
    reportNumber = "JLAB-THY-23-3791",
    doi = "10.1103/PhysRevD.109.034513",
    journal = "Phys. Rev. D",
    volume = "109",
    number = "3",
    pages = "034513",
    year = "2024"
}

@article{Woss:2020ayi,
    author = "Woss, Antoni J. and Dudek, Jozef J. and Edwards, Robert G. and Thomas, Christopher E. and Wilson, David J.",
    collaboration = "Hadron Spectrum",
    title = "{Decays of an exotic $1{-+}$ hybrid meson resonance in QCD}",
    eprint = "2009.10034",
    archivePrefix = "arXiv",
    primaryClass = "hep-lat",
    reportNumber = "JLAB-THY-20-3249",
    doi = "10.1103/PhysRevD.103.054502",
    journal = "Phys. Rev. D",
    volume = "103",
    number = "5",
    pages = "054502",
    year = "2021"
}

@article{Briceno:2017qmb,
    author = "Briceno, Raul A. and Dudek, Jozef J. and Edwards, Robert G. and Wilson, David J.",
    title = "{Isoscalar $\pi\pi, K\overline{K}, \eta\eta$ scattering and the $\sigma, f_0, f_2$ mesons from QCD}",
    eprint = "1708.06667",
    archivePrefix = "arXiv",
    primaryClass = "hep-lat",
    reportNumber = "JLAB-THY-17-2534",
    doi = "10.1103/PhysRevD.97.054513",
    journal = "Phys. Rev. D",
    volume = "97",
    number = "5",
    pages = "054513",
    year = "2018"
}

@article{Dudek:2016cru,
    author = "Dudek, Jozef J. and Edwards, Robert G. and Wilson, David J.",
    collaboration = "Hadron Spectrum",
    title = "{An $a_0$ resonance in strongly coupled $\pi \eta$, $K\overline{K}$ scattering from lattice QCD}",
    eprint = "1602.05122",
    archivePrefix = "arXiv",
    primaryClass = "hep-ph",
    reportNumber = "JLAB-THY-16-2216, DAMTP-2016-18",
    doi = "10.1103/PhysRevD.93.094506",
    journal = "Phys. Rev. D",
    volume = "93",
    number = "9",
    pages = "094506",
    year = "2016"
}

@article{BaryonScattering:2025ziz,
    author = "Bulava, John and others",
    collaboration = "Baryon Scattering",
    title = "{Di-nucleons do not form bound states at heavy pion mass}",
    eprint = "2505.05547",
    archivePrefix = "arXiv",
    primaryClass = "hep-lat",
    reportNumber = "LLNL-JRNL-2005660",
    doi = "10.1103/d2hg-h6d4",
    journal = "Phys. Rev. C",
    volume = "113",
    number = "2",
    pages = "024002",
    year = "2026"
}

@article{Horz:2020zvv,
    author = {H{\"o}rz, Ben and others},
    title = "{Two-nucleon S-wave interactions at the $SU(3)$ flavor-symmetric point with $m_{ud}\simeq m_s^{\rm phys}$: A first lattice QCD calculation with the stochastic Laplacian Heaviside method}",
    eprint = "2009.11825",
    archivePrefix = "arXiv",
    primaryClass = "hep-lat",
    reportNumber = "LLNL-JRNL-813871, RIKEN-iTHEMS-Report-20, MITP/20-055",
    doi = "10.1103/PhysRevC.103.014003",
    journal = "Phys. Rev. C",
    volume = "103",
    number = "1",
    pages = "014003",
    year = "2021"
}

@article{Amarasinghe:2021lqa,
    author = "Amarasinghe, Saman and Baghdadi, Riyadh and Davoudi, Zohreh and Detmold, William and Illa, Marc and Parreno, Assumpta and Pochinsky, Andrew V. and Shanahan, Phiala E. and Wagman, Michael L.",
    title = "{Variational study of two-nucleon systems with lattice QCD}",
    eprint = "2108.10835",
    archivePrefix = "arXiv",
    primaryClass = "hep-lat",
    reportNumber = "FERMILAB-PUB-21-354-T, MIT-CTP/5320, UMD-PP-021-06",
    doi = "10.1103/PhysRevD.107.094508",
    journal = "Phys. Rev. D",
    volume = "107",
    number = "9",
    pages = "094508",
    year = "2023",
    note = "[Erratum: Phys.Rev.D 110, 119904 (2024)]"
}

@article{Detmold:2024iwz,
    author = "Detmold, William and Illa, Marc and Jay, William I. and Parre{\~n}o, Assumpta and Perry, Robert J. and Shanahan, Phiala E. and Wagman, Michael L.",
    collaboration = "NPLQCD",
    title = "{Constraints on the finite volume two-nucleon spectrum at m{\ensuremath{\pi}}{\ensuremath{\approx}}806{\,}{\,}MeV}",
    eprint = "2404.12039",
    archivePrefix = "arXiv",
    primaryClass = "hep-lat",
    reportNumber = "FERMILAB-PUB-24-0126-T, MIT-CTP/5700",
    doi = "10.1103/PhysRevD.111.114501",
    journal = "Phys. Rev. D",
    volume = "111",
    number = "11",
    pages = "114501",
    year = "2025"
}

@article{Jackura:2023qtp,
    author = "Jackura, Andrew W. and Brice{\~n}o, Ra{\'u}l A.",
    title = "{Partial-wave projection of the one-particle exchange in three-body scattering amplitudes}",
    eprint = "2312.00625",
    archivePrefix = "arXiv",
    primaryClass = "hep-ph",
    doi = "10.1103/PhysRevD.109.096030",
    journal = "Phys. Rev. D",
    volume = "109",
    number = "9",
    pages = "096030",
    year = "2024"
}

@article{HadStruc:2021qdf,
    author = "Egerer, Colin and others",
    collaboration = "HadStruc",
    title = "{Transversity parton distribution function of the nucleon using the pseudodistribution approach}",
    eprint = "2111.01808",
    archivePrefix = "arXiv",
    primaryClass = "hep-lat",
    reportNumber = "JLAB-THY-21-3521",
    doi = "10.1103/PhysRevD.105.034507",
    journal = "Phys. Rev. D",
    volume = "105",
    number = "3",
    pages = "034507",
    year = "2022"
}

@article{HadStruc:2024rix,
    author = "Dutrieux, Herv{\'e} and Edwards, Robert G. and Egerer, Colin and Karpie, Joseph and Monahan, Christopher and Orginos, Kostas and Radyushkin, Anatoly and Richards, David and Romero, Eloy and Zafeiropoulos, Savvas",
    collaboration = "HadStruc",
    title = "{Towards unpolarized GPDs from pseudo-distributions}",
    eprint = "2405.10304",
    archivePrefix = "arXiv",
    primaryClass = "hep-lat",
    reportNumber = "JLAB-THY-24-4059, JLAB-THY-24-4059",
    doi = "10.1007/JHEP08(2024)162",
    journal = "JHEP",
    volume = "08",
    pages = "162",
    year = "2024"
}

@article{Lin:2021brq,
    author = "Lin, Huey-Wen",
    title = "{Nucleon helicity generalized parton distribution at physical pion mass from lattice QCD}",
    eprint = "2112.07519",
    archivePrefix = "arXiv",
    primaryClass = "hep-lat",
    reportNumber = "MSUHEP-21-024",
    doi = "10.1016/j.physletb.2021.136821",
    journal = "Phys. Lett. B",
    volume = "824",
    pages = "136821",
    year = "2022"
}

@article{Alexandrou:2019ali,
    author = "Alexandrou, C. and others",
    title = "{Moments of nucleon generalized parton distributions from lattice QCD simulations at physical pion mass}",
    eprint = "1908.10706",
    archivePrefix = "arXiv",
    primaryClass = "hep-lat",
    doi = "10.1103/PhysRevD.101.034519",
    journal = "Phys. Rev. D",
    volume = "101",
    number = "3",
    pages = "034519",
    year = "2020"
}

@article{Bhattacharya:2023ays,
    author = "Bhattacharya, Shohini and Cichy, Krzysztof and Constantinou, Martha and Gao, Xiang and Metz, Andreas and Miller, Joshua and Mukherjee, Swagato and Petreczky, Peter and Steffens, Fernanda and Zhao, Yong",
    title = "{Moments of proton GPDs from the OPE of nonlocal quark bilinears up to NNLO}",
    eprint = "2305.11117",
    archivePrefix = "arXiv",
    primaryClass = "hep-lat",
    doi = "10.1103/PhysRevD.108.014507",
    journal = "Phys. Rev. D",
    volume = "108",
    number = "1",
    pages = "014507",
    year = "2023"
}

@article{Bhattacharya:2024wtg,
    author = "Bhattacharya, Shohini and Cichy, Krzysztof and Constantinou, Martha and Gao, Xiang and Metz, Andreas and Miller, Joshua and Mukherjee, Swagato and Petreczky, Peter and Steffens, Fernanda and Zhao, Yong",
    title = "{Moments of axial-vector GPD from lattice QCD: quark helicity, orbital angular momentum, and spin-orbit correlation}",
    eprint = "2410.03539",
    archivePrefix = "arXiv",
    primaryClass = "hep-lat",
    reportNumber = "LA-UR-24-29020",
    doi = "10.1007/JHEP01(2025)146",
    journal = "JHEP",
    volume = "01",
    pages = "146",
    year = "2025"
}

@article{Chu:2025kew,
    author = "Chu, Min-Huan and Cola{\c{c}}o, Manuel and Bhattacharya, Shohini and Cichy, Krzysztof and Constantinou, Martha and Metz, Andreas and Steffens, Fernanda",
    title = "{Generalized parton distributions from lattice QCD with asymmetric momentum transfer: Unpolarized quarks at nonzero skewness}",
    eprint = "2508.17998",
    archivePrefix = "arXiv",
    primaryClass = "hep-lat",
    doi = "10.1103/ts5s-hvb1",
    journal = "Phys. Rev. D",
    volume = "112",
    number = "9",
    pages = "094510",
    year = "2025"
}

@article{Hackett:2023rif,
    author = "Hackett, Daniel C. and Pefkou, Dimitra A. and Shanahan, Phiala E.",
    title = "{Gravitational Form Factors of the Proton from Lattice QCD}",
    eprint = "2310.08484",
    archivePrefix = "arXiv",
    primaryClass = "hep-lat",
    reportNumber = "MIT-CTP/5630, FERMILAB-PUB-23-592-T",
    doi = "10.1103/PhysRevLett.132.251904",
    journal = "Phys. Rev. Lett.",
    volume = "132",
    number = "25",
    pages = "251904",
    year = "2024"
}

@article{Green:2021qol,
    author = "Green, Jeremy R. and Hanlon, Andrew D. and Junnarkar, Parikshit M. and Wittig, Hartmut",
    title = "{Weakly bound $H$ dibaryon from SU(3)-flavor-symmetric QCD}",
    eprint = "2103.01054",
    archivePrefix = "arXiv",
    primaryClass = "hep-lat",
    reportNumber = "MITP-21-009, CERN-TH-2021-024",
    doi = "10.1103/PhysRevLett.127.242003",
    journal = "Phys. Rev. Lett.",
    volume = "127",
    number = "24",
    pages = "242003",
    year = "2021"
}

@article{Dawid:2024oey,
    author = "Dawid, Sebastian M. and Jackura, Andrew W. and Szczepaniak, Adam P.",
    title = "{Finite-volume quantization condition from the N/D representation}",
    eprint = "2411.15730",
    archivePrefix = "arXiv",
    primaryClass = "hep-lat",
    reportNumber = "JLAB-THY-24-4216",
    doi = "10.1016/j.physletb.2025.139442",
    journal = "Phys. Lett. B",
    volume = "864",
    pages = "139442",
    year = "2025"
}

@article{Rodas:2026zmm,
    author = "Rodas, Arkaitz and Qiu, Lin and Fern{\'a}ndez-Ram{\'\i}rez, C{\'e}sar and Mathieu, Vincent and Monta{\~n}a, Gl{\`o}ria and Pilloni, Alessandro and Szczepaniak, Adam P.",
    title = "{Finite-volume analysis of the $H$-dibaryon including left-hand-cut effects}",
    eprint = "2605.22957",
    archivePrefix = "arXiv",
    primaryClass = "hep-lat",
    reportNumber = "JLAB-THY-26-4756",
    month = "5",
    year = "2026"
}

@article{Jackura:2020bsk,
    author = "Jackura, Andrew W. and Brice{\~n}o, Ra{\'u}l A. and Dawid, Sebastian M. and Islam, Md Habib E. and McCarty, Connor",
    title = "{Solving relativistic three-body integral equations in the presence of bound states}",
    eprint = "2010.09820",
    archivePrefix = "arXiv",
    primaryClass = "hep-lat",
    reportNumber = "JLAB-THY-20-3272",
    doi = "10.1103/PhysRevD.104.014507",
    journal = "Phys. Rev. D",
    volume = "104",
    number = "1",
    pages = "014507",
    year = "2021"
}

@article{Lebed:2016hpi,
    author = "Lebed, Richard F. and Mitchell, Ryan E. and Swanson, Eric S.",
    title = "{Heavy-Quark QCD Exotica}",
    eprint = "1610.04528",
    archivePrefix = "arXiv",
    primaryClass = "hep-ph",
    doi = "10.1016/j.ppnp.2016.11.003",
    journal = "Prog. Part. Nucl. Phys.",
    volume = "93",
    pages = "143--194",
    year = "2017"
}

@article{LHCb:2021auc,
    author = "Aaij, Roel and others",
    collaboration = "LHCb",
    title = "{Study of the doubly charmed tetraquark $T_{cc}^{+}$}",
    eprint = "2109.01056",
    archivePrefix = "arXiv",
    primaryClass = "hep-ex",
    reportNumber = "CERN-EP-2021-169, LHCb-PAPER-2021-032",
    doi = "10.1038/s41467-022-30206-w",
    journal = "Nature Commun.",
    volume = "13",
    number = "1",
    pages = "3351",
    year = "2022"
}

@article{LHCb:2021vvq,
    author = "Aaij, Roel and others",
    collaboration = "LHCb",
    title = "{Observation of an exotic narrow doubly charmed tetraquark}",
    eprint = "2109.01038",
    archivePrefix = "arXiv",
    primaryClass = "hep-ex",
    reportNumber = "CERN-EP-2021-165, LHCb-PAPER-2021-031",
    doi = "10.1038/s41567-022-01614-y",
    journal = "Nature Phys.",
    volume = "18",
    number = "7",
    pages = "751--754",
    year = "2022"
}

@article{Alexandrou:2018jbt,
    author = "Alexandrou, Constantia and Leskovec, Luka and Meinel, Stefan and Negele, John and Paul, Srijit and Petschlies, Marcus and Pochinsky, Andrew and Rendon, Gumaro and Syritsyn, Sergey",
    title = "{$\pi\gamma \to \pi\pi$ transition and the $\rho$ radiative decay width from lattice QCD}",
    eprint = "1807.08357",
    archivePrefix = "arXiv",
    primaryClass = "hep-lat",
    reportNumber = "RBRC-1289",
    doi = "10.1103/PhysRevD.98.074502",
    journal = "Phys. Rev. D",
    volume = "98",
    number = "7",
    pages = "074502",
    year = "2018",
    note = "[Erratum: Phys.Rev.D 105, 019902 (2022)]"
}

@article{Feng:2014gba,
    author = "Feng, Xu and Aoki, Sinya and Hashimoto, Shoji and Kaneko, Takashi",
    title = "{Timelike pion form factor in lattice QCD}",
    eprint = "1412.6319",
    archivePrefix = "arXiv",
    primaryClass = "hep-lat",
    reportNumber = "CU-TP-1206, KEK-CP-316, YITP-14-96",
    doi = "10.1103/PhysRevD.91.054504",
    journal = "Phys. Rev. D",
    volume = "91",
    number = "5",
    pages = "054504",
    year = "2015"
}

@article{Andersen:2018mau,
    author = {Andersen, Christian and Bulava, John and H{\"o}rz, Ben and Morningstar, Colin},
    title = "{The $I=1$ pion-pion scattering amplitude and timelike pion form factor from $N_{\rm f} = 2+1$ lattice QCD}",
    eprint = "1808.05007",
    archivePrefix = "arXiv",
    primaryClass = "hep-lat",
    reportNumber = "CP3-Origins-2018-029 DNRF90 MITP/18-073",
    doi = "10.1016/j.nuclphysb.2018.12.018",
    journal = "Nucl. Phys. B",
    volume = "939",
    pages = "145--173",
    year = "2019"
}

@article{Erben:2019nmx,
    author = "Erben, Felix and Green, Jeremy R. and Mohler, Daniel and Wittig, Hartmut",
    title = "{Rho resonance, timelike pion form factor, and implications for lattice studies of the hadronic vacuum polarization}",
    eprint = "1910.01083",
    archivePrefix = "arXiv",
    primaryClass = "hep-lat",
    reportNumber = "MITP/19-062, DESY-19-165",
    doi = "10.1103/PhysRevD.101.054504",
    journal = "Phys. Rev. D",
    volume = "101",
    number = "5",
    pages = "054504",
    year = "2020"
}

@article{Radhakrishnan:2022ubg,
    author = "Radhakrishnan, Archana and Dudek, Jozef J. and Edwards, Robert G.",
    collaboration = "Hadron Spectrum",
    title = "{Radiative decay of the resonant K* and the {\ensuremath{\gamma}}K{\textrightarrow}K{\ensuremath{\pi}} amplitude from lattice QCD}",
    eprint = "2208.13755",
    archivePrefix = "arXiv",
    primaryClass = "hep-lat",
    reportNumber = "JLAB-THY-22-3709",
    doi = "10.1103/PhysRevD.106.114513",
    journal = "Phys. Rev. D",
    volume = "106",
    number = "11",
    pages = "114513",
    year = "2022"
}

@article{Ortega-Gama:2024rqx,
    author = "Ortega-Gama, Felipe G. and Dudek, Jozef J. and Edwards, Robert G.",
    collaboration = "Hadron Spectrum",
    title = "{Timelike meson form factors beyond the elastic region from lattice QCD}",
    eprint = "2407.20617",
    archivePrefix = "arXiv",
    primaryClass = "hep-lat",
    reportNumber = "JLAB-THY-24-4132",
    doi = "10.1103/PhysRevD.110.094505",
    journal = "Phys. Rev. D",
    volume = "110",
    number = "9",
    pages = "094505",
    year = "2024"
}

@article{Padmanath:2022cvl,
    author = "Padmanath, M. and Prelovsek, S.",
    title = "{Signature of a Doubly Charm Tetraquark Pole in DD* Scattering on the Lattice}",
    eprint = "2202.10110",
    archivePrefix = "arXiv",
    primaryClass = "hep-lat",
    reportNumber = "MITP/22-018",
    doi = "10.1103/PhysRevLett.129.032002",
    journal = "Phys. Rev. Lett.",
    volume = "129",
    number = "3",
    pages = "032002",
    year = "2022"
}

@article{Du:2023hlu,
    author = "Du, Meng-Lin and Filin, Arseniy and Baru, Vadim and Dong, Xiang-Kun and Epelbaum, Evgeny and Guo, Feng-Kun and Hanhart, Christoph and Nefediev, Alexey and Nieves, Juan and Wang, Qian",
    title = "{Role of Left-Hand Cut Contributions on Pole Extractions from Lattice Data: Case Study for Tcc(3875)+}",
    eprint = "2303.09441",
    archivePrefix = "arXiv",
    primaryClass = "hep-ph",
    doi = "10.1103/PhysRevLett.131.131903",
    journal = "Phys. Rev. Lett.",
    volume = "131",
    number = "13",
    pages = "131903",
    year = "2023"
}

@article{Whyte:2024ihh,
    author = "Whyte, Travis and Wilson, David J. and Thomas, Christopher E.",
    collaboration = "Hadron Spectrum",
    title = "{Near-threshold states in coupled DD*-D*D* scattering from lattice QCD}",
    eprint = "2405.15741",
    archivePrefix = "arXiv",
    primaryClass = "hep-lat",
    doi = "10.1103/PhysRevD.111.034511",
    journal = "Phys. Rev. D",
    volume = "111",
    number = "3",
    pages = "034511",
    year = "2025"
}

@article{Lyu:2023xro,
    author = "Lyu, Yan and Aoki, Sinya and Doi, Takumi and Hatsuda, Tetsuo and Ikeda, Yoichi and Meng, Jie",
    title = "{Doubly Charmed Tetraquark Tcc+ from Lattice QCD near Physical Point}",
    eprint = "2302.04505",
    archivePrefix = "arXiv",
    primaryClass = "hep-lat",
    reportNumber = "RIKEN-iTHEMS-Report-23, YITP-23-14",
    doi = "10.1103/PhysRevLett.131.161901",
    journal = "Phys. Rev. Lett.",
    volume = "131",
    number = "16",
    pages = "161901",
    year = "2023"
}

@article{Hansen:2024ffk,
    author = "Hansen, Maxwell T. and Romero-L{\'o}pez, Fernando and Sharpe, Stephen R.",
    title = "{Incorporating DD{\ensuremath{\pi}} effects and left-hand cuts in lattice QCD studies of the T$_{cc}$(3875)$^{+}$}",
    eprint = "2401.06609",
    archivePrefix = "arXiv",
    primaryClass = "hep-lat",
    reportNumber = "MIT-CTP/5667",
    doi = "10.1007/JHEP06(2024)051",
    journal = "JHEP",
    volume = "06",
    pages = "051",
    year = "2024"
}

@article{Raposo:2023oru,
    author = "Bai\~{a}o-Raposo, Andr{\'e} and Hansen, Maxwell T.",
    title = "{Finite-volume scattering on the left-hand cut}",
    eprint = "2311.18793",
    archivePrefix = "arXiv",
    primaryClass = "hep-lat",
    doi = "10.1007/JHEP08(2024)075",
    journal = "JHEP",
    volume = "08",
    pages = "075",
    year = "2024"
}

@article{Lellouch:2000pv,
    author = "Lellouch, Laurent and Luscher, Martin",
    title = "{Weak transition matrix elements from finite volume correlation functions}",
    eprint = "hep-lat/0003023",
    archivePrefix = "arXiv",
    reportNumber = "CERN-TH-2000-091, CPT-2000-PE-3984, LAPTH-788-00",
    doi = "10.1007/s002200100410",
    journal = "Commun. Math. Phys.",
    volume = "219",
    pages = "31--44",
    year = "2001"
}

@article{Christ:2005gi,
    author = "Christ, Norman H. and Kim, Changhoan and Yamazaki, Takeshi",
    title = "{Finite volume corrections to the two-particle decay of states with non-zero momentum}",
    eprint = "hep-lat/0507009",
    archivePrefix = "arXiv",
    reportNumber = "RBRC-530, SHEP-0520, CU-TP-1131",
    doi = "10.1103/PhysRevD.72.114506",
    journal = "Phys. Rev. D",
    volume = "72",
    pages = "114506",
    year = "2005"
}

@article{Briceno:2014uqa,
    author = "Brice{\~n}o, Ra{\'u}l A. and Hansen, Maxwell T. and Walker-Loud, Andr{\'e}",
    title = "{Multichannel 1 $\rightarrow$ 2 transition amplitudes in a finite volume}",
    eprint = "1406.5965",
    archivePrefix = "arXiv",
    primaryClass = "hep-lat",
    reportNumber = "NT@WM-14-04, JLAB-THY-14-1909",
    doi = "10.1103/PhysRevD.91.034501",
    journal = "Phys. Rev. D",
    volume = "91",
    number = "3",
    pages = "034501",
    year = "2015"
}

@article{Briceno:2015csa,
    author = "Brice{\~n}o, Ra{\'u}l A. and Hansen, Maxwell T.",
    title = "{Multichannel 0 $\to$ 2 and 1 $\to$ 2 transition amplitudes for arbitrary spin particles in a finite volume}",
    eprint = "1502.04314",
    archivePrefix = "arXiv",
    primaryClass = "hep-lat",
    reportNumber = "JLAB-THY-15-2009",
    doi = "10.1103/PhysRevD.92.074509",
    journal = "Phys. Rev. D",
    volume = "92",
    number = "7",
    pages = "074509",
    year = "2015"
}

@article{Agadjanov:2016fbd,
    author = "Agadjanov, Andria and Bernard, V{\'e}ronique and Mei{\ss}ner, Ulf-G. and Rusetsky, Akaki",
    title = "{The $B\to K^*$ form factors on the lattice}",
    eprint = "1605.03386",
    archivePrefix = "arXiv",
    primaryClass = "hep-lat",
    doi = "10.1016/j.nuclphysb.2016.07.005",
    journal = "Nucl. Phys. B",
    volume = "910",
    pages = "387--409",
    year = "2016"
}

@article{Briceno:2016kkp,
    author = "Brice{\~n}o, Ra{\'u}l A. and Dudek, Jozef J. and Edwards, Robert G. and Shultz, Christian J. and Thomas, Christopher E. and Wilson, David J.",
    title = "{The $\pi\pi\to\pi\gamma^\star$ amplitude and the resonant $\rho\to\pi\gamma^\star$ transition from lattice QCD}",
    eprint = "1604.03530",
    archivePrefix = "arXiv",
    primaryClass = "hep-ph",
    reportNumber = "JLAB-THY-16-2237, DAMTP-2016-24",
    doi = "10.1103/PhysRevD.93.114508",
    journal = "Phys. Rev. D",
    volume = "93",
    number = "11",
    pages = "114508",
    year = "2016",
    note = "[Erratum: Phys.Rev.D 105, 079902 (2022)]"
}

@article{Moscoso:2026wmz,
    author = "Moscoso, Joseph and Ortega-Gama, Felipe G. and Brice{\~n}o, Ra{\'u}l A. and Jackura, Andrew W. and Kacir, Charles and Nicholson, Amy N.",
    title = "{Resolving the structure of bound states using lattice quantum field theories}",
    eprint = "2602.20373",
    archivePrefix = "arXiv",
    primaryClass = "hep-lat",
    month = "2",
    year = "2026"
}

@article{Wang:2026kab,
    author = "Wang, Zi-Yu and Feng, Xu and Jian, Bo-Hao and Jin, Lu-Chang and Liu, Chuan",
    title = "{First-Principles Determination of the Proton-Proton Fusion Matrix Element from Lattice QCD}",
    eprint = "2603.09554",
    archivePrefix = "arXiv",
    primaryClass = "hep-lat",
    month = "3",
    year = "2026"
}

@article{Vujmilovic:2025czt,
    author = "Vujmilovic, Ivan and Collins, Sara and Leskovec, Luka and Prelovsek, Sasa",
    title = "{Electromagnetic form factors and structure of the $T_{bb}$ tetraquark from lattice QCD}",
    eprint = "2510.17549",
    archivePrefix = "arXiv",
    primaryClass = "hep-lat",
    month = "10",
    year = "2025"
}

@article{Briceno:2019nns,
    author = "Brice{\~n}o, Ra{\'u}l A. and Hansen, Maxwell T. and Jackura, Andrew W.",
    title = "{Consistency checks for two-body finite-volume matrix elements: I. Conserved currents and bound states}",
    eprint = "1909.10357",
    archivePrefix = "arXiv",
    primaryClass = "hep-lat",
    reportNumber = "JLAB-THY-19-3040, CERN-TH-2019-149",
    doi = "10.1103/PhysRevD.100.114505",
    journal = "Phys. Rev. D",
    volume = "100",
    number = "11",
    pages = "114505",
    year = "2019"
}

@article{Briceno:2020xxs,
    author = "Brice{\~n}o, Ra{\'u}l A. and Hansen, Maxwell T. and Jackura, Andrew W.",
    title = "{Consistency checks for two-body finite-volume matrix elements: II. Perturbative systems}",
    eprint = "2002.00023",
    archivePrefix = "arXiv",
    primaryClass = "hep-lat",
    reportNumber = "JLAB-THY-19-3113, CERN-TH-2020-015",
    doi = "10.1103/PhysRevD.101.094508",
    journal = "Phys. Rev. D",
    volume = "101",
    number = "9",
    pages = "094508",
    year = "2020"
}

@article{Baroni:2018iau,
    author = "Baroni, Alessandro and Brice{\~n}o, Ra{\'u}l A. and Hansen, Maxwell T. and Ortega-Gama, Felipe G.",
    title = "{Form factors of two-hadron states from a covariant finite-volume formalism}",
    eprint = "1812.10504",
    archivePrefix = "arXiv",
    primaryClass = "hep-lat",
    reportNumber = "JLAB-THY-18-2878, CERN-TH-2018-263",
    doi = "10.1103/PhysRevD.100.034511",
    journal = "Phys. Rev. D",
    volume = "100",
    number = "3",
    pages = "034511",
    year = "2019"
}

@article{Briceno:2015tza,
    author = "Brice{\~n}o, Ra{\'u}l A. and Hansen, Maxwell T.",
    title = "{Relativistic, model-independent, multichannel $2\to 2$ transition amplitudes in a finite volume}",
    eprint = "1509.08507",
    archivePrefix = "arXiv",
    primaryClass = "hep-lat",
    reportNumber = "JLAB-THY-15-2140",
    doi = "10.1103/PhysRevD.94.013008",
    journal = "Phys. Rev. D",
    volume = "94",
    number = "1",
    pages = "013008",
    year = "2016"
}

@article{Raposo:2025dkb,
    author = "Raposo, Andr{\'e} Bai{\~a}o and Brice{\~n}o, Ra{\'u}l A. and Hansen, Maxwell T. and Jackura, Andrew W.",
    title = "{Extracting scattering amplitudes for arbitrary two-particle systems with one-particle left-hand cuts via lattice QCD}",
    eprint = "2502.19375",
    archivePrefix = "arXiv",
    primaryClass = "hep-lat",
    doi = "10.1007/JHEP06(2025)186",
    journal = "JHEP",
    volume = "06",
    pages = "186",
    year = "2025"
}

@article{Briceno:2020vgp,
    author = "Brice\~no, Ra\'ul A. and Jackura, Andrew W. and Ortega-Gama, Felipe G. and Sherman, Keegan H.",
    title = "{On-shell representations of two-body transition amplitudes: Single external current}",
    eprint = "2012.13338",
    archivePrefix = "arXiv",
    primaryClass = "hep-lat",
    reportNumber = "JLAB-THY-20-3298",
    doi = "10.1103/PhysRevD.103.114512",
    journal = "Phys. Rev. D",
    volume = "103",
    number = "11",
    pages = "114512",
    year = "2021"
}

@article{Dawid:2023jrj,
    author = "Dawid, Sebastian M. and Islam, Md Habib E. and Brice{\~n}o, Ra{\'u}l A.",
    title = "{Analytic continuation of the relativistic three-particle scattering amplitudes}",
    eprint = "2303.04394",
    archivePrefix = "arXiv",
    primaryClass = "nucl-th",
    doi = "10.1103/PhysRevD.108.034016",
    journal = "Phys. Rev. D",
    volume = "108",
    number = "3",
    pages = "034016",
    year = "2023"
}

@article{Francis:2024fwf,
    author = "Francis, Anthony",
    title = "{Lattice perspectives on doubly heavy tetraquarks}",
    eprint = "2502.04701",
    archivePrefix = "arXiv",
    primaryClass = "hep-lat",
    doi = "10.1016/j.ppnp.2024.104143",
    journal = "Prog. Part. Nucl. Phys.",
    volume = "140",
    pages = "104143",
    year = "2025"
}

@article{Brambilla:2019esw,
    author = "Brambilla, Nora and Eidelman, Simon and Hanhart, Christoph and Nefediev, Alexey and Shen, Cheng-Ping and Thomas, Christopher E. and Vairo, Antonio and Yuan, Chang-Zheng",
    title = "{The $XYZ$ states: experimental and theoretical status and perspectives}",
    eprint = "1907.07583",
    archivePrefix = "arXiv",
    primaryClass = "hep-ex",
    reportNumber = "TUM-EFT 125/19",
    doi = "10.1016/j.physrep.2020.05.001",
    journal = "Phys. Rept.",
    volume = "873",
    pages = "1--154",
    year = "2020"
}

@article{Meng:2023bmz,
    author = "Meng, Lu and Baru, Vadim and Epelbaum, Evgeny and Filin, Arseniy A. and Gasparyan, Ashot M.",
    title = "{Solving the left-hand cut problem in lattice QCD: Tcc(3875)+ from finite volume energy levels}",
    eprint = "2312.01930",
    archivePrefix = "arXiv",
    primaryClass = "hep-lat",
    doi = "10.1103/PhysRevD.109.L071506",
    journal = "Phys. Rev. D",
    volume = "109",
    number = "7",
    pages = "L071506",
    year = "2024"
}

@article{Dawid:2024dgy,
    author = "Dawid, Sebastian M. and Romero-L{\'o}pez, Fernando and Sharpe, Stephen R.",
    title = "{Finite- and infinite-volume study of DD{\ensuremath{\pi}} scattering}",
    eprint = "2409.17059",
    archivePrefix = "arXiv",
    primaryClass = "hep-lat",
    reportNumber = "MIT-CTP/5774",
    doi = "10.1007/JHEP01(2025)060",
    journal = "JHEP",
    volume = "01",
    pages = "060",
    year = "2025"
}

@article{Yu:2025gzg,
    author = "Yu, Kang and Wang, Guang-Juan and Wu, Jia-Jun and Yang, Zhi",
    title = "{Finite volume Hamiltonian method for two-particle systems containing long-range potential on the lattice}",
    eprint = "2502.05789",
    archivePrefix = "arXiv",
    primaryClass = "hep-lat",
    doi = "10.1007/JHEP04(2025)108",
    journal = "JHEP",
    volume = "04",
    pages = "108",
    year = "2025"
}

@article{Bubna:2024izx,
    author = {Bubna, Rishabh and Hammer, Hans-Werner and M{\"u}ller, Fabian and Pang, Jin-Yi and Rusetsky, Akaki and Wu, Jia-Jun},
    title = {{L{\"u}scher equation with long-range forces}},
    eprint = "2402.12985",
    archivePrefix = "arXiv",
    primaryClass = "hep-lat",
    doi = "10.1007/JHEP05(2024)168",
    journal = "JHEP",
    volume = "05",
    pages = "168",
    year = "2024"
}

@article{Abolnikov:2026sik,
    author = "Abolnikov, Michael and Meng, Lu and Baru, Vadim and Epelbaum, Evgeny and Filin, Arseniy A. and Gasparyan, Ashot M.",
    title = "{Shallow $T_{bc}$ states from an EFT analysis of $B^{(*)} \bar D^{(*)}$ scattering on the lattice}",
    eprint = "2602.02176",
    archivePrefix = "arXiv",
    primaryClass = "hep-ph",
    month = "2",
    year = "2026"
}

@inproceedings{Alharazin:2026lno,
    author = "Alharazin, Herzallah and Raposo, Andr{\'e} Bai{\~a}o and Bulava, John and Dawid, Sebastian and Green, Jeremy R. and Morningstar, Colin and Romero-L{\'o}pez, Fernando and Salg, Miguel and Sharpe, Stephen R. and Stump, Andres",
    title = "{Three-body study of the $T_{cc}(3875)^+$ from lattice QCD}",
    booktitle = "{42th International Symposium on Lattice Field Theory}",
    eprint = "2602.17204",
    archivePrefix = "arXiv",
    primaryClass = "hep-lat",
    reportNumber = "DESY-26-021, HU-EP-26/10-RTG",
    month = "2",
    year = "2026"
}

@article{Nicholson:2016byl,
    author = "Nicholson, Amy and Berkowitz, Evan and Chang, Chia Cheng and Clark, M. A. and Joo, Balint and Kurth, Thorsten and Rinaldi, Enrico and Tiburzi, Brian and Vranas, Pavlos and Walker-Loud, Andre",
    title = "{Neutrinoless double beta decay from lattice QCD}",
    eprint = "1608.04793",
    archivePrefix = "arXiv",
    primaryClass = "hep-lat",
    reportNumber = "LLNL-PROC-700398",
    doi = "10.22323/1.256.0017",
    journal = "PoS",
    volume = "LATTICE2016",
    pages = "017",
    year = "2016"
}

@article{Meng:2024kkp,
    author = "Meng, Lu and Ortiz-Pacheco, Emmanuel and Baru, Vadim and Epelbaum, Evgeny and Padmanath, M. and Prelovsek, Sasa",
    title = "{Doubly charm tetraquark channel with isospin 1 from lattice QCD}",
    eprint = "2411.06266",
    archivePrefix = "arXiv",
    primaryClass = "hep-lat",
    reportNumber = "MSUHEP-24-013",
    doi = "10.1103/PhysRevD.111.034509",
    journal = "Phys. Rev. D",
    volume = "111",
    number = "3",
    pages = "034509",
    year = "2025"
}

@article{Kaplan:1998sz,
    author = "Kaplan, David B. and Savage, Martin J. and Wise, Mark B.",
    title = "{A Perturbative calculation of the electromagnetic form-factors of the deuteron}",
    eprint = "nucl-th/9804032",
    archivePrefix = "arXiv",
    reportNumber = "DOE-ER-40561-2, NT-UW-98-04, CALT-68-2170",
    doi = "10.1103/PhysRevC.59.617",
    journal = "Phys. Rev. C",
    volume = "59",
    pages = "617--629",
    year = "1999"
}

@article{Park:1995pn,
    author = "Park, Tae-Sun and Min, Dong-Pil and Rho, Mannque",
    title = "{Chiral Lagrangian approach to exchange vector currents in nuclei}",
    eprint = "nucl-th/9505017",
    archivePrefix = "arXiv",
    reportNumber = "SNUTP-95-043",
    doi = "10.1016/0375-9474(95)00406-8",
    journal = "Nucl. Phys. A",
    volume = "596",
    pages = "515--552",
    year = "1996"
}

@article{Pastore:2011ip,
    author = "Pastore, S. and Girlanda, L. and Schiavilla, R. and Viviani, M.",
    title = "{The two-nucleon electromagnetic charge operator in chiral effective field theory ($\chi$EFT) up to one loop}",
    eprint = "1106.4539",
    archivePrefix = "arXiv",
    primaryClass = "nucl-th",
    reportNumber = "JLAB-THY-11-1383",
    doi = "10.1103/PhysRevC.84.024001",
    journal = "Phys. Rev. C",
    volume = "84",
    pages = "024001",
    year = "2011"
}

@article{Kolling:2011mt,
    author = "Kolling, S. and Epelbaum, E. and Krebs, H. and Meissner, U. -G.",
    title = "{Two-nucleon electromagnetic current in chiral effective field theory: One-pion exchange and short-range contributions}",
    eprint = "1107.0602",
    archivePrefix = "arXiv",
    primaryClass = "nucl-th",
    doi = "10.1103/PhysRevC.84.054008",
    journal = "Phys. Rev. C",
    volume = "84",
    pages = "054008",
    year = "2011"
}

\clearpage
\appendix

\section{
Singularities and partial-wave projections of $\Ec^A$
}
\label{app:PWA}

In this appendix, we discuss in more detail the partial-wave projection and corresponding singularities of the single- and double-pole diagrams with a current insertion, namely the diagrams contributing to $\Ec^A$ depicted in Fig.~\ref{fig:iEA}, which were previously discussed in Sec.~\ref{sec:Sing_ECA_v0}. In the partial-wave projections, we will restrict our attention to the $\ell=\ell'=0$ (S-wave) case, and we will only consider a scalar current. In most of the cases considered, we will further assume that the external particles are all degenerate for simplicity. Furthermore, we focus our attention on the $t$-channel contributions for these kernels. One can obtain the $u$-channel contributions using the same procedures outlined here.

All formulae below use the same convention as the main text: a superscript $\star_i$ ($\star_f$) denotes a quantity evaluated in the initial-state (final-state) CMF, while $q_i^\star$ and $q_f^\star$ denote the on-shell relative momenta defined by Eq.~\eqref{eq:qstar}.
First, we will describe our kinematic variables, then analyze the singularities, and finally discuss how to project the kernel onto the S-wave for the simplified case of identical external particles.

\subsection{Kinematics}

\begin{figure}
    \centering
    \includegraphics[width=0.75\linewidth]{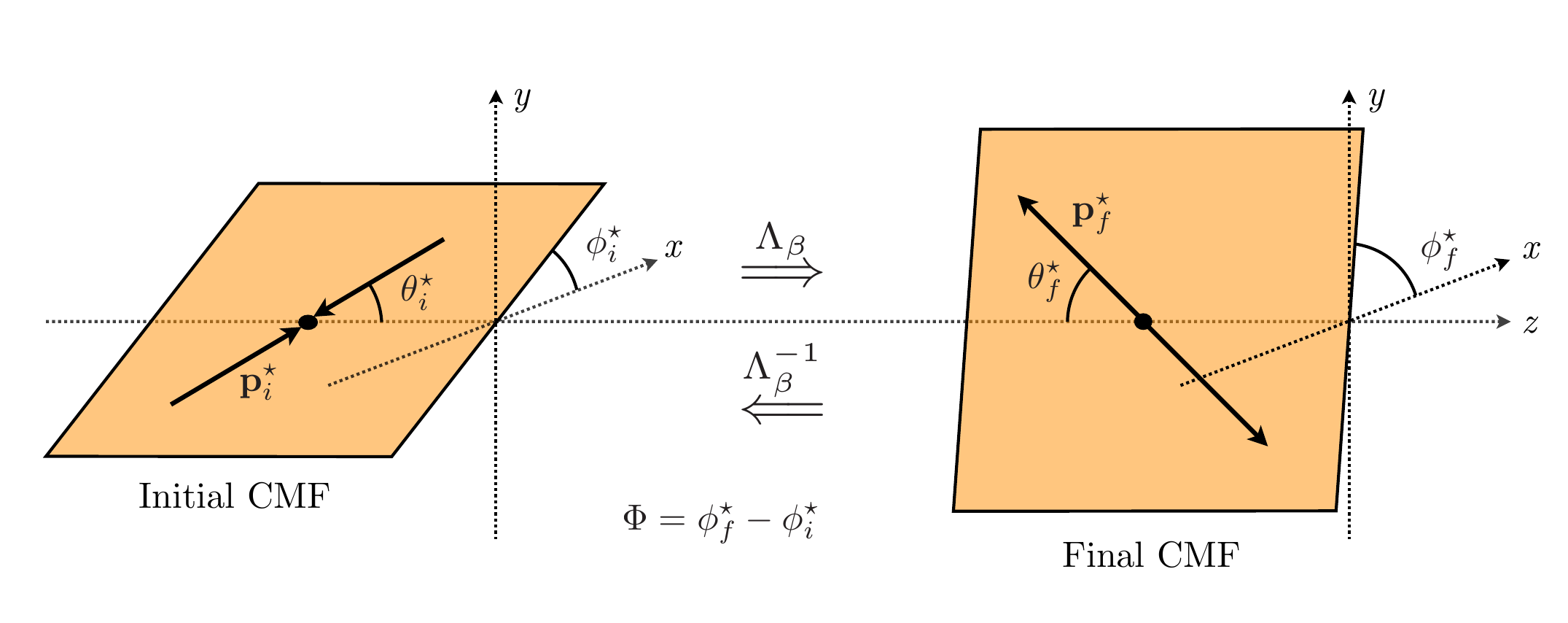}
    \caption{Kinematic variables defined in the initial and final CMF, respectively.}
    \label{fig:frames_diag}
\end{figure}

As illustrated in Fig.~\ref{fig:frames_diag}, the four-momenta in the initial CMF
are 
\begin{align}
p_{1,i}^{\mu\star}&=
\begin{pmatrix}
\omega_{1,i}^\star\\
\vb{p}_i^\star
\end{pmatrix},
\hspace{3cm}
p_{2,i}^{\mu,\star}=
\begin{pmatrix}\omega_{2,i}^\star\\-\vb{p}_i^\star
\end{pmatrix},
\end{align} 
where $\vb{p}_i^\star = q_i^\star\vu{p}_i^\star$ and 
\begin{align}
\vu{p}_i^\star = \begin{pmatrix}
\sin\theta^\star_i\cos\phi_i^\star\\ \sin\theta^\star_i\sin\phi_i^\star\\ \cos\theta^\star_i
\end{pmatrix}
.
\end{align}
The four-momenta in the final CM frame are 
\begin{align}
    p_{3,f}^{\mu\star}&=
    \begin{pmatrix}
    \omega_{3,f}^\star\\ \vb{p}_f^\star
    \end{pmatrix},
    \hspace{3cm}
    p_{4,f}^{\mu,\star}=
    \begin{pmatrix}\omega_{4,f}^\star\\
    -\vb{p}_f^\star
    \end{pmatrix},
\end{align}
where $\vb{p}_f^\star=q_f^\star\vu{p}_f^\star$ and, 
\begin{align}
\vu{p}_f^\star = \begin{pmatrix}
\sin\theta^\star_f\cos\phi_f^\star\\ \sin\theta^\star_f\sin\phi_f^\star\\
- \cos\theta^\star_f
\end{pmatrix}.
\end{align}
The minus sign in the last component is a convention. The angle $\theta_f^\star$ is defined so that particle~4 has positive $z$ component when $\cos\theta_f^\star>0$.  With this convention, the linear terms in $\cos\theta_i^\star$ and $\cos\theta_f^\star$ have the same sign in the exchange denominators below.
When the initial and final state particles are on shell, $p_a^2=m_a^2$, the zero-th component of the momenta are given by $\omega_{a,i/f}^\star=\sqrt{q_{i/f}^{\star2}+m_a^2}$. 

As done in Secs.~\ref{sec:triangles_2d} and \ref{sec:Sing_ECA_v0}, we choose the $z$-axis to be parallel to the boost relating the two frames defined by the boost vector $\vb*{\beta}=\beta \vu{z}$, and one can obtain the boost factor using 
\begin{align}
    \gamma = (1-\beta^{2})^{-1/2}= \frac{P_f\cdot P_i }{  \sqrt{s_fs_i}}=
\frac{s_f+s_i-q^2}{2\sqrt{s_fs_i}}.
\end{align}
The boost matrix $\Lambda_\beta$ acting on a four-momentum defined in the initial CMF gives its value on the final CMF, i.e. 
\begin{align}
   \Lambda_\beta\cdot(\sqrt{s_i},\vb{0})
   & = \begin{pmatrix} \gamma & 0 & 0 & \beta\gamma \\ 0 & 1 & 0 & 0 \\ 0 & 0 & 1 & 0 \\ \beta\gamma & 0 & 0 & \gamma \end{pmatrix} \cdot(\sqrt{s_i},\vb{0})
   \\
    &=
(\gamma\sqrt{s_i},0,0,\gamma\beta\sqrt{s_i})
    \\&\equiv(E_{i,f}^\star,\vb{P}_{i,f}^\star)
\end{align}
where $E_{i,f}^\star$ and $\vb{P}_{i,f}^\star$ are the energy and momentum of the initial state measured in the final CMF. 

For example, the value of $p_1$ and $p_2$ in the final CMF is given by
\begin{equation}
    p^{\mu\star}_{1,f}=
    \begin{pmatrix}
    \gamma \omega_{1,i}^\star + \gamma\beta\, q_i^\star\cos\theta_i^\star\\
    \vb{p}_{1,f}^\star
    \end{pmatrix}
\quad
    p^{\mu\star}_{2,f}=
    \begin{pmatrix}
    \gamma \omega_{2,i}^\star - \gamma\beta\, q_i^\star\cos\theta_i^\star\\
    \vb{p}_{2,f}^\star
    \end{pmatrix}
    \label{eq:apx_fourmom}
\end{equation}
where
\begin{equation}
    \vb{p}_{1,f}^\star = 
    \begin{pmatrix}
    q_i^\star\sin\theta^\star_i\cos\phi_i^\star \\
    q_i^\star\sin\theta^\star_i\sin\phi_i^\star\\
    \gamma \, q_i^\star\cos\theta_i^\star + \gamma\beta \omega_{1,i}^\star
    \end{pmatrix}\,,
\quad
    \vb{p}_{2,f}^\star = 
    \begin{pmatrix}
    -q_i^\star\sin\theta^\star_i\cos\phi_i^\star \\
    -q_i^\star\sin\theta^\star_i\sin\phi_i^\star\\
    -\gamma \, q_i^\star\cos\theta_i^\star +\gamma\beta \omega_{2,i}^\star
    \end{pmatrix}\,.
    \label{eq:apx_threemom}
\end{equation}
We will frequently need the dot products of four-momenta $p_1 \cdot p_3$ and $p_2 \cdot p_4$ expressed in terms of polar coordinates in the initial and final CMF. They are given by
\begin{align}
p_1\cdot p_3 &= \gamma \omega_{3,f}^\star
        \left(\omega_{1,i}^\star   +  \beta   q_i^\star \cos\theta_i^\star \right)
        - \vb{p}_{1,f}^\star \cdot \vb{p}_{f}^\star \\
        &= \gamma \omega_{1,i}^\star \omega_{3,f}^\star  + \gamma\beta (\omega_{3,f}^\star q_i^\star \cos\theta_i^\star + \omega_{1,i}^\star q_f^\star \cos\theta_f^\star) + \gamma q_i^\star  q_f^\star \cos\theta_i^\star  \cos\theta_f^\star - \vb{p}_{i,\perp}^\star \cdot \vb{p}_{f,\perp}^\star\,,
    \label{eq:p1p3_gen}
        \\
p_2\cdot p_4 &= \gamma \omega_{4,f}^\star\left(\omega_{2,i}^\star   -  \beta   q_i^\star \cos\theta_i^\star\right)
    + \vb{p}_{2,f}^\star \cdot \vb{p}_{f}^\star 
    \\
    &= \gamma \omega_{2,i}^\star \omega_{4,f}^\star  - \gamma\beta (\omega_{4,f}^\star q_i^\star \cos\theta_i^\star + \omega_{2,i}^\star q_f^\star \cos\theta_f^\star) + \gamma q_i^\star  q_f^\star \cos\theta_i^\star  \cos\theta_f^\star - \vb{p}_{i,\perp}^\star \cdot \vb{p}_{f,\perp}^\star\,,
    \label{eq:p2p4_gen}
\end{align}
with $\vb{p}_{i/f,\perp}^\star = \vb{p}_{i/f}^\star - \vb*{\beta}(\vb*{\beta}\cdot \vb{p}_{i/f}^\star)/\beta^2$ being the transverse components of $\vb{p}_{i/f}$ with respect to the boost. In our case, these are simply the first and second components of $\vb{p}_{i/f}^\star$.
It is straightforward to see that the dot product between the transverse components is
\begin{equation}
    \vb{p}_{i,\perp}^\star \cdot \vb{p}_{f,\perp}^\star
    = q_i^\star  q_f^\star \sin\theta_i^\star  \sin\theta_f^\star \cos(\phi_f^\star - \phi_i^\star).
\end{equation}

\subsection{Partial wave projection of $\Ec_{t1}$ and $\Ec_{t2}$}
\label{app:PW_2D}

In this section, we consider the simplest possible partial wave projection of $\mathcal{E}^{A}_{t1k}$, and $\mathcal{E}^{A}_{t2}$, namely the case where both the initial and final states have been projected to an S-wave. Furthermore, we will assume that the current is a Lorentz scalar that does not introduce any angular dependence. This simple example is still non-trivial because the initial and final states can, in general, be in different CMFs. Here we explain how the partial-wave projections can be reduced to 2D integrals, and give explicit results for their co-moving limits. 

\subsubsection{Projection of $\mathcal{E}^A_{t1k;00,00}$}

With this in mind, we can start with the definition of the $\ell=\ell'=0$ partial wave projection of $\mathcal{E}^{A}_{t1k}$, 
\begin{align}
    i\mathcal{E}^{A}_{t1k;00,00} &=
   \frac{1}{4\pi} \int d\Omega_f^\star d\Omega_i^\star 
    \, Y_{00}(\Omega_f^\star)Y^\star_{00}(\Omega_i^\star)
    i\mathcal{E}^{A}_{t1k}\,,
\end{align}
where $\mathcal{E}^{A}_{t1k}$ is either of the one-pole contributions of Eq.~\eqref{eq:Ec_At}.
Although this integral seemingly runs over four different angles, one of the azimuthal angular integrals is trivial. To make this more explicit, let us write the angular dependence of $\mathcal{E}^{A}_{t1k}$ more explicitly. This is all encoded in its denominator, which, using Eqs.~\eqref{eq:p1p3_gen} and~\eqref{eq:p2p4_gen} and $z^\star_{i,f} = \cos \theta^\star_{i,f}$, can be written as
\begin{equation}
    D_k(z_i^\star,z_f^\star,\Phi)=A_k(z_i^\star,z_f^\star)+B(z_i^\star,z_f^\star)\cos\Phi\,,\qquad k=1,2\,,
    \label{eq:app_denom}
\end{equation}
where $\Phi =\phi_f^\star - \phi_i^\star$. A simple variable transformation of $\phi_f^\star\to \phi_f^\star + \phi_i^\star $ gets rid of the $\phi_i^\star$ dependence, supporting the claim above. 
The rest of the building blocks of $D_k$ are defined as
\begin{align}
    A_1 &= m_1^2+m_3^2-m_e^2+i\epsilon
    -2\gamma\omega_{1,i}^\star\omega_{3,f}^\star
    -2\gamma\beta\left(\omega_{3,f}^\star q_i^\star z_i^\star+\omega_{1,i}^\star q_f^\star z_f^\star\right)
    -2\gamma q_i^\star q_f^\star z_i^\star z_f^\star\,, \label{eq:app_A1}\\
    A_2 &= m_2^2+m_4^2-m_e^2+i\epsilon
    -2\gamma\omega_{2,i}^\star\omega_{4,f}^\star
    +2\gamma\beta\left(\omega_{4,f}^\star q_i^\star z_i^\star+\omega_{2,i}^\star q_f^\star z_f^\star\right)
    -2\gamma q_i^\star q_f^\star z_i^\star z_f^\star\,, \label{eq:app_A2}\\
    B &= 2q_i^\star q_f^\star
    \sqrt{1-z_i^{\star2}}\sqrt{1-z_f^{\star2}}\,. \label{eq:app_B}
\end{align}
At this point, we can perform the non-trivial azimuthal integral using the identity 
\begin{align}
    \int_0^{2\pi}\frac{d\Phi}{A_k+B\cos\Phi}
    &=\frac{2\pi  \eta_k}{\sqrt{A_k^2-B^2}}
    =\frac{2\pi \eta_k}{S_k}\,,
    \label{eq:S_k_int}
    \\
S_k(z_i^\star,z_f^\star)&=\sqrt{A_k^2(z_i^\star,z_f^\star)-B^2(z_i^\star,z_f^\star)},
\end{align}
where 
\begin{equation}
    \eta_k  =\begin{cases}
+1, & \text{if } \left|\frac{A_k-S_k}{B}\right|  \le 1 \text{ and } \left|\frac{A_k+S_k}{B}\right|  > 1, \\
-1, & \text{if } \left|\frac{A_k-S_k}{B}\right|  > 1 \text{ and } \left|\frac{A_k+S_k}{B}\right|  \le 1 , \\
0, & \text{otherwise } .
\end{cases}
\end{equation}
We can obtain 
\begin{align}
    i\mathcal{E}^{A}_{t1a;00,00}
    &= -\frac{i g_{13} h^A_{24,\rm on}}{8\pi}
    \int dz_i^\star\int dz_f^\star
    \int_0^{2\pi}d\Phi\,
    \frac{1}{A_1+B\cos\Phi}\,
     =-\frac{i g_{13} h^A_{24,\rm on}}{4}
    \int dz_i^\star\int dz_f^\star
    \frac{\eta_1}{S_1(z_i^\star,z_f^\star)}\,,
    \label{eq:Et1a_phi_int}
\\
    i\mathcal{E}^{A}_{t1b;00,00}
    &= -\frac{i h^A_{13,\rm on} g_{24}}{8\pi}
    \int dz_i^\star\int dz_f^\star
    \int_0^{2\pi}d\Phi\,
    \frac{1}{A_2+B\cos\Phi}\,
    = -\frac{i h^A_{13,\rm on} g_{24}}{4}
    \int dz_i^\star\int dz_f^\star
    \frac{\eta_2}{S_2(z_i^\star,z_f^\star)}
\end{align}
These integrals are compact and can be evaluated numerically relatively quickly for arbitrary kinematics. 

\subsubsection{Projection of $\mathcal{E}^A_{t2;00,00}$}
The S-wave projection of $\mathcal{E}^A_{t2}$, the double-pole diagram, is slightly more involved since we have the product of two denominators, each depending on three angles. For simplicity, we assume a scalar current, $w^A_{e,{\rm on}}=j$, as a generic current would in general have dependence on the external momenta through its Lorentz decomposition, which could lead to additional angular dependence in the numerator. The partial wave projected $\mathcal{E}^A_{t2;00,00} $ can be written as
\begin{align} 
    i\mathcal{E}^A_{t2;00,00} &=
   \frac{1}{4\pi} \int d\Omega_f^\star d\Omega_i^\star 
    \, Y_{00}(\Omega_f^\star)Y^\star_{00}(\Omega_i^\star)
    i\mathcal{E}^A_{t2}\,
    \\
    &=\frac{i g_{13}g_{24}j}{8\pi}
    \int dz_i^\star\int dz_f^\star\,
    \int_0^{2\pi}d\Phi\,
    \frac{1}{(A_1+B\cos\Phi)(A_2+B\cos\Phi)}\,,
\end{align}
where we have performed one of the azimuthal integrals in the second equality. 

Here, we can evaluate the remaining $\Phi$ integral by considering two distinct scenarios. If $A_1\neq A_2$, then we can use the simple identity
\begin{equation}
    \frac{1}{(A_1+B\cos\Phi)(A_2+B\cos\Phi)}
    =
    \frac{1}{A_2-A_1}
    \left[
    \frac{1}{A_1+B\cos\Phi}
    -
    \frac{1}{A_2+B\cos\Phi}
    \right]\,,
\end{equation}
to perform the $\Phi$ integral 
\begin{equation}
    \int_0^{2\pi}d\Phi\,
    \frac{1}{(A_1+B\cos\Phi)(A_2+B\cos\Phi)}
    =
    \frac{2\pi}{A_2-A_1}
    \left(\frac{\eta_1}{S_1}-\frac{\eta_2}{S_2}\right)\,.
\end{equation}
This then results in $\mathcal{E}^{A}_{t2;00,00}$ being equal to
\begin{equation}
    i\mathcal{E}^{A}_{t2;00,00}
    =
    \frac{i g_{13}g_{24}j}{4}
    \int dz_i^\star\int dz_f^\star\,
    \frac{1}{A_2-A_1}
    \left(\frac{\eta_1}{S_1}-\frac{\eta_2}{S_2}\right)\,.
    \label{eq:Et2_phi_int}
\end{equation}

Alternatively, if $A_1=A_2$, we can write the integrand as
\begin{equation}
    \frac{1}{(A_1+B\cos\Phi)(A_2+B\cos\Phi)}
    =
    \frac{1}{(A_1+B\cos\Phi)^2}\,.
\end{equation}
We can then evaluate the corresponding integral by taking a derivative of Eq.~\eqref{eq:S_k_int} with respect to $A_1$ to find, 
\begin{equation}
    \int_0^{2\pi}d\Phi\,
    \frac{1}{(A_1+B\cos\Phi)^2}
    =
    \frac{2\pi \eta_1 A_1}{S_1^3}\,,
\end{equation}
so that in this case $\mathcal{E}^{A}_{t2;00,00}$ is equal to
\begin{equation}
    i\mathcal{E}^{A}_{t2;00,00}
    =
    \frac{i g_{13}g_{24}j}{4}
    \int dz_i^\star\int dz_f^\star\,
    \frac{\eta_1A_1}{S_1^3}\,.
\end{equation}
Again, we have reduced $\mathcal{E}^{A}_{t2;00,00}$ to a 2D integral. 

\subsubsection{Co-moving limit [$\beta =0$]}
A particularly simple limit occurs when $\beta=0$. In this case, the current injects no spatial momentum in the common CMF, and the denominators depend only on the relative angle
\begin{equation}
    z=\hat{\vb p}_i^\star\cdot\hat{\vb p}_f^\star\,.
\end{equation}
In this limit, rotational invariance ensures that 
\begin{equation}
    \frac{1}{(4\pi)^2}
    \int d\Omega_i^\star d\Omega_f^\star\,\Ec_t(z)
    =
    \frac{1}{2}\int_{-1}^{1}dz\,\Ec_t(z)\,,
\end{equation}
which holds for all types of $\Ec_t$ considered here.
If we write the single exchange denominator as $a_k+bz$, where 
\begin{align}
    a_1 &= m_1^2 + m_3^2 - m_e^2 -2\omega_{1}^\star \omega_{3}^\star \label{eq:app_a1}\\
    a_2 &= m_2^2 + m_4^2 - m_e^2 -2\omega_{2}^\star \omega_{4}^\star \label{eq:app_a2}\\
    b &= 2 {q_i^\star q_f^\star} \;,
    \label{eq:app_b}
\end{align}
we can then write the partial-wave projected single-pole diagrams as
\begin{align}
    i\mathcal{E}^{A}_{t1a;00,00}
    &=
    -\frac{i g_{13}h^A_{24,\rm on}}{2}
    \int_{-1}^{1}\frac{dz}{a_1+bz}
    =
    -\frac{i g_{13}h^A_{24,\rm on}}{4q_i^\star q_f^\star}
    \log\!\left(\frac{a_1+b}{a_1-b}\right)\,,
    \label{eq:app_et1a_beta0}
    \\
    i\mathcal{E}^{A}_{t1b;00,00}
    &=
    -\frac{i h^A_{13,\rm on}g_{24}}{2}
    \int_{-1}^{1}\frac{dz}{a_2+bz}
    =
    -\frac{i h^A_{13,\rm on}g_{24}}{4q_i^\star q_f^\star}
    \log\!\left(\frac{a_2+b}{a_2-b}\right)\,
    .
    \label{eq:app_et1b_beta0}
\end{align}
With identical masses and $s_i = s_f$ this reduces to the well-known result for the logarithmic left-hand cut, given by
\begin{align}
    i\mathcal{E}^A_{t1a;00,00} &= \frac{i g_{13} h^A_{24,\rm on} }{s-4m^2} \ln \Big[ 1 + \frac{s-4m^2}{m_e^2} - i\epsilon \Big],
\\
    i\mathcal{E}^A_{t1b;00,00} &= \frac{i g_{24} h^A_{13,\rm on} }{s-4m^2} \ln \Big[ 1 + \frac{s-4m^2}{m_e^2} - i\epsilon \Big],
\end{align}
where we have added back the $i \epsilon$.

For the double-pole diagram, one finds that in general
\begin{align}
    i\Ec^A_{t2;00,00}
    &=
    \frac{i g_{13}g_{24}j}{2}
    \int_{-1}^{1}\frac{dz}{(a_1+bz)(a_2+bz)}
    \notag\\
    &=
    \frac{i g_{13}g_{24}j}{2b(a_2-a_1)}
    \log\!\left[
    \frac{(a_1+b)(a_2-b)}
         {(a_1-b)(a_2+b)}
    \right]\,.
    \label{eq:app_et2_beta0_general}
\end{align}
Thus, in general $\Ec^A_{t2;00,00}$ has the same logarithmic branch points as the two single-pole denominators. 
In the limit that $a_1=a_2=a$, this simplifies to 
\begin{align}
    i\mathcal{E}^A_{t2;00,00}
    &=
    \frac{i g_{13}g_{24}j}{a^2-b^2}
\end{align}

Thus, in this special co-moving degenerate limit, $\mathcal{E}^{A}_{t2;00,00}$ has exchange poles instead of a logarithmic left-hand cut. 
In the case of degenerate masses, this pole can be expressed as
\begin{align}
 i\mathcal{E}^A_{t2;00,00} &=
    \frac{i g_{13}g_{24}j}
    {(2m^2-m_e^2+i\epsilon-2\omega_i^\star\omega_f^\star)^2
    -4{q_i^\star}^2{q_f^\star}^2}\,.
    \label{eq:app_et2_beta0}
\end{align}
If we further restrict ourselves to the  diagonal $s_i=s_f=s$, this further simplifies to,
\begin{align}
    i\mathcal{E}^A_{t2;00,00}(s,s,q^2=0)
    &=
    \frac{i g_{13}g_{24}j}{m_e^2(m_e^2+4{q^\star}^2+i\epsilon)}
    \\
    &=
    \frac{i g_{13}g_{24}j}{m_e^2(s+m_e^2-4m^2 +i\epsilon)}\,.
    \label{eq:res_pw_et2}
\end{align}
We see that in this case, the pole occurs at the same kinematic location as the branch point of the projected single-pole amplitude.

\subsection{Singularities}
\label{app:sing}
Whenever poles of the integrand appear within the integration region, the amplitudes are dominated by the leading singular term, which is independent of the behavior of the integrand numerator as long as it is smooth and non-zero.
That implies that the singularities in the scalar S-wave amplitude will in general also occur in other partial-wave projections and for currents with different Lorentz structures.
From the explicit calculations at $\beta = 0$ (Eqs.~\ref{eq:app_et1a_beta0}, \ref{eq:app_et1b_beta0} and \ref{eq:app_et2_beta0}) we can immediately read off the location of the singularities, which lie on a contour below the initial- and final-state thresholds, given by
\begin{equation}
    \omega_{i}^\star \omega_{f}^\star \pm q_i^\star q_f^\star = \frac{1}{2} (m_i^2 + m_f^2 - m_e^2).
	\label{eq:app_betazero}
\end{equation}
Here we have introduced $m_i = m_1,m_2$ and $m_f = m_3,m_4$ depending on the diagram. The projected diagrams $\mathcal{E}^A_{t1a;00,00}$ and $\mathcal{E}^A_{t1b;00,00}$ contribute a logarithmic branch point whereas $\mathcal{E}^A_{t2;00,00}$ has a simple pole. The latter occurs only when $m_1 = m_2$ and $m_3 = m_4$.
Note that the projected single-pole diagrams also exhibit a discontinuous sign-flip in the real part along the line
\begin{equation}
    m_i^2 + m_f^2 - m_e^2 = 2\omega_{i}^\star \omega_{f}^\star
\end{equation}
when $q_i^{\star2} < 0 < q_f^{\star2}$ or $q_f^{\star2} < 0 < q_i^{\star2}$. This follows from the argument of the logarithm in
\begin{equation}
    \mathcal{E}^A_{t1k;00,00} \sim -\frac{1}{b} \log \left( \frac{a+b}{a-b} \right) \;
\end{equation}
when $a$ vanishes and $b$ is purely imaginary (cf.\ Eqs.~\ref{eq:app_a1} and~\ref{eq:app_b}).
Lastly note that the apparent singularity at $q_i^{\star2} = q_f^{\star2} = 0$ in $\mathcal{E}^A_{t1k;00,00}$ is removed by the zero of the logarithm.

When $\beta \neq 0$ we do not in general have analytic results available for the projected amplitudes, even in the simplest $\ell = \ell' = 0$ case. However, we can still identify and classify the singularities based on an expansion around the zeros of the integrand denominator~\cite{Landau:1959fi}.
For this type of integral over a compact domain to be singular, the following conditions need to hold for the denominators
$D_k$ of the integrands defined in Eq.~\ref{eq:app_denom}:
\begin{enumerate}
	\item They need to vanish in the integration region,
	\begin{equation}
		\alpha_k D_k = 0 \; \text{ for $k = 1,2$},
	\end{equation}
    where $\alpha_k \ge 0$ is a real parameter.
	\item The resulting pole needs to get pinched by an \emph{integration endpoint}, in our case
	\begin{equation}
		\begin{aligned}
			z_i^\star &= \pm 1 \,,\\
			z_f^\star &= \pm 1 \,,
            \label{eq:app_endpoint}
		\end{aligned}
	\end{equation}
	or in the \emph{bulk},
    \begin{equation}
        \sum_k \alpha_k \nabla_{(z_i^\star, z_f^\star,\Phi)} D_k = \mathbf{0} \,.
        \label{eq:app_dDdPhi}
    \end{equation}
\end{enumerate}

End-point and bulk pinches can in principle mix, for instance a zero of the denominator in some integration variable might get constrained by an endpoint and for other variables by a vanishing derivative. 
Note that our integrands are periodic in $\Phi$ such that there is no end-point condition in this variable. 
Hence, singularities appear when $\partial D_k / \partial \Phi = 0$, requiring $\cos \Phi = \pm1$, which we substitute in all subsequent expressions.
The other two partial derivatives are then given by
\begin{equation}
    \frac{\partial D_k}{\partial z_i^\star} = -2\left[(-1)^{k-1}\gamma \beta \omega_{b,f}^\star q_i^\star + \gamma q_i^\star q_f^\star z_f^\star \pm q_i^\star q_f^\star z_i^\star \sqrt{(1 - z_f^{\star 2}) / (1-z_i^{\star2})}\right]
\end{equation}
and
\begin{equation}
    \frac{\partial D_k}{\partial z_f^\star} = -2\left[(-1)^{k-1}\gamma \beta \omega_{a,i}^\star q_f^\star + \gamma q_i^\star q_f^\star z_i^\star \pm q_i^\star q_f^\star z_f^\star \sqrt{(1-z_i^{\star2}) / (1 - z_f^{\star 2})}\right] \;.
\end{equation}

\subsubsection{Endpoint solutions}

The endpoint conditions~\eqref{eq:app_endpoint} immediately give $B(z_i^\star, z_f^\star) = 0$.
Note that at the corners of the integration region $D_1$ is obtained from $D_2$ simply by exchanging corners $(z_i^\star, z_f^\star) = (1, 1) \leftrightarrow (-1, -1)$ and $(1, -1) \leftrightarrow (-1, 1)$ if the masses are identical. Consequently, in the case of degenerate particles, $\mathcal{E}_{t1a}^A$ and $\mathcal{E}_{t1b}^A$ share the same endpoint solutions. The analytic behavior derived in this section applies to all three diagrams.

\emph{When $\beta \neq 0$} there are three terms in \eqref{eq:app_A1} and~\eqref{eq:app_A2} that can become purely imaginary, depending on whether $s_i$ and $s_f$ are below threshold separately or simultaneously. These need to cancel for the denominator to have a zero. All three have the same sign which means that all solutions have to come from the $z_i^\star = -z_f^\star = \pm1$ corners. It is easiest to separate cases by the kinematic region and solve the imaginary and real parts individually. 
Below the initial- and final-state thresholds, that is, $q_i^{\star2},q_f^{\star2} < 0$, the two terms proportional to $\gamma \beta$ are imaginary. A zero would require
\begin{align}
    2\gamma \left(\omega_i^\star \omega_f^\star - q_i^\star q_f^\star\right) &= m_i^2 + m_f^2 - m_e^2 \;,\\
    {\omega_i^\star}/{\omega_f^\star} &= {q_i^\star}/{q_f^\star} \;,
\end{align}
which for degenerate particles has a solution when $s_f=s_i$, and $\gamma = 1-m_e^2/(2m^2)$, which can only be true for $m_e=0$ as $\gamma\ge 1$.
Hence, for this case we conclude that there is no zero in $D_k$ at the endpoints for $\beta \neq 0$ when both Mandelstams are below threshold. In other words, for degenerate systems, where both channels are below threshold, $\Ec^A_{t1a}$ and $\Ec^A_{t1b}$ do not exhibit singularities.

When $q_i^{\star2} < 0 < q_f^{\star2}$ we do get a solution subject to the conditions
\begin{align}
    2\gamma \omega_i^\star \left( \omega_f^\star - \beta q_f^\star\right) &= m_i^2 + m_f^2 - m_e^2 \;,\\
    \beta {\omega_f^\star} &= {q_f^\star} \;,
    \label{eq:app_epsol1}
\end{align}
coming from the integration region near $z_i^\star = 1 = -z_f^\star$.
Analogously, for $q_f^{\star2} < 0 < q_i^{\star2}$, we get
\begin{align}
    2\gamma \omega_f^\star \left( \omega_i^\star - \beta q_i^\star\right) &= m_i^2 + m_f^2 - m_e^2 \,,\\
    \beta {\omega_i^\star} &= {q_i^\star} \,,
    \label{eq:app_epsol2}
\end{align}
where $z_i^\star = -1 = -z_f^\star$. When all masses are degenerate these solutions become 
\begin{align}
    s_i &= \frac{1}{m^2}\left( 2m^2 - m_e^2 \right)^2 \\
    s_f &= 4 \gamma^2 m^2
\end{align}
and analogously the mirrored one with $s_i$ and $s_f$ interchanged. These are marked in Figures~\ref{fig:singularities_single} and~\ref{fig:singularities_double} in the middle panels.

To understand the class of singularity, we compute the integral only in the region of the endpoints by expanding the denominator. We do this for $z_i = 1 = -z_f$. The other corner is completely analogous. Care must be taken with the derivatives of the square roots, which are not defined at the endpoints. 

Without loss of generality, let us consider the consequence of this analysis for $\Ec_{t1a}$. To simplify the derivation, we will first perform the integral over $\Phi$ analytically, resulting in Eq.~\eqref{eq:Et1a_phi_int}. Doing this, the quantity we need to study is $S_1^2(z_i^\star,z_f^\star) = A_1^2(z_i^\star,z_f^\star) - B^2(z_i^\star,z_f^\star)$, which now only depends on two kinematic variables. Note, $A_1$ and $B$  were previously defined in Eqs.~\eqref{eq:app_A1} and \eqref{eq:app_B}, respectively.

Following the arguments first laid out by Landau~\cite{Landau:1959fi}, our goal is to identify the scaling of the remaining integral with respect to
\begin{align}
\delta_{t1a}(s_i,s_f)
&\equiv S_1^2(1,-1)
=A_1^2(1,-1)
\notag\\
&=\big[
m_i^2+m_f^2-m_e^2+i\epsilon
-2\gamma\omega_i^\star\omega_f^\star
-2\gamma\beta
 \big(\omega_f^\star q_i^\star-\omega_i^\star q_f^\star\big)
+2\gamma q_i^\star q_f^\star
\big]^2 .
\label{eq:delta_t1a}
\end{align}
Here $m_i=m_1$ and $m_f=m_3$ for $\Ec_{t1a}$. The second equality follows because $B^2(1,-1)=0$. Thus $\delta_{t1a}$ vanishes at the endpoint singularity specified by Eq.~\eqref{eq:app_epsol1}.

With this in mind, we proceed to rewrite these quantities as follows,
\begin{align}
A_1(z_i^\star,z_f^\star)
&=A_1(1,-1)
-2\gamma q_i^\star
 \left(\beta\omega_f^\star-q_f^\star\right)(z_i^\star-1)
\nonumber\\
&\hspace{1cm}-2\gamma q_f^\star
 \left(\beta\omega_i^\star+q_i^\star\right)(z_f^\star+1)
-2\gamma q_i^\star q_f^\star
 (z_i^\star-1)(z_f^\star+1)\,,
\label{eq:A1_sing_v1}
\\
B^2(z_i^\star,z_f^\star)
&=4q_i^{\star2}q_f^{\star2}
 (1-z_i^{\star2})(1-z_f^{\star2}),
\end{align}
For $q_i^{\star 2}<0$, the endpoint singularity occurs at $
q_f^\star=\beta\omega_f^\star\,$, which allows us to simplify the expression for $A_1(z_i^\star,z_f^\star)$ further,
\begin{align}
A_1(z_i^\star,z_f^\star)
={}&A_1(1,-1)-a x-bxy\,,
\end{align}
where we have introduced the following variables, 
\begin{align}
x&=z_f^\star+1\,, &y=z_i^\star-1\,,
\\
a={}&2\gamma q_f^\star
 \left(\beta\omega_i^\star+q_i^\star\right)\,,\qquad
&b=2\gamma q_i^\star q_f^\star\,.
\label{eq:endpoint_ab}
\end{align}
Using these variables, and introducing $c =- 16q_i^{\star2}q_f^{\star2}$, we can rewrite $B^2(z_i^\star,z_f^\star)$ as,
\begin{equation}
B^2(z_i^\star,z_f^\star)
=\frac{c}{4}
 xy(2-x)(2+y)\,.
\end{equation}
With this, we find that 
\begin{align}
S_1^2(x,y)
&=\left(\sqrt{\delta_{t1a}}-ax-bxy\right)^2-cxy(2-x)(2+y)
\\
&=
\delta_{t1a} \times \left(
\left(1-a\frac{x}{\sqrt{\delta_{t1a}}}-b\frac{x}{\sqrt{\delta_{t1a}}}y\right)^2-c\frac{xy}{4{\delta_{t1a}}}(2-x)(2+y)
\right)
\\
&\equiv 
\delta_{t1a} \times \widetilde{S}_1^2(x,y)\,,
\label{eq:endpoint_S}
\end{align}
where in the last equality we have introduced a new function, $\widetilde{S}_1$. 
For this to scale as $\delta_{t1a}$ in the limit as $\delta_{t1a}\to0$, one has two options. One either scales $x$ by $\delta_{t1a}$, leaving $y$ unscaled, or scales both symmetrically by $\sqrt{\delta_{t1a}}$, i.e., $u=x/\sqrt{\delta_{t1a}}$ and $v=-y/\sqrt{\delta_{t1a}}$. This ensures that the new function is finite and non-zero at the endpoints. Either choice leads to the same conclusion. 

Using a symmetric scaling for both variables, we see that near the endpoints, we can simplify $\widetilde{S}_1^2$ as
\begin{align}
\widetilde{S}_1^2 \approx
(1-au)^2+cvu.
\end{align}
Let us define $u_0 = 2/\sqrt{\delta_{t1a}}$ and 
$v_0 = 2/\sqrt{\delta_{t1a}}$ as the cutoff of the integration needed to be evaluated below. With this, we can obtain the scaling of $\Ec_{t1a;00,00}^A$ by performing the integrals over these variables as follows
\begin{align}
\Ec_{t1a;00,00}^A
&\propto
\int dx\,dy\,\frac{1}{\sqrt{S_1^2(x,y)}}
\\
&\propto
\sqrt{\delta_{t1a}}\int^{u_0}_0 du\,
\int^{v_0}_0dv\, \frac{1}{\sqrt{(1-au)^2+cuv}}
\label{eq:endpoint_scaled_integral}
\\
&\propto
\sqrt{\delta_{t1a}}
\int^{u_0}_0 \frac{du}{u}\,
\left[
\sqrt{(1-au)^2+cuv_0}
-\sqrt{(1-au)^2}
\right].
\label{eq:endpoint_v_integral}
\end{align}

If we expand the integral for large $v_0$ and perform the integral over $u$, the leading-order term will be finite. 

The first square root in Eq.~\eqref{eq:endpoint_v_integral}
contributes a finite term as $\delta_{t1a}\to0$.
The leading nonanalytic contribution instead arises from the second
square root. For $u\gg 1/|a|$, its large-$u$ expansion is
\begin{equation}
\sqrt{(1-au)^2}
=
au-1+\mathcal O(u^{-1}).
\end{equation}
After including
the factor $1/u$ in Eq.~\eqref{eq:endpoint_v_integral}, the term
proportional to $a$ contributes only to the finite part of $\Ec_{t1a;00,00}^{A}$,
whereas the non-analytic term gives
\begin{align}
\Ec_{t1a;00,00}^{A,\mathrm{nonan}}
&\propto
\sqrt{\delta_{t1a}}
\int^{u_0}\frac{du}{u}
\notag\\
&\sim
\sqrt{\delta_{t1a}}\log u_0
\sim
-\frac{1}{2}
\sqrt{\delta_{t1a}}\log\delta_{t1a}\,,
\label{eq:endpoint_log}
\end{align}
where we used $u_0\propto1/\sqrt{\delta_{t1a}}$ in the final step.
As a result, we find that near this singularity
\begin{equation}
\Ec_{t1a;00,00}^{A}
=
C_1+C_2\sqrt{\delta_{t1a}}\log\delta_{t1a}
,\end{equation}
where the $C$'s are non-singular kinematic functions that do not vanish near the singularity. As a result, this function remains finite
at the endpoint, but it does have a branch cut.

\subsubsection{Bulk solutions}

\emph{At non-zero $\beta$}, the derivatives cannot vanish individually. However, in the double-pole amplitude, the linear combinations
\begin{align}
    \sum_k \alpha_k \frac{\partial D_k}{\partial z_i^\star} \Big|_{z_i^\star = z_f^\star = 0} &= -2\gamma \beta q_i^\star \left( \alpha_1 \omega_{3,f}^\star - \alpha_2 \omega_{4,f}^\star \right) \; , \\
    \sum_k \alpha_k \frac{\partial D_k}{\partial z_f^\star} \Big|_{z_i^\star = z_f^\star = 0} &= -2\gamma \beta q_f^\star \left( \alpha_1 \omega_{1,i}^\star - \alpha_2 \omega_{2,i}^\star \right) 
\end{align}
do vanish for $\alpha_2 / \alpha_1 = \omega_{3,f}^\star / \omega_{4,f}^\star  = \omega_{1,i}^\star / \omega_{2,i}^\star$. This is true when the initial and final-state particles are degenerate, that is, $m_1 = m_2 \equiv m_i$ and $m_3 = m_4 \equiv m_f$. In this case, the two denominators vanish simultaneously and the projected double-pole amplitude has a singularity on the contour
\begin{equation}\label{eq:bulk_sing}
    \gamma \omega_i^\star \omega_f^\star \pm q_i^\star q_f^\star = \frac{1}{2} (m_i^2 + m_f^2 - m_e^2) \;,
\end{equation}
where we defined $\omega_i^\star \equiv \omega_{1,i}^\star = \omega_{2,i}^\star$ and $\omega_f^\star \equiv \omega_{3,f}^\star = \omega_{4,f}^\star$.

To determine the class of this singularity, it is more useful to return to the original three-dimensional angular integral, before carrying out the $\Phi$ integration. This keeps the product of propagators explicit,
\begin{equation}
\Ec_{t2;00,00}^{A}
\propto
\int dz_i^\star\,dz_f^\star\,d\Phi\,
\frac{1}{D_1D_2}\,.
\label{eq:Et2_bulk_3D}
\end{equation}
Let $\Phi_0=0$ or $\pi$ denote the azimuthal location of the singularity, and define
\begin{equation}
\sigma\equiv\cos\Phi_0=\pm1\,,
\qquad
\overline{\Phi}\equiv\Phi-\Phi_0\,.
\end{equation}
For degenerate initial and final pairs, both denominators have the same value at the bulk point. We use this value as the distance from the singular contour,
\begin{align}
\delta_{t2}(s_i,s_f)
&\equiv D_1(0,0,\Phi_0)
=D_2(0,0,\Phi_0)
\notag\\
&=m_i^2+m_f^2-m_e^2+i\epsilon
-2\gamma\omega_i^\star\omega_f^\star
+2\sigma q_i^\star q_f^\star\,.
\label{eq:delta_t2}
\end{align}
Thus $\delta_{t2}=0$ gives us the location of the bulk singularity as a function of energies and $\gamma$, which corresponds to Eq.~\eqref{eq:bulk_sing} and coincides with Eq.~\eqref{eq:app_betazero} in the $\gamma\to 1$ limit.

Expanding the two $D_k$ around
$(z_i^\star,z_f^\star,\overline{\Phi})=(0,0,0)$ gives
\begin{align}
D_k
={}&\delta_{t2}
+(-1)^k\left(Bz_i^\star+Cz_f^\star\right)
-2\gamma q_i^\star q_f^\star z_i^\star z_f^\star
\notag\\
&-\sigma q_i^\star q_f^\star
\left(
z_i^{\star2}+z_f^{\star2}+\overline{\Phi}^{2}
\right)
+\mathcal O\!\left(
z^4,z^2\overline{\Phi}^{2},\overline{\Phi}^{\,4}
\right).
\label{eq:Dk_bulk_expansion}
\end{align}
where
\begin{equation}
  B=2\gamma\beta\omega_f^\star q_i^\star\,,
\qquad
  C=2\gamma\beta\omega_i^\star q_f^\star\,.
\end{equation}
With this in mind,  let us define a new set of variables that simplifies the product of $D_1$ and $D_2$. These new variables can be defined as,
\begin{equation}
u=  B z_i^\star+  C z_f^\star\,,
\qquad
v=-  C z_i^\star+  B z_f^\star\,,
\label{eq:Et2_uv}
\end{equation}
from which one finds, 
\begin{align}
\begin{pmatrix}z_i^\star\\ z_f^\star\end{pmatrix}
=\frac{1}{  B^2+  C^2}
\begin{pmatrix}  B&-  C\\
  C&  B\end{pmatrix}
\begin{pmatrix}u\\v\end{pmatrix}.
\end{align}
Using these new variables, the product of $D_1$ and $D_2$ simplifies to,
\begin{align}
D_1D_2
=&
\left[
\delta_{t2}
+c_{vv}v^2+c_{\Phi\Phi}\overline{\Phi}^{2}
+\mathcal O\left(u^2,uv,v^3,\overline{\Phi}^{\,4}\right)
\right]^2-u^2\,
\\
=&
\delta_{t2}^2
\left(
\left[1
+c_{vv}\frac{v^2}{\delta_{t2}}+c_{\Phi\Phi}\frac{\overline{\Phi}^{2}}{\delta_{t2}}
+\mathcal O\left(u^2,uv,v^3,\overline{\Phi}^{\,4}\right)
\right]^2-\frac{u^2}{\delta_{t2}^2}
\right)\,,
\\
&\equiv
\delta_{t2}^2 \widetilde{D}_{12}
\label{eq:D1D2_bulk}
\end{align}
where, 
\begin{equation}
c_{vv}
=q_i^\star q_f^\star
\left(\frac{2\gamma  B  C}{  N^2}
-\frac{\sigma}{  N}\right),
\qquad
c_{\Phi\Phi}=-\sigma q_i^\star q_f^\star\,,
\label{eq:Et2_bulk_coeffs}
\end{equation}
and with $  N=  B^2+  C^2$.

From Eq.~\eqref{eq:D1D2_bulk}, one can see that in order for $D_1D_2$ to scale as $\delta_{t2}^2$, the integration variables must scale as
\begin{equation}
u\sim\delta_{t2}\,,
\qquad
v\sim\sqrt{\delta_{t2}}\,,
\qquad
\overline{\Phi}\sim\sqrt{\delta_{t2}}\,.
\label{eq:Et2_bulk_scaling}
\end{equation}
With this in mind, we introduce
\begin{align}
u\equiv \widetilde{u}
\delta_{t2}\,,
\qquad
v\equiv\widetilde{v}
\sqrt{\frac{\delta_{t2}}{c_{vv}}}\,,
\qquad
\overline{\Phi}\equiv
\widetilde{\Phi}
\sqrt{\frac{\delta_{t2}}{c_{\Phi\Phi}}}
\,.
\end{align}
Putting all of the pieces together,
we find that the $\Ec_{t2}$ integral in the vicinity of the singularity as 
\begin{align}
 \Ec_{t2;00,00}^A
&\propto
\int du\,dv\,d\overline{\Phi}
\frac{1}{\delta_{t2}^2 \widetilde{D}_{12}}
\\
&\propto
\frac{\delta_{t2}^2}{\delta_{t2}^2}
\int d\widetilde{u}\,d\widetilde{v}\,d\widetilde{\Phi}
\frac{1}{\widetilde{D}_{12}}
\\
&\propto
\int 
\frac{\,d\widetilde{v}\,d\widetilde{\Phi}}{1
+\widetilde{v}^2+
\widetilde{\Phi}^{2}}.
\end{align}
where in the last line we carried out the $\widetilde u$ integration
and kept only the part relevant to the singular scaling. Introducing
polar coordinates,
\begin{equation}
\widetilde{ v}=R\cos\theta\,,
\qquad
\widetilde{\Phi}=R\sin\theta\,,
\end{equation}
we obtain
\begin{align}
\Ec_{t2;00,00}^{A}
&\propto
\int_0^{\#/ \sqrt{\delta_{t2}}}
\frac{R\,dR}{1+R^2}
\nonumber\\
&=
\frac{1}{2}
\log\left(
1+\frac{\#}{\delta_{t2}}
\right)
\sim
-\frac{1}{2}\log\delta_{t2}\,.
\label{eq:bulk_log}
\end{align}

Therefore, a generic bulk solution at nonzero $\beta$ produces a
logarithmic branch point in $\Ec_{t2;00,00}^{A}$.

\end{document}